\documentclass[preprint2]{aastex}

\newcommand{\HII}{H {\scshape{ii}} }
\newcommand{\OII}{[O {\scshape{ii}}] }
\newcommand{\OIII}{[O {\scshape{iii}}] }
\newcommand{\SII}{[S {\scshape{ii}}] }
\newcommand{\NII}{[N {\scshape{ii}}] }
\newcommand{\Ha}{H$\alpha$ }
\newcommand{\hh}{$^{\mathrm{h}}$}
\newcommand{\mm}{$^{\mathrm{m}}$}
\newcommand{\scn}{$^{\mathrm{s}}$}
\newcommand{\dd}{$^{\mathrm{d}}$}
\newcommand{\B}{\mathrm{M_B}}

\slugcomment{Not to appear in Nonlearned J., 45.}
\shorttitle{Spectroscopy of SF galaxies in the Hercules cluster}
\shortauthors{Petropoulou et al.}
\usepackage{pifont}

\begin{document}

\title{Spatially resolved spectroscopy and chemical history of star-forming galaxies in the Hercules cluster: the effects of the environment}

\author{Petropoulou V.\altaffilmark{1}, V\'ilchez J.\altaffilmark{1}, Iglesias-P\'aramo J.\altaffilmark{1,2}, Papaderos P.\altaffilmark{3},  Magrini L.\altaffilmark{4}, Cedr\'es B.\altaffilmark{1,5}, Reverte D.\altaffilmark{6}}
\affil{\altaffilmark{1}Instituto de Astrof\'isica de Andaluc\'ia- C.S.I.C., Glorieta de la Astronom\'ia, 18008 Granada, Spain;\\
\altaffilmark{2}Centro Astron\'omico Hispano Alem\'an, C/ Jes\'us Durb\'an Rem\'on 2-2, 04004 Almer\'ia, Spain;\\
\altaffilmark{3}Centro de Astrof\'isica da Universidade do Porto, Rua das Estrelas, 4150-762 Porto, Portugal;\\
\altaffilmark{4}INAF - Osservatorio Astrofisico di Arcetri, Largo E. Fermi 5, 50125 Firenze, Italy;\\
\altaffilmark{5}Instituto de Astrof\'isica de Canarias, C/ V\'ia L\'actea s/n, 38205  La Laguna (Tenerife), Spain;\\
\altaffilmark{6} GRANTECAN S.A., Centro de Astrof\'isica de La Palma, C/ Cuesta de San Jos\'e s/n, 38712 Bre\~na Baja (La Palma), Spain}

\begin{abstract}

Spatially resolved spectroscopy has been obtained for a sample of 27 star-forming (SF) galaxies selected from our deep H$\alpha$ survey of the Hercules cluster. We have applied spectral synthesis models to all emission-line spectra of this sample using the population synthesis code STARLIGHT and we have obtained fundamental parameters of the stellar components, as the mean metallicity and age. The emission-line spectra were corrected for underlying stellar absorption using these spectral synthesis models. Line fluxes were measured and O/H and N/O gas chemical abundances were obtained using the latest empirical calibrations. We have derived masses and total luminosities of the galaxies using available SDSS broadband photometry. The effects of cluster environment on the chemical evolution of galaxies and on their mass-metallicity (MZ) and luminosity-metallicity (LZ) relations were studied combining the derived gas metallicities, the mean stellar metallicities and ages, the masses and luminosities of galaxies and their existing HI data. We have found that our Hercules SF galaxies divide into three main subgroups: a) chemically evolved spirals with truncated ionized-gas disks and nearly flat oxygen gradients, witnessing the effect of ram-pressure stripping,  b) chemically evolved dwarfs/irregulars populating the highest local densities, possible products of tidal interactions in preprocessing events, or c) less metallic dwarf galaxies which appear to be  ``newcomers'' to the cluster, experiencing pressure-triggered star-formation. Most Hercules SF galaxies follow well defined MZ and LZ sequences (for both O/H and N/O); though the dwarf/irregular galaxies located at the densest regions appear to be  outliers to these global relations, suggesting a physical reason for the dispersion in these fundamental relations. The Hercules cluster appears to be currently assembling via the merger of smaller substructures, providing an ideal laboratory where the local environment has been found to be a key parameter to understand the chemical history of galaxies.

\end{abstract}


\section{INTRODUCTION}\label{INTRO}

The star formation history (SFH), gas-content and the mass exchange with the environment (infall of metal-poor gas and/or outflow of enriched material) are  fundamental variables  that regulate the chemical evolution of a galaxy. Clusters of galaxies host large numbers of galaxies of various sizes, luminosities and morphologies as well as a large mass of gas confined within a given volume of space. Thus galaxy clusters can provide an excellent place to study the impact of the environment on the aforementioned fundamental variables regulating the metal content of a galaxy. 

The relation between stellar mass and metallicity is fundamental  to constrain galaxy evolution scenarios in dense environments.  From the observational point of view, \citet{Lequeux1979} was the first to point out that metallicity is strongly correlated with galaxy stellar mass or luminosity, with the more massive galaxies being metal richer. Since then a lot of work has been done, extending these correlations to large samples of galaxies \citep[][among others]{Skillman1989, Zaritsky1994, Lamareille2004}. This holds at all redshifts, although for a given stellar mass, more distant galaxies appear to have a lower metal content than local galaxies \citep[e.g][]{Savaglio2005, Erb2006}. \citet{Maiolino2008} argue that both the zero point and the slope of the MZ relation are a function of redshift.

Many different scenarios have been proposed for the physical mechanisms underlying such a relation. According to \citet{Finlator2008}, the chemical evolution of galaxies is governed by momentum-driven gas outflows, which are more efficient in expelling metal-enriched gas in dwarf galaxies than in giant galaxies. \citet{Spitoni2010} also support the ejection scenario caused by galactic winds. \citet{Koppen2007} claim that the MZ relation is linked to the initial mass function, which in turn  presents a higher upper mass cutoff in more massive galaxies. \citet{Calura2009} advocate that the MZ relation is intrinsically related to the mass and morphological type of the galaxies. They support an increasing efficiency of star formation with mass without any need to invoke galactic outflows. The inflows scenario, whether smooth or in the form of gas-rich satellites, has been explored by \citet{Dave2010}.

In this context, several cluster-related environmental processes can affect the star formation history or the gas exchange between the galaxy and its environment and, as a consequence, could alter the chemical evolution of cluster galaxies.
Some processes  affect mainly the gaseous content of a galaxy, such as the ram-pressure stripping \citep{Gunn1972, Kenney2004, vanGorkom2004},
re-accretion of the stripped gas \citep{Vollmer2001}, turbulence and viscosity \citep[e.g.][]{Quilis2000} and starvation/strangulation \citep{Larson1980}. 
Other processes can affect both the gaseous and the stellar properties of a galaxy, ranging from low-velocity tidal interactions and mergers \citep[e.g.][]{Mamon1996, Barnes1996, Conselice2006} to high-velocity interactions between galaxies and/or clusters \citep[``harassment'',][]{Moore1998, Moore1999, Struck1999, Mihos2004}; see  \citet{Boselli2006} for an extensive review on this subject. 

A number of works have been devoted to searching for the possible effects of cluster environment on the chemical properties of spiral galaxies in Virgo cluster \citep{Shields1991, Henry1992, Henry1994, Henry1996, Skillman1996, Pilyugin2002}. The spirals at the periphery of the cluster are indistinguishable from the field galaxies, whereas for the HI deficient Virgo spirals near the core it has been claimed that their abundances are higher than for the spirals located at the periphery of the cluster or in the field with comparable luminosity or Hubble type. On the other hand, several works devoted to the study of the star-formation and the metal content of dwarfs in nearby clusters such as Virgo, Coma, and Hydra have revealed that some of them show a different behavior with respect to field galaxies, and this fact has been attributed to the effect of the cluster environment \citep{Vilchez1995, Duc2001, Iglesias2003, Vilchez2003, Lee2003,  Vaduvescu2007, Smith2009, Mahajan2010}.

In the era of the large surveys, observational clues which constrain  galaxy evolution scenarios in dense environments have lately started to be investigated on the basis of large samples of galaxies.  \citet{Mouhcine2007} suggested that the evolution of SF galaxies is driven primarily by their intrinsic properties and is largely independent of their environment over a large range of local galaxy density. On the contrary,  \citet{Cooper2008} find a strong relationship between metallicity and environment, such that more metal-rich galaxies favor regions of higher overdensity. \citet{Ellison2009} also support the environmental influence, and specially of the local density, on the MZ relation. It is clear that a full  understanding of the role played by the environment on galaxy evolution still remains a major issue. All these works based on SDSS spectroscopic data present a deficit of important observational information on dwarf cluster galaxies. In addition, SDSS provides aperture spectroscopy, thus missing spatially resolved information that can be much needed, e.g. to analyze the spatial properties of galaxies in clusters.    

In this paper we focus our attention on the nearby Hercules (A2151) cluster of galaxies and we investigate the impact of the cluster environment on the chemical history of galaxies. The Hercules cluster shows an amazing variety of different environments found in different parts of the cluster, which makes it one of the more interesting nearby (at a distance of 158.3 Mpc) dense environments, never studied before from the chemical evolution point of view. It is the richest and densest among the three members of the Hercules supercluster  \citep[A2151, A2147, and A2152,][]{Barmby1998} and appears to be a collection of groups, gravitationally bound but far from dynamical relaxation \citep{Bird1995,Maccagni1995}. \citet{Sanchez2005} report the presence of at least three distinct regions with a varying dwarf to giant galaxy ratio within the A2151 cluster.  \citet{Giovanelli1981} studied the supercluster neutral hydrogen content and found that A2147 is clearly deficient in contrast with A2151 which shows normal to mildly gas deficient galaxy population, concluding that A2151 may still be in the process of formation.    \citet{Dickey1997} reported the presence of strong environmental effects on the mass of HI in the A2151 cluster.  High-Resolution ROSAT data \citep{Huang1996, Bird1995} reveal an irregular X-ray emission pattern which divides this cluster into two main components, which correspond to two groups of galaxies in the region. The maximum of the cluster galaxy distribution is not coincident with the primary X-ray maximum, but with a secondary one located 10-15 arcmin to the East of the primary. This implies that most A2151 galaxies are not expected to be embedded in a hot X-ray intra-cluster medium (ICM). Thus, Hercules cluster  provides an ideal case in which  the different effects of the cluster environment on the evolution of the galaxies can be studied ``caught in the act''. 

In this work we have benefited from our deep search for H$\alpha$ emitting galaxies in the Hercules cluster \citep[C09 from now on]{Cedres2009} to define a new sample of SF galaxies. Tracing H$\alpha$ emission in order to study the population of galaxies with recent star formation has two important advantages: 1) it is not aperture biased as many of the spectroscopic studies using fibers and 2) it reveals star forming dwarfs which would not have been classified as SF galaxies with an optical flux criterion. We investigate gas-phase metallicities and stellar  properties of galaxies which span a large range of magnitudes ($-21 \leq \mathrm{M_B} \leq -16.5$ mag) and  morphologies (from grand design spirals to blue compacts and tidal dwarfs).

The present paper is organized as follows: in \S2 we present the characteristics of our spectroscopic sample, the observations and the data reduction. In \S3 we present the spectral synthesis techniques performed to compute characteristic properties of the underlying stellar population. In \S4 we derive the chemical and physical properties of our sample of galaxies  and in \S5 we discuss the environmental effects on the Hercules cluster galaxies. \S6 summarizes the overall picture of the Hercules cluster as revealed by the chemical evolution point of view.

\section{OBSERVATIONS}\label{OBS}

\subsection{The galaxy sample}\label{SAMP}

\begin{figure}[h]
\medskip 
\center 
\includegraphics[height=7cm]{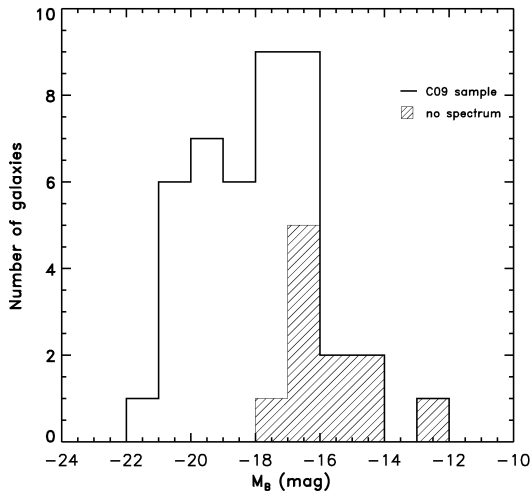}
\caption{The histogram of the C09 sample of SF galaxies as a function of their $\mathrm{M_B}$ absolute magnitude where we show the fraction of galaxies with no spectroscopic data (filled region). Our spectroscopic follow-up of the C09 Hercules SF galaxies is complete up to absolute magnitude $\mathrm{M_B} = -17$ mag.  \label{sample}}
\end{figure}

Spatially resolved spectroscopy has been obtained for 27 Hercules SF galaxies of the C09 Hercules sample\footnote{From our spectroscopic data it is shown that the radial velocity of one (SDSS J160547.17+173501.6)  of the C09 galaxies is out of the velocity range of the Hercules cluster, rendering its membership to the cluster unlikely. The various SDSS photo-z estimates were consistent with the criteria adopted by C09 for cluster membership.   More details are given in Appendix.}. We complement our sample with two more galaxies of the C09 sample which have SDSS spectroscopy. In addition, two more galaxies are considered from \citet{Iglesias2003} (IP03 from now on) A2151 sample of dwarf galaxies (see section \ref{COMPARE} for more details). 
Thus our spectroscopic sample comprises a total of 31 galaxies and our spectroscopic follow-up of the C09 sample is completed up to an absolute magnitude $\B = -17$ mag. To illustrate this, in Fig.~\ref{sample} we show the histogram of the C09 sample of SF galaxies as a function of their $\mathrm{M_B}$ absolute magnitude and the fraction of the galaxies with no spectroscopic data. 

Our sample is composed of 15 luminous SF galaxies ($\mathrm{M_B} \leq -19$) and 16 dwarf and Magellanic type irregular galaxies ($\mathrm{M_B} > -19$). Among the spiral/irregular galaxies, 9 show discernible spatial structure and therefore their different parts  have been analyzed independently (see also \S\ref{DR} and Fig.~\ref{SLITS}). 
For three luminous galaxies and three dwarfs/irregulars we have gathered together evidences that they are affected/perturbed from galaxy-galaxy interactions, either being members of mergers, tidal dwarf candidates, or their HI morphology showing evidences of interaction (we give details of these objects in the Appendix). 

\begin{figure*}
\center
\includegraphics[width=10cm]{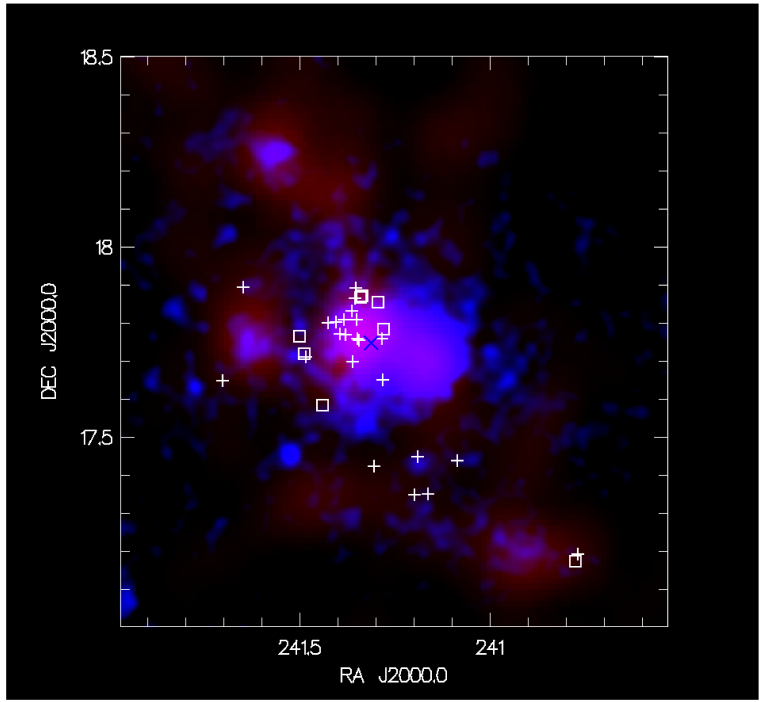}
\caption{Superposition of the map of galaxy density of  the Hercules cluster (red) and the ROSAT All-Sky Survey X-ray map (blue). The density map includes all Hercules galaxies with  SDSS spectroscopy.  The SF galaxies of the present sample are indicated by pluses and squares.  Open squares indicate dwarf galaxies with  abs($\Delta \mathrm{V}) > \sigma_V$. The surface density of galaxies reveals the presence of subclusters. The extended X-ray emission coming from the hot intra-cluster gas shows  an irregular pattern which divides the cluster into two components. Most \Ha emitters  are located around the secondary X-ray maximum, in the regions of higher density of the A2151 galaxies. The blue cross indicates the position of the cluster center given in NED. See the text for details.\label{XRAY}}
\end{figure*}

The Hercules cluster shows abundant substructures seen in X-ray emission and broadband imaging \citep{Huang1996}. The galaxy surface density as well as the velocity space reveals the presence of subclusters in the galaxy distribution \citep{Maccagni1995,Bird1995}.  The extended X-ray emission coming from the hot intra-cluster gas shows  an irregular pattern which divides the cluster into two components.  Fig.~\ref{XRAY} illustrates the positions of our spectroscopic sample galaxies (pluses and squares) with respect to the density of the  ICM (as seen in X-ray surface brightness map from the ROSAT All-Sky Survey; in blue) and the galaxy surface density of the cluster (derived counting Hercules members\footnote{within the velocity range of the subclusters of A2151 defined in \citet{Bird1995}} with  SDSS spectroscopy; red color). There is a primary X-ray maximum concentric with the the brighter cluster galaxy NGC6041A \citep{Huang1996}. Though, most \Ha emitters  are located around the secondary X-ray maximum, in the regions of higher density of  A2151 galaxies (indicated by the NED position of the mass center; blue cross) within the central region $<R_{200}$ of the Hercules cluster.\footnote{$R_{200}=1.4$ Mpc for the Hercules cluster, following \citet{Finn2005} equation 8.} Open squares mark the dwarf/irregular galaxies with radial velocity different from the radial velocity of  NGC6041A $\Delta V$ larger than $\mathrm{\sigma{V}}$, the radial velocity dispersion\footnote{The radial velocity dispersion of the Hercules cluster  $\mathrm{\sigma_V}= 752$ km s$^{-1}$ \citep{Bird1995}.} of Hercules cluster, i.e. abs($\Delta \mathrm{V}) > \mathrm{\sigma_V}$, a possible indication of an infalling population as we will discuss in \S\ref{LD} and \S\ref{CBOX}.

\subsection{Optical spectroscopy}
Long-slit spectroscopy was carried out at the 4.2m WHT and the 2.5m INT at the Observatorio del Roque de los Muchachos (ORM), in Spain. 
Table~\ref{LOG} shows the journal of the spectroscopic observations. We present for each galaxy the object name as quoted by C09, RA, DEC, observation date, telescope,  effective wavelength range, slit position angle and total exposure time. Most of the galaxies were observed as closer to the zenith as possible, always with air-mass less than 1.3, in order to minimize any differential atmospheric refraction effects; except in four cases, as indicated in Table~\ref{LOG}, for which the paralactic angle was used. 

Observations at the WHT were performed during two observing runs on June and July 2009, using the double arm spectrograph ISIS with slightly different instrumental set-ups. On June 2009 the dichroic splitting the beam was set at 5300 \AA\ and the gratings used were R300B and R158R giving nominal dispersions of 0.86 \AA\ pix$^{-1}$  and 1.82 \AA\ pix$^{-1}$ respectively and a spectral coverage of $\sim$3600-9230 \AA. On July 2009 the dichroic was set at 5700 \AA\ and  the gratings used were R300B and R316R giving respective dispersions 0.83 and 0.89 \AA\ pix$^{-1}$ and a spectral coverage $\sim$3160-8263 \AA. During both runs the CCD set-up was the same: on the blue arm  the EEV12 detector, providing a spatial scale 0.19 arcsec pix$^{-1}$, and on the red arm the RED+ detector sampling 0.22 arcsec pix$^{-1}$. Observations at the INT were also performed during two observational runs, on June 2008 and May 2009, using the Intermediate Dispersion Spectrograph (IDS) with the EEV10 detector and R300V grating, giving a nominal dispersion of 1.86 \AA\ pix$^{-1}$ and a spatial scale of 0.4 arcsec pix$^{-1}$. The total wavelength coverage free of vignetting with this set-up was $\sim$3400-7700 \AA. In both ISIS and IDS spectra the slit width was set to 1.5 arcsec. The orientation of the slit, as listed in Table~\ref{LOG}, was selected to cover the maximum surface of the H$\alpha$ emission detected by \cite{Cedres2009}, with few exceptions that were observed on paralactic angle.   

\subsection{Data reduction}\label{DR}

The spectra were reduced in the standard manner using
IRAF routines. First, the two-dimensional spectra were bias subtracted and flat-field corrected. For this, dome flat exposures were used to remove pixel-to-pixel variations in response. Wavelength calibration was achieved using spectra of  CuNe and CuAr comparison lamps, reaching an accuracy of $\sim$0.3 \AA\ for IDS and less than 0.1 \AA\ for ISIS spectra. Several exposures were taken for each object in order to remove cosmic ray hits. The spectra were corrected for atmospheric extinction using an extinction curve for ORM \citep{king1985} and all observing nights were photometric, according to Carlsberg Meridian Telescope records. Enough slit length free of any object permitted an adequate sky-subtraction. For the flux calibration, during the observational runs each night at least 3 spectrophotometric standard stars were observed, varying in color, except during the 26-27 June 2009 when we observed two and one stars respectively.  The expected spectrophotometric error is estimated 10\% (June 2008) 8\% (May 2009) 3\% (June 2009) and 2\% (July 2009). ISIS matching between red and blue arm spectra was successful as can be seen in the spectra shown in Fig.~\ref{SPE} in the Appendix. 

From the 2D frames an integrated  1D spectrum was produced for those galaxies showing no significant spatial structure. Conversely for galaxies showing rich spatial structure, their 2D spectrum is divided into different 1D spectra corresponding to their different sub-regions. For these galaxies, in Fig.~\ref{SLITS} in the Appendix we show g-i color maps produced using  SDSS images and  \Ha equivalent-width (EW) maps using our data, with the precise position of the slit overploted on both maps. We also give the spatial profiles of the H$\alpha$ line and the nearby continuum emission along the slit position extracted from the 2D-spectra. On these profiles the different galaxy parts considered in our analysis are  highlighted. All the 1D spectra extracted in this study  are presented in Fig.~\ref{SPE} in the Appendix. 

\begin{deluxetable}{lccccccc}
\tabletypesize{\scriptsize}
\tablecaption{Log of Observations\label{LOG}}
\tablewidth{0pt}
\tablehead{
\colhead{Galaxy} & \colhead{RA} &\colhead{DEC} & \colhead{Date} & \colhead{Tel.} & \colhead{Range}& \colhead{PA} & \colhead{Exp.}\\
\colhead{} & \colhead{J2000.0} &\colhead{J2000.0} & \colhead{} & \colhead{} & \colhead{}& \colhead{} & \colhead{}\hspace{-1mm}\\
\colhead{(1)} & \colhead{(2)} &\colhead{(3)} & \colhead{(4)} & \colhead{(5)} & \colhead{(6)}& \colhead{(7)} & \colhead{(8)}\hspace{-1mm}}
\startdata
PGC057185 			& 16 06 48.2 & 17 38 51.6 & 2008 Jun 06 & INT & 3500-7730 & 13  & 3600\\
IC1173 				& 16 05 12.6 & 17 25 22.3 & 2008 Jun 06 & INT & 3500-7730   & 45  & 3600\\
KUG1603+179A 			& 16 05 30.6 & 17 46 07.2 & 2008 Jun 06 & INT & 3500-7730  & 30  & 4500\\
NGC6050A,B\tablenotemark{\ddagger} 			& 16 05 23.4 & 17 45 25.8 & 2008 Jun 06 & INT &  3500-7730 & 67\tablenotemark{\dagger} & 3600\\
LEDA1543586 			& 16 05 10.5 & 17 51 16.1 & 2008 Jun 07 & INT &  3500-7730 & 297 & 5300\\
NGC6045 			& 16 05 07.9 & 17 45 27.6 & 2008 Jun 07 & INT & 3500-7730  & 75  & 4500\\
KUG1602+174A 			& 16 04 39.0 & 17 20 59.9 & 2008 Jun 07 & INT &  3500-7730 & 59\tablenotemark{\dagger}  & 5400\\
LEDA084719 			& 16 05 07.1 & 17 38 57.8 & 2008 Jun 07 & INT &  3500-7730 & 67\tablenotemark{\dagger}  & 4200\\
PGC057077 			& 16 05 34.2 & 17 46 11.8 & 2008 Jun 08 & INT &  3500-7730 & 297 & 4500\\
UGC10190 			& 16 05 26.3 & 17 41 48.6 & 2008 Jun 08 & INT &  3500-7730 & 150 & 4500\\
LEDA140568 			& 16 06 00.1 & 17 45 52.0 & 2008 Jun 08 & INT &  3500-7730 & 10  & 5400\\
$[D97]$ce-200 			& 16 05 06.8 & 17 47 02.0 & 2008 Jun 08 & INT &  3500-7730 & 66\tablenotemark{\dagger}  & 5400\\
PGC057064 			& 16 05 27.2 & 17 49 51.6 & 2008 Jun 09 & INT &  3500-7730 & 115 & 5400\\
LEDA084703 			& 16 03 05.7 & 17 10 20.4 & 2008 Jun 09 & INT &  3500-7730 & 82  & 5400\\
KUG1602+175 			& 16 04 45.4 & 17 26 54.3 & 2009 May 26 & INT & 3400-7630 & 315 & 3600\\
LEDA084710 			& 16 04 20.4 & 17 26 11.2 & 2009 May 26 & INT & 3400-7630 & 105 & 3600\\
CGCG108-149 			& 16 06 35.3 & 17 53 33.2 & 2009 May 26 & INT & 3400-7630 & 172 & 3600\\
KUG1602+174B 			& 16 04 47.6 & 17 20 52.0 & 2009 May 26 & INT & 3400-7630 & 67 & 2400\\
LEDA084724 			& 16 05 45.5 & 17 34 56.7 & 2009 Jun 26 & WHT & 3600-9230 & 0 & 3000 \\
SDSS J160556.98+174304.1 	& 16 05 57.0 & 17 43 04.1 & 2009 Jun 26 & WHT  &  3600-9230 & 90 & 3600 \\
$[$DKP87$]$160310.21+175956.7 	& 16 05 25.0 & 17 51 50.6 & 2009 Jun 26 & WHT &  3600-9230 & 114 & 3600 \\
SDSS J160524.27+175329.3 	& 16 05 24.3 & 17 53 29.4 & 2009 Jun 27 & WHT &  3600-9230 & 102 & 3600 \\
SDSS J160547.17+173501.6 	& 16 05 47.2 & 17 35 01.7 & 2009 Jun 27 & WHT &  3600-9230 & 170 & 3600 \\
SDSS J160523.66+174832.3 	& 16 05 23.7 & 17 48 32.3 & 2009 Jun 27 & WHT &  3600-9230 & 140 & 3600 \\
IC1182				& 16 05 36.8 & 17 48 07.5 & 2009 Jul 19 & WHT & 3160-8263 & 127 & 3000 \\
IC1182:[S72]d 			& 16 05 41.9 & 17 47 58.4 & 2009 Jul 19 & WHT  & 3160-8263 & 96  & 3600 \\
SDSS J150531.84+174826.1 	& 16 05 31.8 & 17 48 26.2 & 2009 Jul 19 & WHT  & 3160-8263 & 66 & 3600 \\
\enddata
\tablecomments{Column 1: Galaxy name following \citet{Cedres2009}; Column 2: right ascension in hours, minutes, and seconds; Column 3: declination, in degrees, arcminutes, and arcseconds; Column 4: observation date; Column 5: telescope used; Column 6: wavelength  range coverage in \AA; Column 7: position angle in degrees; Column 8: exposure time in seconds.}
\tablenotetext{\dagger}{Observed at paralactic angle through air-mass $ \geq 1.3$.}
\tablenotetext{\ddagger}{This pair is also known as Arp 272.}
\end{deluxetable}

\section{SPECTRAL SYNTHESIS MODEL FITTING}\label{STL}

We use the population synthesis code STAR-\nextline
LIGHT to fit spectral synthesis models to all our spectra and we compute some characteristic properties of the underlying stellar component as the luminosity-weighted and mass-weighted stellar age and metallicity \citep[see, e.g.][]{Asari2007}. We also used the  model fits to correct our spectra for underlying stellar absorption. Our galaxy sample spans a considerable range in its emission-line properties, from systems with Balmer emission lines with large equivalent widths to systems where the  H$\beta$ is just detectable above a high level of stellar continuum.  Especially in the latter cases, an accurate correction for underlying stellar absorption is crucial. The usual practice of adopting an EW$_{\rm abs}$ of $\sim$2 \AA\ for stellar Balmer lines, though generally justifiable, bears the risk of introducing significant systematic errors in the analysis of systems with faint emission lines. This is because the exact value for EW$_{\rm abs}$ depends on the stellar population mixture i.e. the star formation history (SFH) in a galaxy. For example, \cite{GonzalezDelgado1999} infer for a single stellar population (SSP) a monotonous increase of the EW$_{\rm abs}$'s for the Balmer H$\beta$ through H$\delta$ lines from $\la$3 \AA\ at an age $\leq$3 Myr to $>$10 \AA\ at ages $\sim$0.5 Gyr, followed  by a decrease for older ages. Star-forming galaxies, consisting of different SSPs of young to intermediate age, in addition to a (significantly) older underlying host galaxy, likely follow a  considerably more complex time evolution in their luminosity-weighted EW$_{\rm abs}$. This issue has been explored in much detail by e.g. \citet{Guseva2001} who have considered a variety of SFHs and mass proportions between the young and older stellar components in SF dwarf galaxies. Obviously, for systems with low emission-line EWs, improper correction for underlying stellar absorption may propagate into significant errors in e.g., the intrinsic extinction or spectroscopic classification using BPT \citep{BPT81} diagrams. For example, for an intermediate luminosity-weighted age of $0.2-0.6$ Gyr, where 
Balmer EW$_{\rm abs}$ are largest, the underestimation of Balmer line fluxes due to incomplete correction for underlying stellar absorption may shift a star forming galaxy on the log([O{\sc iii}]$\lambda$5007/H$\beta$) vs. log([N{\sc ii}]$\lambda$6583/H$\alpha$) BPT diagram into the zone of active galactic nuclei (AGNs). 

The determination of the EW$_{\rm abs}$ may be carried out by simultaneous fitting of two Gaussians to each emission and absorption Balmer line profile. 
Another adequate, and arguably more efficient, technique for correcting for underlying stellar absorption is the fitting and subsequent subtraction of the observed spectrum of a synthetic stellar spectrum in order to isolate the net nebular line emission fluxes. To this end, we used the population synthesis code STARLIGHT \citep{CidFernandes04,CidFernandes05a,CidFernandes05b,GarciaRissmann05}. This code synthesizes the observed stellar continuum 
as due to the superposition of SSPs of different age and metallicity. The SSP library used is based on stellar models by \citet{Bruzual2003} for a 
Salpeter {\it initial mass function} (IMF) for three metallicities ($Z_{\odot}/5$, $Z_{\odot}/2.5$ and $Z_{\odot}$)\footnote{The solar metallicity is assumed $Z_{\sun}=0.019$, which is translated to an oxygen abundance $12+\log (O/H)_{\sun}=8.69$ \citep{Asplund2009}.} and 59 ages (between 0.25 Myr and 13 Gyr). Output of the model is the list of those SSPs with a mass contribution $M_i>0$ (\%) and their luminosity contribution $L_i$ (\%) to the wavelength interval to which the input spectra have been normalized (5100--5300 \AA\ here).

Prior to modeling the observed spectra were de-redshifted and resampled to 1 \AA/pixel. Additionally, emission lines and residuals from the night sky subtraction were identified and interactively flagged using a code developed by us. Throughout, we preferred not to correct the input spectra for intrinsic extinction or to strongly constrain its allowed range in the STARLIGHT models for a twofold reason: first, the extinction coefficient C(H$\beta$) computed prior to correction of Balmer emission line fluxes for underlying stellar absorption is, especially for galaxies with faint nebular emission, uncertain.
An a priori de-redenning of the input spectra using this approximate C(H$\beta$) value would therefore yield no advantage towards reliability and computational expense. For the same reason, and in order to minimize biases in the STARLIGHT solutions, we have allowed the intrinsic extinction to vary between $+4.5$ and $-0.7$ V mag. The allowance for a negative $A_V$ is justified by possible uncertainties in flux calibration, the (partial) coverage of the physical spatial area of the galaxy, as well as the incompleteness of the SSP library. As evident from Table~\ref{stlout}, solutions with $A_V\la 0$ are the exception in our analysis. Secondly, fixing the intrinsic extinction in the STARLIGHT models to the observed C(H$\beta$) would imply that ionized gas and stars are subject to equal extinction. No compelling evidence exists for this as yet. On the other hand, for the sake of simplicity and to keep computational time low, we have throughout assumed a single foreground extinction model, although provision is given in STARLIGHT to solve for a different extinction in the young ($\leq$10 Myr) and older stellar population.

\begin{figure*}
\medskip 
\center \includegraphics[height=15cm,angle=-90]{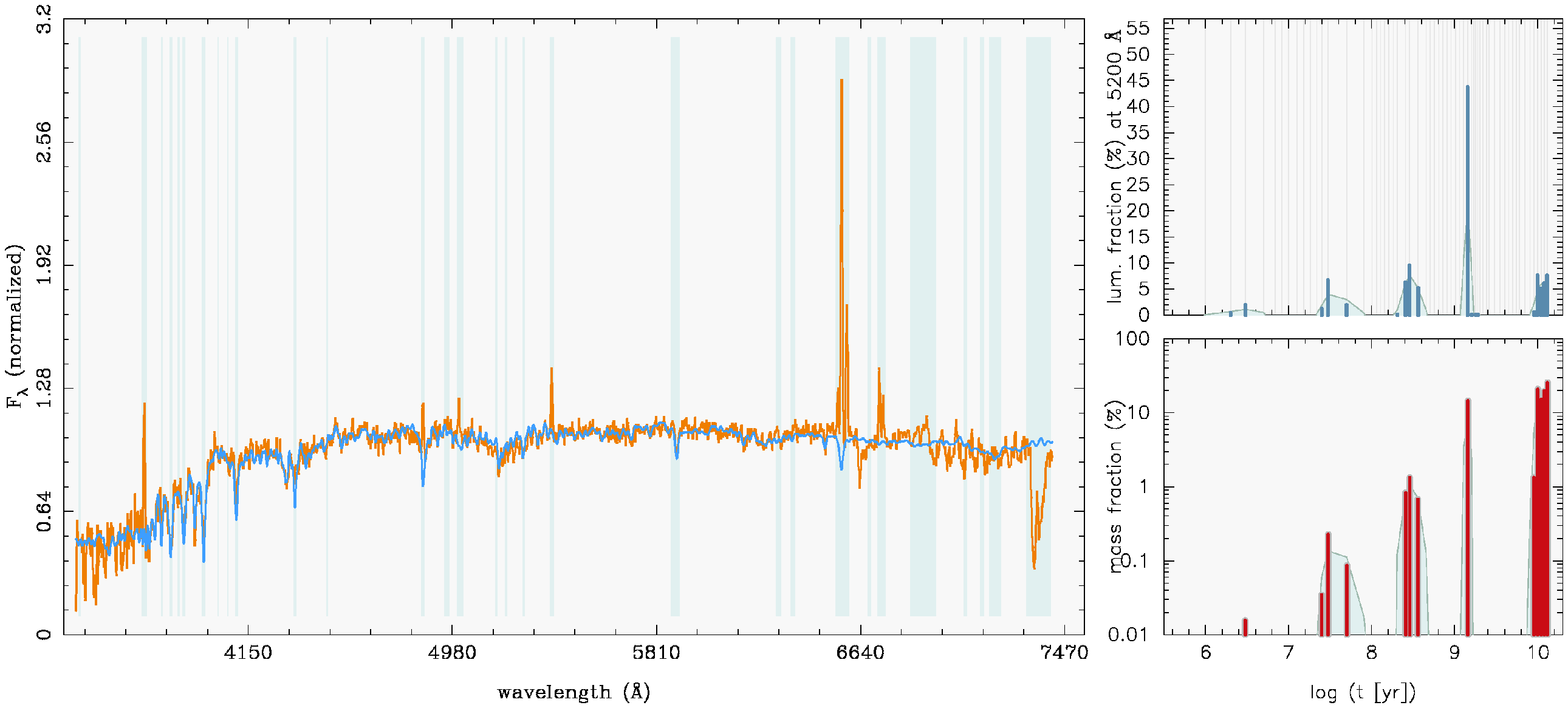}\medskip
\caption{A typical example of spectral synthesis with STARLIGHT for the galaxy KUG1602+175. 
The left panel shows the observed, normalized spectrum (orange color) with 
the best-fitting synthetic spectrum overlaid  (blue color).  
Flagged intervals in each spectrum are marked with shaded vertical bars.
The smaller panels to the upper-right and lower-right show the age distribution 
of the SSPs evaluated by STARLIGHT as a function of, respectively, their luminosity 
and mass fraction in \%. Thin vertical lines in the upper right panel indicate the ages of the library SSPs. 
The shaded area in the lower diagram shows the smoothed $M_i$ distribution and is meant 
as schematic illustration of the SFH.\label{Fig:starlight_fit}}
\end{figure*}

A typical example of the spectral synthesis for a galaxy with relatively faint nebular emission in our sample is shown in Fig.~\ref{Fig:starlight_fit}. 
The left panel shows the observed, normalized spectrum (orange color) with the best-fitting synthetic spectrum overlaid  (blue color). Flagged intervals in each spectrum are marked with shaded vertical bars. The smaller panels to the upper-right and lower-right show the age distribution of the SSPs evaluated by STARLIGHT as a function of, respectively, their luminosity and mass fraction in \%. Thin vertical lines in the upper right panel indicate the ages of the library SSPs. The shaded area in the lower diagram shows the smoothed $M_i$ distribution and is meant as schematic illustration of the SFH. It may be seen from the left panel that synthetic spectra fit well broad stellar absorption features, thus permitting an adequate determination and correction  for underlying absorption. The corrected emission line fluxes, as measured from the observed spectra after subtraction of the synthetic stellar spectra are listed in Table~\ref{LINES} (see \S\ref{RES} for a detailed discussion).

Although an elaborated study of the SFH of the sample galaxies is beyond the scope of this paper, we took advantage of the STARLIGHT models to compute some characteristic properties of our sample galaxies. These include the luminosity-weighted and mass-weighted metallicity $Z_{\star,L}$ and $Z_{\star,M}$ and stellar age $\tau_{\star,L}$ and $\tau_{\star,M}$ along with their  standard deviations \citep[see, e.g.,][]{Asari2007}. These quantities and the intrinsic extinction $A_V$ of the stellar component, are listed in Table ~\ref{stlout} and will be further discussed in \S\ref{DIS}.

\begin{deluxetable}{lccccc}
\tabletypesize{\scriptsize}
\tablecaption{STARLIGHT Output\label{stlout}}
\tablewidth{0pt}
\tablehead{
\colhead{Galaxy}  &\colhead{$Z_{\star,L}$} & \colhead{$Z_{\star,M}$} & \colhead{$\tau_{\star,L}$} & \colhead{$\tau_{\star,M}$} & \colhead{$A_V$}\\
\colhead{}  &\colhead{} & \colhead{} & \colhead{Gyr} & \colhead{Gyr} & \colhead{mag}}
\startdata
           PGC05185a   &$0.014  \pm0.008  $&$0.012  \pm0.008  $&$ 0.94  \pm 0.30  $&$ 1.08  \pm 0.28  $&$ 0.98  $\\
           PGC05185b   &$0.009  \pm0.005  $&$0.007  \pm0.004  $&$ 2.02  \pm 1.02  $&$ 7.28  \pm 3.07  $&$ 1.68  $\\
           PGC05185c   &$0.014  \pm0.007  $&$0.009  \pm0.004  $&$ 2.64  \pm 1.24  $&$ 8.54  \pm 2.07  $&$ 1.32  $\\
             IC1173a   &$0.016  \pm0.006  $&$0.019  \pm0.009  $&$ 7.73  \pm 1.92  $&$11.10  \pm 0.80  $&$-0.21  $\\
             IC1173e   &$0.018  \pm0.009  $&$0.016  \pm0.007  $&$ 2.09  \pm 0.72  $&$ 5.66  \pm 1.86  $&$ 0.63  $\\
       KUG1603+179Aa   &$0.014  \pm0.007  $&$0.012  \pm0.005  $&$ 3.69  \pm 1.35  $&$ 6.60  \pm 1.81  $&$-0.56  $\\
       KUG1603+179Ab   &$0.015  \pm0.006  $&$0.018  \pm0.011  $&$ 7.27  \pm 2.25  $&$11.62  \pm 0.76  $&$ 0.91  $\\
       KUG1603+179Ac   &$0.012  \pm0.004  $&$0.008  \pm0.005  $&$ 4.80  \pm 1.98  $&$11.07  \pm 0.77  $&$ 1.09  $\\
           NGC6050Aa   &$0.011  \pm0.004  $&$0.014  \pm0.005  $&$ 8.10  \pm 1.93  $&$11.36  \pm 0.58  $&$ 0.47  $\\
            NGC6050Ab  &$0.013  \pm0.003  $&$0.012  \pm0.005  $&$ 5.67  \pm 1.87  $&$11.58  \pm 0.54  $&$ 0.41  $\\
            NGC6050B   &$0.007  \pm0.002  $&$0.004  \pm0.002  $&$ 5.82  \pm 1.35  $&$ 8.54  \pm 0.97  $&$-0.12  $\\
         LEDA1543586   &$0.006  \pm0.002  $&$0.008  \pm0.003  $&$ 1.98  \pm 0.91  $&$ 6.66  \pm 1.71  $&$ 0.42  $\\
            NGC6045a   &$0.015  \pm0.005  $&$0.018  \pm0.010  $&$ 4.87  \pm 1.94  $&$10.82  \pm 1.20  $&$ 0.57  $\\
            NGC6045b   &$0.013  \pm0.004  $&$0.017  \pm0.008  $&$ 5.42  \pm 1.74  $&$10.10  \pm 1.31  $&$ 0.27  $\\
           NGC6045c    &$0.016  \pm0.008  $&$0.017  \pm0.013  $&$ 8.30  \pm 2.45  $&$12.68  \pm 0.28  $&$ 2.13  $\\
            NGC6045d   &$0.013  \pm0.005  $&$0.015  \pm0.008  $&$ 4.25  \pm 1.78  $&$11.14  \pm 1.38  $&$ 1.33  $\\
            NGC6045e   &$0.011  \pm0.007  $&$0.010  \pm0.008  $&$ 1.06  \pm 0.30  $&$ 2.71  \pm 1.60  $&$ 0.93  $\\
        KUG1602+174A   &$0.007  \pm0.002  $&$0.008  \pm0.003  $&$ 4.66  \pm 1.34  $&$ 8.34  \pm 0.87  $&$ 0.19  $\\
          LEDA084719   &$0.008  \pm0.004  $&$0.012  \pm0.004  $&$ 4.45  \pm 1.41  $&$ 8.34  \pm 1.53  $&$ 0.15  $\\
          PGC057077a   &$0.013  \pm0.006  $&$0.018  \pm0.017  $&$ 4.24  \pm 2.86  $&$12.03  \pm 1.19  $&$ 3.23  $\\
          PGC057077b   &$0.014  \pm0.009  $&$0.013  \pm0.008  $&$ 1.75  \pm 0.41  $&$ 2.10  \pm 0.60  $&$-0.70  $\\
           UGC10190a   &$0.007  \pm0.003  $&$0.005  \pm0.002  $&$ 3.43  \pm 1.46  $&$ 8.14  \pm 1.91  $&$ 0.22  $\\
           UGC10190b   &$0.007  \pm0.002  $&$0.006  \pm0.002  $&$ 5.85  \pm 1.22  $&$10.04  \pm 0.71  $&$ 0.79  $\\
           UGC10190c   &$0.014  \pm0.005  $&$0.009  \pm0.005  $&$ 5.45  \pm 2.16  $&$11.63  \pm 0.64  $&$ 0.82  $\\
          LEDA140568   &$0.008  \pm0.004  $&$0.006  \pm0.003  $&$ 0.86  \pm 0.45  $&$ 4.84  \pm 2.17  $&$ 0.56  $\\
            D97ce200   &$0.007  \pm0.003  $&$0.008  \pm0.005  $&$ 0.73  \pm 0.25  $&$ 1.09  \pm 0.11  $&$ 0.33  $\\
          PGC057064a   &$0.014  \pm0.005  $&$0.017  \pm0.008  $&$ 5.06  \pm 1.36  $&$ 8.13  \pm 0.85  $&$ 0.74  $\\
          PGC057064b   &$0.011  \pm0.004  $&$0.017  \pm0.007  $&$ 6.56  \pm 1.75  $&$10.52  \pm 0.80  $&$ 0.36  $\\
       LEDA084703int   &$0.009  \pm0.002  $&$0.009  \pm0.002  $&$ 1.22  \pm 0.26  $&$ 2.42  \pm 0.50  $&$ 0.31  $\\
         LEDA084703a   &$0.007  \pm0.005  $&$0.008  \pm0.007  $&$ 0.71  \pm 0.26  $&$ 1.00  \pm 0.02  $&$-0.34  $\\
         LEDA084703b   &$0.009  \pm0.003  $&$0.012  \pm0.004  $&$ 1.97  \pm 0.52  $&$ 5.93  \pm 1.60  $&$ 0.50  $\\
         LEDA084703c   &$0.009  \pm0.005  $&$0.008  \pm0.005  $&$ 0.94  \pm 0.29  $&$ 1.75  \pm 0.63  $&$ 0.43  $\\
         KUG1602+175   &$0.013  \pm0.006  $&$0.017  \pm0.007  $&$ 3.76  \pm 1.54  $&$ 9.74  \pm 1.55  $&$ 0.55  $\\
          LEDA084710   &$0.015  \pm0.007  $&$0.017  \pm0.008  $&$ 2.86  \pm 1.24  $&$ 9.05  \pm 2.05  $&$ 1.02  $\\
         CGCG108-149   &$0.018  \pm0.008  $&$0.019  \pm0.010  $&$ 7.68  \pm 2.46  $&$11.11  \pm 1.08  $&$ 0.40  $\\
        KUG1602+174B   &$0.016  \pm0.007  $&$0.018  \pm0.008  $&$ 2.91  \pm 1.11  $&$ 7.76  \pm 1.31  $&$ 0.48  $\\
          LEDA084724   &$0.007  \pm0.003  $&$0.008  \pm0.003  $&$ 3.15  \pm 1.24  $&$ 7.57  \pm 0.92  $&$ 0.37  $\\
SDSSJ160556.98+174304.1&$0.008  \pm0.006  $&$0.008  \pm0.007  $&$ 0.82  \pm 0.49  $&$ 1.26  \pm 0.21  $&$ 0.61  $\\
$[$DKP87$]$160310.21+175956.7&$0.006  \pm0.001  $&$0.006  \pm0.002  $&$ 3.11  \pm 1.48  $&$ 7.72  \pm 0.69  $&$ 0.31  $\\
SDSSJ160524.27+175329.3 &$0.004  \pm0.004  $&$0.004  \pm0.004  $&$ 1.06  \pm 0.21  $&$ 1.06  \pm 0.10  $&$-0.01  $\\
SDSSJ160523.66+174832.3 &$0.007  \pm0.004  $&$0.004  \pm0.002  $&$ 2.25  \pm 0.99  $&$ 6.16  \pm 2.52  $&$ 0.38  $\\
              IC1182    &$0.012  \pm0.004  $&$0.014  \pm0.008  $&$ 6.37  \pm 1.97  $&$11.56  \pm 0.88  $&$ 1.47  $\\
  IC1182:[S72]d         &$0.019  \pm0.011  $&$0.018  \pm0.011  $&$ 1.92  \pm 1.49  $&$11.51  \pm 0.76  $&$ 0.42  $\\
SDSSJ160531.84+174826.1 &$0.007  \pm0.003  $&$0.008  \pm0.004  $&$ 6.11  \pm 2.46  $&$11.10  \pm 1.13  $&$ 1.11  $\\
SDSSJ160304.20+171126.7 &$0.007  \pm0.002  $&$0.006  \pm0.002  $&$ 3.88  \pm 0.71  $&$ 5.78  \pm 0.63  $&$ 0.48  $\\
SDSSJ160520.58+175210.6 &$0.008  \pm0.003  $&$0.008  \pm0.003  $&$ 5.76  \pm 0.65  $&$ 7.11  \pm 0.46  $&$ 0.50  $\\
\enddata
\tablecomments{The luminosity-weighted and mass-weighted stellar metallicity (e.g. the mass fraction of metals over H, where the solar value is assumed to be $Z_{\sun}=0.019$) $Z_{\star,L}$ and $Z_{\star,M}$ and stellar age $\tau_{\star,L}$ and
$\tau_{\star,M}$ along with their the standard deviations and the intrinsic extinction $A_V$ of the stellar component as given by fitting STARLIGHT models. }
\end{deluxetable}



\section{RESULTS}\label{RES}

\subsection{Line fluxes}\label{LF}

Line fluxes were measured in the spectra after subtracting  the best-fitting spectral energy distribution (SED) of the underlying stellar population (see \S\ref{STL}) using the {\scshape{splot}} task of {\scshape{iraf}}. Line fluxes were measured integrating the flux under the line profile over a linearly fitted continuum. In order to account for the main source of error in our data, driven by the continuum placement, five independent measurements of the emission line flux were carried out for each line and we assume the mean and the standard deviation as the flux measurement and the corresponding statistical error.  

The reddening coefficients c(H$\beta$) were calculated adopting the galactic extinction law of \citet{miller1972}  with $R_V=3.2$, as presented in \citet{hagele2008}, using the expression:
\begin{equation}\label{RED}
\footnotesize{
\mathrm {log \left[ \frac{I(\lambda)}{I(H\beta)}\right] =log \left[ \frac{F(\lambda)}{F(H\beta)}\right] +c(H\beta)f(\lambda)}}
\end{equation}  
where $\mathrm{I(\lambda)/I(H\beta)}$ is the theoretical and $\mathrm{F(\lambda)/F(H\beta)}$ the observed Balmer decrements. The theoretical Balmer decrement depends on electron temperature and density and we used the data for Case B and assuming low-density limit and 10,000K (Storey \& Hummer 1995). 
In those cases where we measured at least three Balmer lines with $S/N>5$ we have performed a least-squares fit to derive c(H$\beta$) and the corresponding error was taken to be the uncertainty of the fit.
When only H$\alpha$ and H$\beta$ were available, c(H$\beta$) was derived applying expression (\ref{RED}) and its error was calculated as the rms of a gaussian error distribution produced by a random sampling simulation, taking into account the errors in the line fluxes of H$\alpha$ and H$\beta$. 

Reddening-corrected line fluxes and reddening coefficients c(H$\beta$)  are presented in Table~\ref{LINES}. The errors quoted account for the uncertainties in the line flux measurement and in the extinction coefficient.  
Table~\ref{LINES} also presents the H$\beta$ flux not corrected for extinction and equivalent width, as well as the galaxy recessional velocity, as derived by our spectroscopic data. The H$\beta$ equivalent width quoted was determined from the net emission spectrum, i.e. the observed spectrum after subtraction of the best-fitting stellar SED, and the level of the adjacent continuum in the original spectrum. 

\begin{figure}
\includegraphics[height=7cm,angle=0]{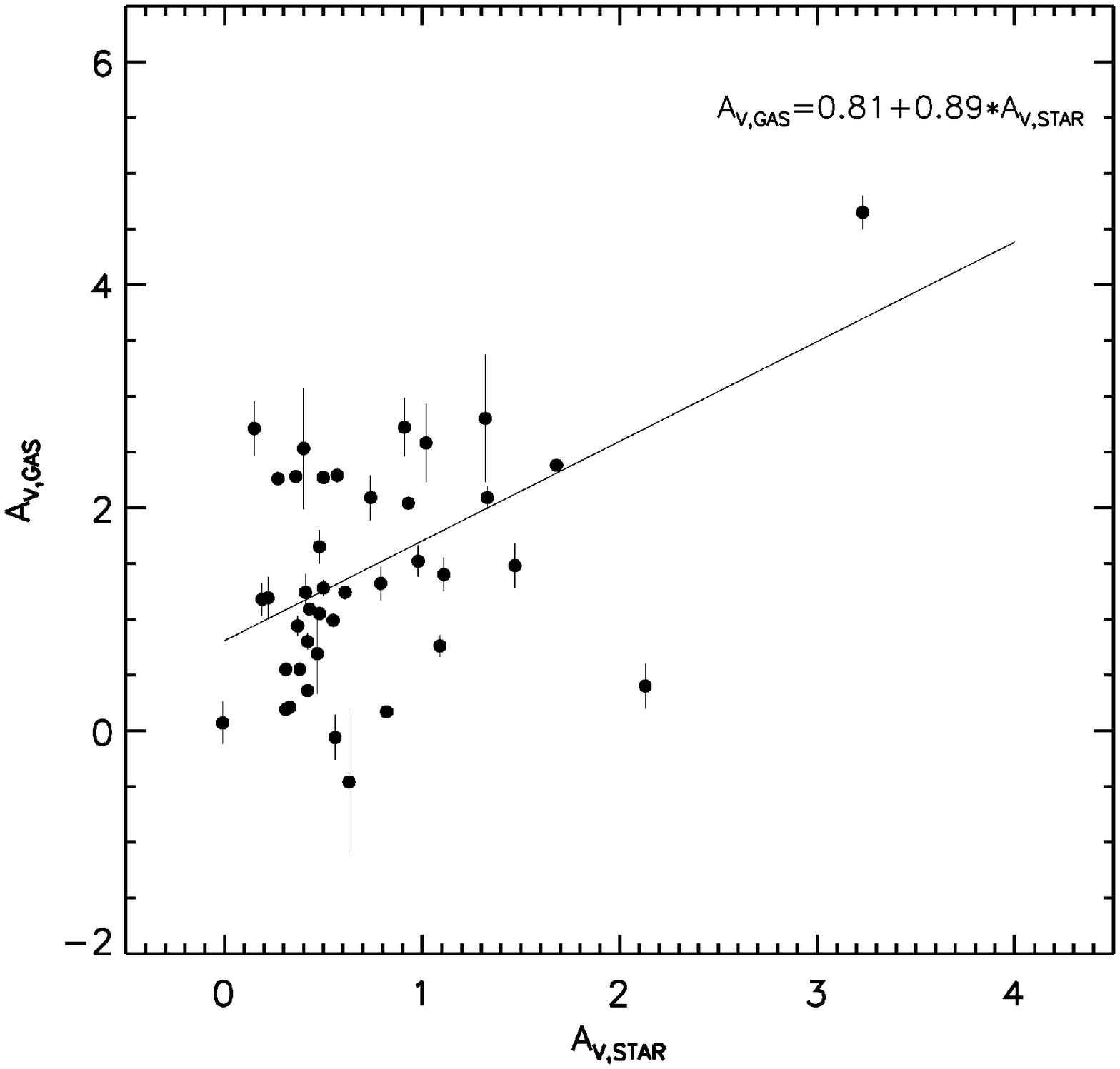}
\caption{The gas-phase  extinction estimated using the Balmer emission lines versus the extinction of the stellar component derived by STARLIGHT model fits on the continuum emission.  A weighted linear fit to the data (solid line) yields the relation $A_{V,gas}=0.81+0.89A_{V,star}$.\label{EXT}}
\end{figure}

Fig.~\ref{EXT} shows the extinction of the gas-phase component $A_{V,gas}=2.18\times \mathrm{c(H}\beta)$ as a function of the intrinsic extinction of the stellar component derived by STARLIGHT in the previous section.  A linear fit, considering  (the reliable range, see \S\ref{STL}) $A_{V,star}> -0.1$ and  weighted by the errors of the c(H$\beta$) estimation, yield the relation $A_{V,gas}=0.81+0.89A_{V,star}$.  We find that stars suffer less extinction than gas, that  can be explained if the dust has a larger covering factor for the ionized gas than for the stars \citep{Calzetti1994}.
In the same line are recent results presented by \citet{Monreal2010a} and \citet{Alonso2010}, who explore the relation between the extinction suffered by the gas and by the stellar populations in the central regions of SF galaxies using IFU data. 

\begin{deluxetable}{lccccccccc}
\tablecolumns{9}
\tabletypesize{\scriptsize}
\tablecaption{Reddening-corrected line fluxes\label{LINES}}
\tablewidth{0pt}
\tablehead{
\colhead{Line\hspace{2in}} 
& \colhead{$f(\lambda)$} 
& \multicolumn{3}{c}{PGC057185} & \colhead{} & \multicolumn{3}{c}{IC1173} \\
\cline{3-5}  \cline{7-9}  \\
\colhead{} & \colhead{} &\colhead{a} &  \colhead{b} &  \colhead{c} &  \colhead{} &  \colhead{a} &  \colhead{b} &  \colhead{e}\vspace*{0in}
}
\startdata
3727 $[$O \scshape{ii}$]$\dotfill & 0.271
&$         305\pm          20$
&$         231\pm           5$
&$         239\pm          53$&
&$         206\pm          37$
&$         991\pm         131$
&$         118\pm           9$
\\
3869 $[$Ne \scshape{iii}$]$\dotfill & 0.238
&\nodata
&\nodata
&\nodata&
&\nodata
&\nodata
&\nodata
\\
3889 He \scshape{i}$+$H8\dotfill  & 0.233
&\nodata
&\nodata
&\nodata&
&\nodata
&\nodata
&\nodata
\\
3968 $[$Ne \scshape{iii}$]$$+$H7\dots  & 0.216
&\nodata
&\nodata
&\nodata&
&\nodata
&\nodata
&\nodata
\\
4069 $[$S \scshape{ii}$]$\dotfill & 0.195
&\nodata
&\nodata
&\nodata&
&\nodata
&\nodata
&\nodata
\\
4102 H$\delta$\dotfill & 0.188 
&\nodata
&$          26\pm           5$
&\nodata&
&\nodata
&\nodata
&\nodata
\\
4340 H$\gamma$\dotfill  & 0.142
&$          50\pm          16$
&$          47\pm           8$
&\nodata&
&\nodata
&\nodata
&$          48\pm           2$
\\
4861 H$\beta$\dotfill & 0.00
&$         100\pm           8$
&$         100\pm           9$
&$         100\pm          18$&
&$         100\pm          13$
&$         100\pm          12$
&$         100\pm           7$
\\
4959 $[$O \scshape{iii}$]$\dotfill  & -0.024
&\nodata
&$          12\pm           2$
&\nodata&
&\nodata
&\nodata
&\nodata
\\
5007 $[$O \scshape{iii}$]$\dotfill  & -0.035
&$          57\pm           4$
&$          36\pm           2$
&$          54\pm          10$&
&$          25\pm          15$
&$         236\pm          18$
&$          88\pm           6$
\\
5876 He \scshape{i}\dotfill  & -0.209
&\nodata
&$          11\pm           1$
&\nodata&
&\nodata
&\nodata
&\nodata
\\
6300 $[$O \scshape{i}$]$\dotfill  & -0.276
&$          21\pm           5$
&$          16\pm           1$
&\nodata&
&\nodata
&$          71\pm          15$
&\nodata
\\
6548 $[$N \scshape{ii}$]$\dotfill & -0.311
&$          31\pm           3$
&$          39\pm           6$
&$          30\pm           7$&
&$          62\pm          11$
&$         180\pm          26$
&\nodata
\\
6563 H$\alpha$\dotfill &  -0.313
&$         302\pm          14$
&$         294\pm           8$
&$         282\pm          54$&
&$         285\pm          40$
&$         285\pm          41$
&$         287\pm          15$
\\
6584 $[$N \scshape{ii}$]$\dotfill & -0.316
&$          85\pm           5$
&$         113\pm           7$
&$          94\pm          18$&
&$         140\pm          21$
&$         492\pm          68$
&$         103\pm          10$
\\
6717 $[$S \scshape{ii}$]$\dotfill & -0.334
&$          64\pm           4$
&$          46\pm           4$
&$          50\pm          11$&
&$          38\pm           7$
&$         157\pm          24$
&$          84\pm           3$
\\
6731 $[$S \scshape{ii}$]$\dotfill &  -0.336 
&$          55\pm           4$
&$          41\pm           4$
&$          44\pm          11$&
&$          54\pm          19$
&$         152\pm          24$
&$          63\pm           8$
\\
\tableline
F(H$\beta$) & 
&$   4.8\pm   0.4$
&$  15.7\pm   1.5$
&$   2.1\pm   0.4$&
&$   1.7\pm   0.2$
&$   2.3\pm   0.3$
&$   2.2\pm   0.2$
\\
EW(H$\beta$) & 
&$   6.5\pm   0.8$
&$   5.5\pm   0.6$
&$   3.0\pm   0.6$&
&$   1.9\pm   0.3$
&$   1.6\pm   0.4$
&$   8.1\pm   1.6$
\\
c(H$\beta$) & 
&$   0.70\pm   0.06$
&$   1.09\pm   0.02$
&$   1.29\pm   0.26$&
&$   1.02\pm   0.19$
&$   0.96\pm   0.19$
&$  -0.21\pm   0.29$
\\
velocity & 
&$  11403\pm     27$
&$  11204\pm     19$
&$  11122\pm     30$&
&$  10617\pm     15$
&$  11204\pm     19$
&$  10248\pm     44$
\\
\enddata
\tablecomments{F(H$\beta$) in $10^{-16}$ erg s$^{-1}$ cm$^{-2}$ \AA$^{-1}$; EW(H$\beta$) in \AA; velocity in km s$^{-1}$.}
\end{deluxetable}
\addtocounter{table}{-1}
\begin{deluxetable}{lccccccccc}
\tablecolumns{9}
\tabletypesize{\scriptsize}
\tablecaption{Continued}
\tablewidth{0pt}
\tablehead{
\colhead{Line\hspace{2in}} 
& \colhead{$f(\lambda)$} 
& \multicolumn{3}{c}{KUG1603$+$179A} 
& \colhead{}  
& \multicolumn{2}{c}{NGC6050A} 
& \colhead{NGC6050B} \\
\cline{3-5}  \cline{7-8}  \\
\colhead{} & \colhead{} &\colhead{a} &  \colhead{b} &  \colhead{c} &  \colhead{} &  \colhead{a} &  \colhead{b} &  \colhead{}
}
\startdata
3727 $[$O \scshape{ii}$]$\dotfill & 0.271
&$         205\pm           5$
&$         133\pm          11$
&$         172\pm           7$&
&$         151\pm          16$
&$         128\pm          15$
&$         176\pm          45$
\\
3869 $[$Ne \scshape{iii}$]$\dotfill & 0.238
&\nodata
&\nodata
&\nodata&
&\nodata
&\nodata
&\nodata
\\
3889 He \scshape{i}$+$H8\dotfill  & 0.233
&\nodata
&\nodata
&$          24\pm           1$&
&\nodata
&$          35\pm           2$
&\nodata
\\
3968 $[$Ne \scshape{iii}$]$$+$H7\dots  & 0.216
&$          17\pm           3$
&\nodata
&\nodata&
&\nodata
&$          18\pm           2$
&\nodata
\\
4069 $[$S \scshape{ii}$]$\dotfill & 0.195
&$          14\pm           2$
&\nodata
&\nodata&
&\nodata
&\nodata
&\nodata
\\
4102 H$\delta$\dotfill & 0.188 
&$          25\pm           4$
&\nodata
&$          28\pm           5$&
&\nodata
&$          27\pm           8$
&\nodata
\\
4340 H$\gamma$\dotfill  & 0.142
&$          46\pm           2$
&$          41\pm           5$
&$          49\pm           2$&
&\nodata
&$          43\pm           8$
&$          48\pm          11$
\\
4861 H$\beta$\dotfill & 0.00
&$         100\pm           3$
&$         100\pm           3$
&$         100\pm           3$&
&$         100\pm          10$
&$         100\pm           3$
&$         100\pm           5$
\\
4959 $[$O \scshape{iii}$]$\dotfill  & -0.024
&\nodata
&$          13\pm           3$
&$          17\pm           3$&
&\nodata
&\nodata
&$          38\pm           6$
\\
5007 $[$O \scshape{iii}$]$\dotfill  & -0.035
&$          43\pm           2$
&$          37\pm           4$
&$          48\pm           2$&
&$          61\pm           5$
&$          33\pm           3$
&$         107\pm           6$
\\
5876 He \scshape{i}\dotfill  & -0.209
&$          12\pm           2$
&\nodata
&$           9\pm           1$&
&\nodata
&\nodata
&\nodata
\\
6300 $[$O \scshape{i}$]$\dotfill  & -0.276
&$           8\pm           1$
&$          16\pm           2$
&\nodata&
&$          41\pm           6$
&$           5\pm           1$
&\nodata
\\
6548 $[$N \scshape{ii}$]$\dotfill & -0.311
&$          36\pm           2$
&$          44\pm           5$
&$          40\pm           1$&
&$          53\pm           8$
&$          34\pm           3$
&\nodata
\\
6563 H$\alpha$\dotfill &  -0.313
&$         279\pm           4$
&$         261\pm          23$
&$         306\pm          10$&
&$         286\pm          38$
&$         285\pm          15$
&$         293\pm           9$
\\
6584 $[$N \scshape{ii}$]$\dotfill & -0.316
&$         105\pm           2$
&$         148\pm          13$
&$         116\pm           4$&
&$         184\pm          23$
&$         100\pm           6$
&$          82\pm           7$
\\
6717 $[$S \scshape{ii}$]$\dotfill & -0.334
&$          46\pm           1$
&$          36\pm           4$
&$          47\pm           2$&
&$          36\pm           7$
&$          37\pm           3$
&$          45\pm           4$
\\
6731 $[$S \scshape{ii}$]$\dotfill &  -0.336 
&$          39\pm           1$
&$          47\pm           5$
&$          38\pm           2$&
&$          69\pm          11$
&$          36\pm           3$
&$          65\pm          10$
\\
\tableline
F(H$\beta$)  & 
&$  11.2\pm   0.4$
&$  16.0\pm   0.5$
&$  14.4\pm   0.5$&
&$   8.0\pm   0.8$
&$  10.2\pm   0.4$
&$   7.7\pm   0.4$
\\
EW(H$\beta$) & 
&$   9.3\pm   0.7$
&$   4.9\pm   0.3$
&$  10.4\pm   0.7$&
&$   2.6\pm   0.3$
&$  15.1\pm   1.5$
&$   4.9\pm   0.5$
\\
c(H$\beta$) & 
&$   0.49\pm   0.02$
&$   1.25\pm   0.12$
&$   0.35\pm   0.05$&
&$   0.32\pm   0.17$
&$   0.57\pm   0.07$
&$   0.16\pm   0.03$
\\
velocity & 
&$  11280\pm     30$
&$  11162\pm     20$
&$  11121\pm     17$&
&$   9524\pm     38$
&$   9686\pm     37$
&$  11134\pm     39$
\\
\enddata
\end{deluxetable}
\addtocounter{table}{-1}
\begin{deluxetable}{lcccccccc}
\tablecolumns{8}
\tabletypesize{\scriptsize}
\tablecaption{Continued}
\tablewidth{0pt}
\tablehead{
\colhead{Line\hspace{2in}} 
& \colhead{$f(\lambda)$} 
& \colhead{LEDA1543586}
& \multicolumn{5}{c}{NGC6045}  \\
\cline{4-8}  \\
\colhead{} & \colhead{} &\colhead{} &  \colhead{a} &  \colhead{b} &  \colhead{c} &  \colhead{d} &  \colhead{e} 
}
\startdata
3727 $[$O \scshape{ii}$]$\dotfill & 0.271
&$         281\pm           4$
&$         334\pm          21$
&$         214\pm           8$
&$         141\pm          12$
&$         152\pm           9$
&$         310\pm           5$
\\
3869 $[$Ne \scshape{iii}$]$\dotfill & 0.238
&\nodata
&\nodata
&\nodata
&\nodata
&\nodata
&\nodata
\\
3889 He \scshape{i}$+$H8\dotfill  & 0.233
&\nodata
&\nodata
&\nodata
&\nodata
&\nodata
&\nodata
\\
3968 $[$Ne \scshape{iii}$]$$+$H7\dots  & 0.216
&\nodata
&\nodata
&\nodata
&\nodata
&$          19\pm           4$
&\nodata
\\
4069 $[$S \scshape{ii}$]$\dotfill & 0.195
&\nodata
&\nodata
&\nodata
&\nodata
&\nodata
&\nodata
\\
4102 H$\delta$\dotfill & 0.188 
&\nodata
&\nodata
&\nodata
&\nodata
&$          25\pm           6$
&\nodata
\\
4340 H$\gamma$\dotfill  & 0.142
&$          46\pm           8$
&$          45\pm           6$
&$          47\pm           5$
&\nodata
&$          49\pm           3$
&$          48\pm           9$
\\
4861 H$\beta$\dotfill & 0.00
&$         100\pm           2$
&$         100\pm           8$
&$         100\pm           1$
&$         100\pm           6$
&$         100\pm           4$
&$         100\pm           5$
\\
4959 $[$O \scshape{iii}$]$\dotfill  & -0.024
&$         103\pm           4$
&\nodata
&\nodata
&\nodata
&\nodata
&$          20\pm           4$
\\
5007 $[$O \scshape{iii}$]$\dotfill  & -0.035
&$         310\pm           4$
&$          40\pm           8$
&$          35\pm           3$
&$          44\pm           2$
&$          19\pm           1$
&$          65\pm           3$
\\
5876 He \scshape{i}\dotfill  & -0.209
&\nodata
&\nodata
&\nodata
&\nodata
&$           8\pm           1$
&$           8\pm           1$
\\
6300 $[$O \scshape{i}$]$\dotfill  & -0.276
&$          22\pm           2$
&\nodata
&\nodata
&\nodata
&$          14\pm           2$
&$          23\pm           1$
\\
6548 $[$N \scshape{ii}$]$\dotfill & -0.311
&$          17\pm           4$
&$          44\pm           5$
&$          42\pm           1$
&\nodata
&$          31\pm           1$
&$          37\pm           2$
\\
6563 H$\alpha$\dotfill &  -0.313
&$         286\pm           4$
&$         280\pm           7$
&$         288\pm           2$
&$         287\pm          20$
&$         288\pm          10$
&$         292\pm           6$
\\
6584 $[$N \scshape{ii}$]$\dotfill & -0.316
&$          44\pm           5$
&$         118\pm           6$
&$         109\pm           3$
&$         203\pm          16$
&$          92\pm           3$
&$         110\pm           3$
\\
6717 $[$S \scshape{ii}$]$\dotfill & -0.334
&$          46\pm           2$
&$          46\pm           3$
&$          37\pm           2$
&$          38\pm           4$
&$          28\pm           1$
&$          51\pm           1$
\\
6731 $[$S \scshape{ii}$]$\dotfill &  -0.336 
&$          29\pm           2$
&$          44\pm           4$
&$          27\pm           3$
&$          44\pm           3$
&$          22\pm           1$
&$          38\pm           1$
\\
\tableline
F(H$\beta$)  & 
&$   6.9\pm   0.1$
&$   5.2\pm   0.4$
&$   8.5\pm   0.1$
&$   5.4\pm   0.3$
&$  17.2\pm   0.7$
&$  15.4\pm   0.7$
\\
EW(H$\beta$) & 
&$   8.9\pm   1.0$
&$   3.0\pm   0.3$
&$   3.3\pm   0.1$
&$   2.1\pm   0.2$
&$   7.9\pm   0.4$
&$   5.7\pm   0.4$
\\
c(H$\beta$) & 
&$   0.17\pm   0.01$
&$   1.05\pm   0.03$
&$   1.04\pm   0.01$
&$   0.18\pm   0.09$
&$   0.96\pm   0.05$
&$   0.94\pm   0.02$
\\
velocity & 
&$   9849\pm     13$
&$  10284\pm     11$
&$  10335\pm     12$
&$   9830\pm     71$
&$   9703\pm     15$
&$   9789\pm     23$
\\
\enddata
\end{deluxetable}
\addtocounter{table}{-1}
\begin{deluxetable}{lcccccccccc}
\tablecolumns{10}
\tabletypesize{\scriptsize}
\tablecaption{Continued}
\tablewidth{0pt}
\tablehead{
\colhead{Line\hspace{2in}} 
& \colhead{$f(\lambda)$} 
& \colhead{KUG1602$+$174A}
& \colhead{LEDA084719}
& \multicolumn{2}{c}{PGC057077}  
& \colhead{}
& \multicolumn{3}{c}{UGC10190}  
\\
\cline{5-6} \cline{8-10} \\
\colhead{} & \colhead{} &\colhead{} &  \colhead{} &  \colhead{a} &  \colhead{b} &  \colhead{} &  \colhead{a} &  \colhead{b}  &  \colhead{c}
}
\startdata
3727 $[$O \scshape{ii}$]$\dotfill & 0.271
&$         175\pm          10$
&$         318\pm          97$
&$         328\pm          17$
&$         165\pm           3$&
&$         241\pm          36$
&$         182\pm          18$
&$         230\pm          52$
\\
3869 $[$Ne \scshape{iii}$]$\dotfill & 0.238
&\nodata
&\nodata
&\nodata
&\nodata&
&\nodata
&\nodata
&\nodata
\\
3889 He \scshape{i}$+$H8\dotfill  & 0.233
&$          55\pm           8$
&\nodata
&\nodata
&$          13\pm           1$&
&\nodata
&\nodata
&\nodata
\\
3968 $[$Ne \scshape{iii}$]$$+$H7\dots  & 0.216
&\nodata
&\nodata
&\nodata
&$          23\pm           2$&
&\nodata
&\nodata
&\nodata
\\
4069 $[$S \scshape{ii}$]$\dotfill & 0.195
&\nodata
&\nodata
&\nodata
&\nodata&
&\nodata
&\nodata
&\nodata
\\
4102 H$\delta$\dotfill & 0.188 
&\nodata
&\nodata
&$          26\pm           4$
&$          25\pm           2$&
&\nodata
&\nodata
&\nodata
\\
4340 H$\gamma$\dotfill  & 0.142
&$          43\pm           5$
&\nodata
&$          42\pm           2$
&$          45\pm           1$&
&\nodata
&$          43\pm           8$
&$          46\pm          18$
\\
4861 H$\beta$\dotfill & 0.00
&$         100\pm           5$
&$         100\pm           8$
&$         100\pm           5$
&$         100\pm           1$&
&$         100\pm           6$
&$         100\pm           7$
&$         100\pm          10$
\\
4959 $[$O \scshape{iii}$]$\dotfill  & -0.024
&\nodata
&$          48\pm          11$
&\nodata
&$          20\pm           1$&
&\nodata
&\nodata
&$          24\pm           3$
\\
5007 $[$O \scshape{iii}$]$\dotfill  & -0.035
&$          25\pm           7$
&\nodata
&$          63\pm           3$
&$          55\pm           1$&
&$          36\pm           2$
&$          24\pm           3$
&$          79\pm           3$
\\
5876 He \scshape{i}\dotfill  & -0.209
&\nodata
&\nodata
&$          12\pm           2$
&$          13\pm           1$&
&\nodata
&\nodata
&\nodata
\\
6300 $[$O \scshape{i}$]$\dotfill  & -0.276
&$          14\pm           2$
&\nodata
&$          14\pm           3$
&$           6\pm           1$&
&\nodata
&\nodata
&\nodata
\\
6548 $[$N \scshape{ii}$]$\dotfill & -0.311
&$          31\pm           3$
&$          46\pm           7$
&$          36\pm           2$
&$          33\pm           1$&
&$          35\pm          11$
&$          32\pm           4$
&\nodata
\\
6563 H$\alpha$\dotfill &  -0.313
&$         272\pm          14$
&$         286\pm          23$
&$         277\pm          13$
&$         277\pm           5$&
&$         286\pm          18$
&$         272\pm          14$
&$         282\pm           7$
\\
6584 $[$N \scshape{ii}$]$\dotfill & -0.316
&$          99\pm           6$
&$         112\pm          12$
&$         103\pm           5$
&$          93\pm           2$&
&$         111\pm           8$
&$          98\pm           7$
&$          73\pm           6$
\\
6717 $[$S \scshape{ii}$]$\dotfill & -0.334
&$          49\pm           4$
&$          42\pm           5$
&$          38\pm           2$
&$          34\pm           1$&
&$          64\pm           5$
&$          60\pm           5$
&$          90\pm           8$
\\
6731 $[$S \scshape{ii}$]$\dotfill &  -0.336 
&$          41\pm           5$
&$          40\pm           7$
&$          31\pm           2$
&$          27\pm           1$&
&$          83\pm           7$
&$          38\pm           3$
&$          64\pm           5$
\\
\tableline
F(H$\beta$)  & 
&$   5.9\pm   0.3$
&$   2.7\pm   0.2$
&$   5.9\pm   0.3$
&$  52.0\pm   0.5$&
&$   1.4\pm   0.1$
&$   7.2\pm   0.5$
&$   1.9\pm   0.2$
\\
EW(H$\beta$) & 
&$   4.1\pm   0.3$
&$   1.7\pm   0.2$
&$   7.8\pm   0.6$
&$  17.3\pm   0.7$&
&$   3.5\pm   0.4$
&$   4.1\pm   0.3$
&$   6.8\pm   1.2$
\\
c(H$\beta$) & 
&$   0.54\pm   0.07$
&$   1.24\pm   0.11$
&$   2.14\pm   0.07$
&$   0.13\pm   0.02$&
&$   0.55\pm   0.09$
&$   0.61\pm   0.07$
&$   0.08\pm   0.02$
\\
velocity & 
&$  10614\pm     18$
&$  10155\pm     14$
&$  10209\pm     15$
&$  10200\pm     14$&
&$  11171\pm     30$
&$  11046\pm     13$
&$  10915\pm     25$
\\
\enddata
\end{deluxetable}
\addtocounter{table}{-1}
\begin{deluxetable}{lccccccccc}
\tablecolumns{9}
\tabletypesize{\scriptsize}
\tablecaption{Continued}
\tablewidth{0pt}
\tablehead{
\colhead{Line\hspace{2in}} 
& \colhead{$f(\lambda)$} 
& \colhead{LEDA140568}
& \colhead{$[D97]ce-200$}
& \multicolumn{2}{c}{PGC057064}  
& \colhead{}
& \multicolumn{2}{c}{LEDA084703}  
\\
\cline{5-6} \cline{8-9} \\
\colhead{} & \colhead{} &\colhead{} &  \colhead{} &  \colhead{a} &  \colhead{b} &  \colhead{} &  \colhead{int} &  \colhead{a}  
}
\startdata
3727 $[$O \scshape{ii}$]$\dotfill & 0.271
&$         314\pm          20$
&$         326\pm          23$
&$         256\pm          23$
&$         636\pm          21$&
&$         331\pm           5$
&$         339\pm          10$
\\
3869 $[$Ne \scshape{iii}$]$\dotfill & 0.238
&$          44\pm           9$
&\nodata
&\nodata
&\nodata&
&$          22\pm           1$
&$          22\pm           2$
\\
3889 He \scshape{i}$+$H8\dotfill  & 0.233
&\nodata
&\nodata
&\nodata
&\nodata&
&$          22\pm           1$
&$          15\pm           2$
\\
3968 $[$Ne \scshape{iii}$]$$+$H7\dots  & 0.216
&\nodata
&\nodata
&\nodata
&\nodata&
&$          18\pm           1$
&$          20\pm           3$
\\
4069 $[$S \scshape{ii}$]$\dotfill & 0.195
&\nodata
&\nodata
&\nodata
&$          44\pm           8$&
&\nodata
&\nodata
\\
4102 H$\delta$\dotfill & 0.188 
&\nodata
&\nodata
&\nodata
&\nodata&
&$          26\pm           2$
&$          27\pm           2$
\\
4340 H$\gamma$\dotfill  & 0.142
&$          47\pm          10$
&$          46\pm           3$
&\nodata
&$          46\pm           8$&
&$          49\pm           1$
&$          49\pm           1$
\\
4861 H$\beta$\dotfill & 0.00
&$         100\pm           6$
&$         100\pm           8$
&$         100\pm           7$
&$         100\pm           5$&
&$         100\pm           1$
&$         100\pm           1$
\\
4959 $[$O \scshape{iii}$]$\dotfill  & -0.024
&$          70\pm           4$
&$          53\pm           6$
&\nodata
&$          41\pm          10$&
&$          57\pm           3$
&$          61\pm           2$
\\
5007 $[$O \scshape{iii}$]$\dotfill  & -0.035
&$         214\pm           5$
&$         163\pm           4$
&$          45\pm           4$
&$         134\pm           8$&
&$         177\pm           1$
&$         187\pm           2$
\\
5876 He \scshape{i}\dotfill  & -0.209
&\nodata
&\nodata
&\nodata
&\nodata&
&$          14\pm           1$
&$          14\pm           2$
\\
6300 $[$O \scshape{i}$]$\dotfill  & -0.276
&\nodata
&$          20\pm           1$
&$          13\pm           4$
&$          16\pm           2$&
&$          14\pm           1$
&$          12\pm           1$
\\
6548 $[$N \scshape{ii}$]$\dotfill & -0.311
&\nodata
&$          12\pm           3$
&$          44\pm           4$
&$          56\pm           3$&
&$          16\pm           2$
&$          11\pm           2$
\\
6563 H$\alpha$\dotfill &  -0.313
&$         286\pm          20$
&$         286\pm           2$
&$         286\pm          19$
&$         282\pm           6$&
&$         292\pm           6$
&$         219$\tablenotemark{\dagger}
\\
6584 $[$N \scshape{ii}$]$\dotfill & -0.316
&$          16\pm           5$
&$          35\pm           2$
&$         126\pm          10$
&$         155\pm           5$&
&$          43\pm           2$
&$          31\pm           1$
\\
6717 $[$S \scshape{ii}$]$\dotfill & -0.334
&$          39\pm           3$
&$          60\pm           1$
&$          48\pm           4$
&$          77\pm           5$&
&$          55\pm           2$
&$          39\pm           2$
\\
6731 $[$S \scshape{ii}$]$\dotfill &  -0.336 
&$          23\pm           3$
&$          51\pm           1$
&$          27\pm           2$
&$          64\pm           6$&
&$          40\pm           2$
&$          29\pm           1$
\\
\tableline
F(H$\beta$)  & 
&$   4.1\pm   0.3$
&$   3.8\pm   0.3$
&$   6.0\pm   0.4$
&$   5.5\pm   0.3$&
&$  30.2\pm   0.2$
&$   9.2\pm   0.1$
\\
EW(H$\beta$) & 
&$  10.4\pm  10.1$
&$   9.7\pm   1.4$
&$   2.9\pm   0.2$
&$   1.7\pm   0.1$&
&$  14.4\pm   0.8$
&$  17.1\pm   1.5$
\\
c(H$\beta$) & 
&$  -0.03\pm   0.09$
&$   0.10\pm   0.01$
&$   0.96\pm   0.09$
&$   1.04\pm   0.02$&
&$   0.25\pm   0.03$
&$   0.00\pm   0.05$
\\
velocity & 
&$  11992\pm     25$
&$   9915\pm      8$
&$  10308\pm     24$
&$  10048\pm     21$&
&$  10033\pm      6$
&$  10008\pm      4$
\\
\enddata
\tablenotetext{\dagger}{Uncertain value.}
\end{deluxetable}
\addtocounter{table}{-1}
\begin{deluxetable}{lcccccccc}
\tablecolumns{8}
\tabletypesize{\scriptsize}
\tablecaption{Continued}
\tablewidth{0pt}
\tablehead{
\colhead{Line\hspace{2in}} 
& \colhead{$f(\lambda)$} 
& \multicolumn{2}{c}{LEDA084703}  
& \colhead{KUG1602$+$175}
& \colhead{LEDA084710}
& \colhead{CGCG108$-$149}
& \colhead{KUG1602$+$174B}
\\
\cline{3-4} \\
\colhead{} & \colhead{} &\colhead{b} &  \colhead{c} &  \colhead{} &  \colhead{} &  \colhead{} &  \colhead{}
}
\startdata
3727 $[$O \scshape{ii}$]$\dotfill & 0.271
&$         360\pm           8$
&$         514\pm          10$
&$         159\pm          12$
&$         361\pm         110$
&$         159\pm          53$
&$         164\pm          18$
\\
3869 $[$Ne \scshape{iii}$]$\dotfill & 0.238
&$          21\pm           2$
&\nodata
&\nodata
&\nodata
&\nodata
&\nodata
\\
3889 He \scshape{i}$+$H8\dotfill  & 0.233
&$          12\pm           7$
&\nodata
&\nodata
&\nodata
&\nodata
&\nodata
\\
3968 $[$Ne \scshape{iii}$]$$+$H7\dots  & 0.216
&\nodata
&\nodata
&$          22\pm           4$
&\nodata
&\nodata
&\nodata
\\
4069 $[$S \scshape{ii}$]$\dotfill & 0.195
&\nodata
&\nodata
&\nodata
&\nodata
&\nodata
&\nodata
\\
4102 H$\delta$\dotfill & 0.188 
&$          24\pm           1$
&\nodata
&$          25\pm           3$
&\nodata
&\nodata
&\nodata
\\
4340 H$\gamma$\dotfill  & 0.142
&$          47\pm           1$
&$          47\pm           8$
&$          47\pm           5$
&\nodata
&\nodata
&$          46\pm           7$
\\
4861 H$\beta$\dotfill & 0.00
&$         100\pm           1$
&$         100\pm           2$
&$         100\pm           8$
&$         100\pm          11$
&$         100\pm          17$
&$         100\pm           6$
\\
4959 $[$O \scshape{iii}$]$\dotfill  & -0.024
&$          57\pm           1$
&$          44\pm           2$
&\nodata
&\nodata
&\nodata
&\nodata
\\
5007 $[$O \scshape{iii}$]$\dotfill  & -0.035
&$         186\pm           2$
&$         131\pm           3$
&$          56\pm           7$
&$          82\pm           9$
&$          53\pm           4$
&$          50\pm           5$
\\
5876 He \scshape{i}\dotfill  & -0.209
&$          13\pm           2$
&$          11\pm           1$
&\nodata
&$           7\pm           4$
&\nodata
&\nodata
\\
6300 $[$O \scshape{i}$]$\dotfill  & -0.276
&$          13\pm           1$
&$          20\pm           5$
&\nodata
&$          15\pm           4$
&\nodata
&\nodata
\\
6548 $[$N \scshape{ii}$]$\dotfill & -0.311
&$          14\pm           1$
&$          20\pm           1$
&$          37\pm           2$
&$          30\pm           5$
&\nodata
&$          34\pm           6$
\\
6563 H$\alpha$\dotfill &  -0.313
&$         280\pm           7$
&$         290\pm           3$
&$         282\pm           7$
&$         285\pm          34$
&$         283\pm          52$
&$         283\pm           6$
\\
6584 $[$N \scshape{ii}$]$\dotfill & -0.316
&$          42\pm           1$
&$          49\pm           2$
&$         103\pm           6$
&$          89\pm          12$
&$         165\pm          32$
&$          97\pm           5$
\\
6717 $[$S \scshape{ii}$]$\dotfill & -0.334
&$          48\pm           1$
&$          84\pm           2$
&$          45\pm           3$
&$          63\pm           9$
&$          50\pm          14$
&$          55\pm           4$
\\
6731 $[$S \scshape{ii}$]$\dotfill &  -0.336 
&$          38\pm           1$
&$          48\pm           2$
&$          23\pm           4$
&$          36\pm           6$
&$          54\pm          15$
&$          44\pm           4$
\\
\tableline
F(H$\beta$)  & 
&$  17.0\pm   0.0$
&$   3.7\pm   0.1$
&$  27.2\pm   2.1$
&$  10.0\pm   1.1$
&$   4.1\pm   0.7$
&$  11.6\pm   0.7$
\\
EW(H$\beta$) & 
&$  16.5\pm   0.8$
&$   6.2\pm   0.6$
&$   4.2\pm   0.4$
&$   3.3\pm   0.5$
&$   1.4\pm   0.2$
&$   4.8\pm   0.5$
\\
c(H$\beta$) & 
&$   0.59\pm   0.03$
&$   0.50\pm   0.01$
&$   0.45\pm   0.03$
&$   1.18\pm   0.16$
&$   1.16\pm   0.25$
&$   0.48\pm   0.02$
\\
velocity & 
&$  10032\pm     11$
&$  10087\pm     16$
&$  10574\pm     20$
&$  10781\pm     32$
&$  11121\pm     15$
&$  10622\pm     15$
\\
\enddata
\end{deluxetable}
\addtocounter{table}{-1}
\begin{deluxetable}{lccccccc}
\tablecolumns{7}
\tabletypesize{\scriptsize}
\tablecaption{Continued}
\tablewidth{0pt}
\tablehead{
\colhead{Line\hspace{2in}} 
& \colhead{$f(\lambda)$} 
& \colhead{LEDA084724}
& \colhead{SDSS J160556.98}
& \colhead{$[$DKP87$]$160310.21}
& \colhead{SDSS J160524.27}
& \colhead{SDSS J160523.66}
}
\startdata
3727 $[$O \scshape{ii}$]$\dotfill & 0.271
&$         493\pm          14$
&$         317\pm          19$
&$         281\pm          13$
&$         214\pm          13$
&$         405\pm           5$
\\
3869 $[$Ne \scshape{iii}$]$\dotfill & 0.238
&\nodata
&$          21\pm           6$
&\nodata
&$          30\pm           3$
&\nodata
\\
3889 He \scshape{i}$+$H8\dotfill  & 0.233
&\nodata
&$          19\pm           6$
&\nodata
&$          16\pm           2$
&\nodata
\\
3968 $[$Ne \scshape{iii}$]$$+$H7\dots  & 0.216
&\nodata
&$          36\pm           5$
&\nodata
&$          28\pm           2$
&\nodata
\\
4069 $[$S \scshape{ii}$]$\dotfill & 0.195
&\nodata
&\nodata
&\nodata
&\nodata
&\nodata
\\
4102 H$\delta$\dotfill & 0.188 
&\nodata
&$          25\pm           3$
&\nodata
&$          30\pm           3$
&\nodata
\\
4340 H$\gamma$\dotfill  & 0.142
&$          49\pm           6$
&$          47\pm           4$
&$          46\pm           5$
&$          49\pm           5$
&$          47\pm           3$
\\
4861 H$\beta$\dotfill & 0.00
&$         100\pm           8$
&$         100\pm           6$
&$         100\pm           8$
&$         100\pm           3$
&$         100\pm           4$
\\
4959 $[$O \scshape{iii}$]$\dotfill  & -0.024
&\nodata
&$          95\pm           4$
&\nodata
&$         114\pm           3$
&\nodata
\\
5007 $[$O \scshape{iii}$]$\dotfill  & -0.035
&$          91\pm           5$
&$         282\pm           4$
&$          47\pm           3$
&$         337\pm           4$
&$          69\pm           8$
\\
5876 He \scshape{i}\dotfill  & -0.209
&\nodata
&$          10\pm           1$
&\nodata
&$          13\pm           2$
&\nodata
\\
6300 $[$O \scshape{i}$]$\dotfill  & -0.276
&\nodata
&$          16\pm           4$
&\nodata
&$          10\pm           1$
&\nodata
\\
6548 $[$N \scshape{ii}$]$\dotfill & -0.311
&\nodata
&\nodata
&$          27\pm           2$
&\nodata
&$          29\pm           8$
\\
6563 H$\alpha$\dotfill &  -0.313
&$         296\pm          13$
&$         280\pm           7$
&$         284\pm           4$
&$         320\pm          21$
&$         289\pm           2$
\\
6584 $[$N \scshape{ii}$]$\dotfill & -0.316
&$          62\pm           8$
&$          27\pm           6$
&$          75\pm           3$
&$          13\pm           2$
&$          61\pm           2$
\\
6717 $[$S \scshape{ii}$]$\dotfill & -0.334
&$          74\pm           8$
&$          40\pm           2$
&$          72\pm           2$
&$          35\pm           3$
&$          72\pm           3$
\\
6731 $[$S \scshape{ii}$]$\dotfill &  -0.336 
&$          50\pm           6$
&$          27\pm           2$
&$          66\pm           4$
&$          22\pm           2$
&$          64\pm           3$
\\
\tableline
F(H$\beta$) & 
&$   3.8\pm   0.3$
&$   9.6\pm   0.5$
&$   3.6\pm   0.3$
&$   6.6\pm   0.2$
&$   2.0\pm   0.1$
\\
EW(H$\beta$) & 
&$   2.7\pm   0.4$
&$  18.1\pm   2.3$
&$   4.4\pm   0.5$
&$  19.3\pm   1.9$
&$   3.6\pm   0.4$
\\
c(H$\beta$) & 
&$   0.43\pm   0.04$
&$   0.57\pm   0.03$
&$   0.09\pm   0.01$
&$   0.03\pm   0.09$
&$   0.25\pm   0.01$
\\
velocity & 
&$  11714\pm     60$
&$  11621\pm     24$
&$  10308\pm     56$
&$  11101\pm     38$
&$  10337\pm     35$
\\
\enddata
\end{deluxetable}
\addtocounter{table}{-1}
\begin{deluxetable}{lccccccc}
\tablecolumns{7}
\tabletypesize{\scriptsize}
\tablecaption{Continued}
\tablewidth{0pt}
\tablehead{
\colhead{Line\hspace{2in}} 
& \colhead{$f(\lambda)$} 
& \colhead{IC1182}
& \colhead{IC1182:[S72]d}
& \colhead{SDSS J150531.84}
& \colhead{SDSS J160304.20}
& \colhead{SDSS J160520.58}
}
\startdata
3727 $[$O \scshape{ii}$]$\dotfill & 0.271
&$         379\pm          22$
&$         399\pm           9$
&$          74\pm           4$
&$         455\pm          26$
&$         108\pm          30$
\\
3869 $[$Ne \scshape{iii}$]$\dotfill & 0.238
&$          23\pm           2$
&$          35\pm           1$
&\nodata
&\nodata
&\nodata
\\
3889 He \scshape{i}$+$H8\dotfill  & 0.233
&$          21\pm           2$
&$          20\pm           2$
&\nodata
&\nodata
&\nodata
\\
3968 $[$Ne \scshape{iii}$]$$+$H7\dots  & 0.216
&$          24\pm           3$
&$          24\pm           3$
&\nodata
&\nodata
&\nodata
\\
4069 $[$S \scshape{ii}$]$\dotfill & 0.195
&\nodata
&\nodata
&\nodata
&\nodata
&\nodata
\\
4102 H$\delta$\dotfill & 0.188 
&$          30\pm           2$
&$          27\pm           1$
&\nodata
&\nodata
&\nodata
\\
4340 H$\gamma$\dotfill  & 0.142
&$          48\pm           2$
&$          47\pm           3$
&\nodata
&$          44\pm           3$
&$          46\pm           4$
\\
4861 H$\beta$\dotfill & 0.00
&$         100\pm           2$
&$         100\pm           2$
&$         100\pm           5$
&$         100\pm           2$
&$         100\pm          10$
\\
4959 $[$O \scshape{iii}$]$\dotfill  & -0.024
&$          73\pm           2$
&$          87\pm           1$
&\nodata
&\nodata
&\nodata
\\
5007 $[$O \scshape{iii}$]$\dotfill  & -0.035
&$         218\pm           4$
&$         253\pm           2$
&$          96\pm           3$
&$          66\pm           4$
&$          11\pm           3$
\\
5876 He \scshape{i}\dotfill  & -0.209
&$          11\pm           1$
&\nodata
&\nodata
&\nodata
&\nodata
\\
6300 $[$O \scshape{i}$]$\dotfill  & -0.276
&$          36\pm           3$
&$          14\pm           1$
&$          18\pm           2$
&$          11\pm           1$
&\nodata
\\
6548 $[$N \scshape{ii}$]$\dotfill & -0.311
&\nodata
&$          13\pm           1$
&\nodata
&$          24\pm           2$
&$          35\pm           2$
\\
6563 H$\alpha$\dotfill &  -0.313
&$         317\pm          21$
&$         298\pm           7$
&$         287\pm          15$
&$         272\pm          13$
&$         286\pm           2$
\\
6584 $[$N \scshape{ii}$]$\dotfill & -0.316
&$          83\pm           8$
&$          36\pm           1$
&$          87\pm           5$
&$          73\pm           4$
&$          99\pm           2$
\\
6717 $[$S \scshape{ii}$]$\dotfill & -0.334
&$          78\pm           7$
&$          48\pm           1$
&$          27\pm           8$
&$          65\pm           4$
&$          33\pm           2$
\\
6731 $[$S \scshape{ii}$]$\dotfill &  -0.336 
&$          73\pm          10$
&$          31\pm           1$
&$          14\pm           2$
&$          48\pm           3$
&$          33\pm           2$
\\
\tableline
F(H$\beta$)  & 
&$ 167.5\pm   3.5$
&$  13.1\pm   0.2$
&$   1.9\pm   0.1$
&$   6.3\pm   0.1$
&$   3.0\pm   0.3$
\\
EW(H$\beta$) & 
&$  28.5\pm   0.9$
&$  29.7\pm   4.6$
&$   4.3\pm   0.5$
&$   0.0\pm   0.0$
&$   0.0\pm   0.0$
\\
c(H$\beta$) & 
&$   0.68\pm   0.09$
&$   0.37\pm   0.03$
&$   0.64\pm   0.07$
&$   0.76\pm   0.07$
&$   1.04\pm   0.01$
\\
velocity & 
&$  10191\pm     43$
&$  10186\pm      4$
&$   9485\pm      2$
&$  10830\pm     24$
&$  10320\pm     36$
\\
\enddata
\end{deluxetable}

\subsection{Comparison with previous data}\label{COMPARE}

\begin{figure}
\includegraphics[height=7cm,angle=0]{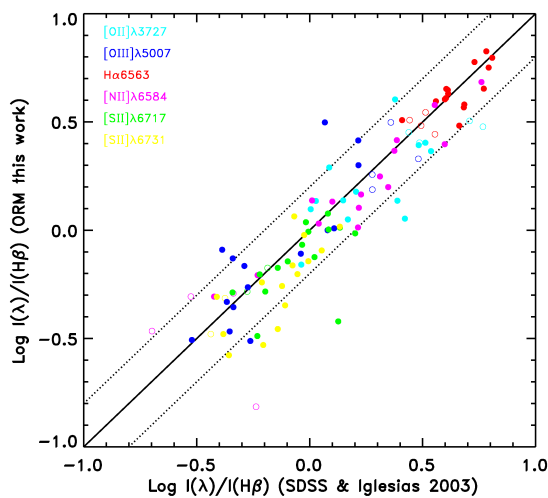}
\caption{Emission line ratios I($\lambda$)/I(H$\beta$) for the principal emission lines (without applying  reddening correction) as derived by the ORM long-slit spectra observed for this work versus corresponding ratios from SDSS (filled points) or IP03 spectra (open cycles) for the same galaxy regions. Cyan: \OII$\lambda3727$, red: H$\alpha$ $6563$, blue: \OIII$\lambda5007$, pink: \NII$\lambda6583$, green: \SII$\lambda6717$ and  yellow: \SII$\lambda6731$). The continuous line is the line of equality and the dotted lines indicate a variation by $\pm0.2$ dex. \label{compare}}
\end{figure}

SDSS spectroscopic data exist for 22 out of the 43 SF galaxies of the C09 sample. In most of the cases the SDSS fiber covers the central 3 arcsec of the galaxy; for  3 galaxies, NGC6050, NGC6045 and PGC057064, SDSS provides 2 different apertures on each of them.
We have applied the same procedure to these spectra as to our spatially resolved spectra, removing the continuum emission by fitting STARLIGHT and measuring line fluxes on the subtracted emission-line spectrum as described in  \S\ref{STL} and ~\ref{LF}. We then have compared these line fluxes  with our line fluxes for 19 galaxies for which we have ORM long-slit spectrum. We have checked first whether the SDSS aperture position matches the long-slit locus, and then we have extracted the spectrum from the corresponding area of the long-slit. The comparison is shown in Fig.~\ref{compare} where the principal emission lines are shown  when measured, without applying any reddening correction to them (cyan: \OII$3727$, red: H$\alpha$ $6563$, blue: \OIII$5007$, pink: \NII$6583$, green: \SII$6717$ and  yellow: \SII$6731$). Despite the fact that the covered regions are not identical  (this could be a problem because of the inhomogeneous nature of star-forming regions), we can see that line ratios show a good agreement. While most of the points lie within the 0.2 dex scatter region from the 1:1 line, there are some emission features that display larger discrepancy. This mostly happens in the cases of galaxies that SDSS aperture is not completely coincident with our long-slit spectrum. 

In addition, IP03 have spectroscopic data for 7 galaxies belonging to C09 sample of SF galaxies in the Hercules cluster. We have compared the line fluxes, before the reddening correction, for 5 galaxies for which we have spectra in common (Fig.~\ref{compare} open circles). We also see a very good agreement, with a small variation  expected given the different position angles and different data analyzes. 

The galaxies {\footnotesize SDSSJ160304.20+171126.7},  {\footnotesize SDSSJ160 520.58+175210.6} and {\footnotesize SDSSJ160305.24+171136.1}, quoted in the C09 sample, only have SDSS spectroscopy. The same procedure described in \S\ref{STL} and \ref{LF} has been applied to these galaxies. For the latter, even after applying STARLIGHT  on the SDSS spectrum, we do not appreciate any emission line, because the SDSS fiber is placed in the center of the galaxy where no  \Ha emission is present. The other two are included in our sample and their STARLIGHT outputs and line flux measurements are also quoted in Table~\ref{stlout} and \ref{LINES} respectively. Finally, in this work are also incorporated 2 galaxies, (LEDA3085054 and [D97]ce-143), which belong to the C09 sample and for which spectroscopic data are available by IP03; all these spectra giving a final coverage of the C09 sample of 72\%. 

\subsection{Abundance derivation }\label{AD}

The radiation emitted by a nebula depends on the abundances of the chemical elements and the physical state of the gas, specially its average temperature and density. So, physical and chemical properties of nebulae can be derived by measuring the intensities of collisionally excited emission lines.  After hydrogen and helium, the most abundant element in the universe is oxygen. The oxygen abundance can be derived from  because it exhibits emission lines in \HII region spectra which are bright and easy to measure. In practice, the metallicity of SF galaxies is quantified via the oxygen abundance,  as 12 + log(O/H). 

\begin{figure}
\includegraphics[height=7cm,angle=0]{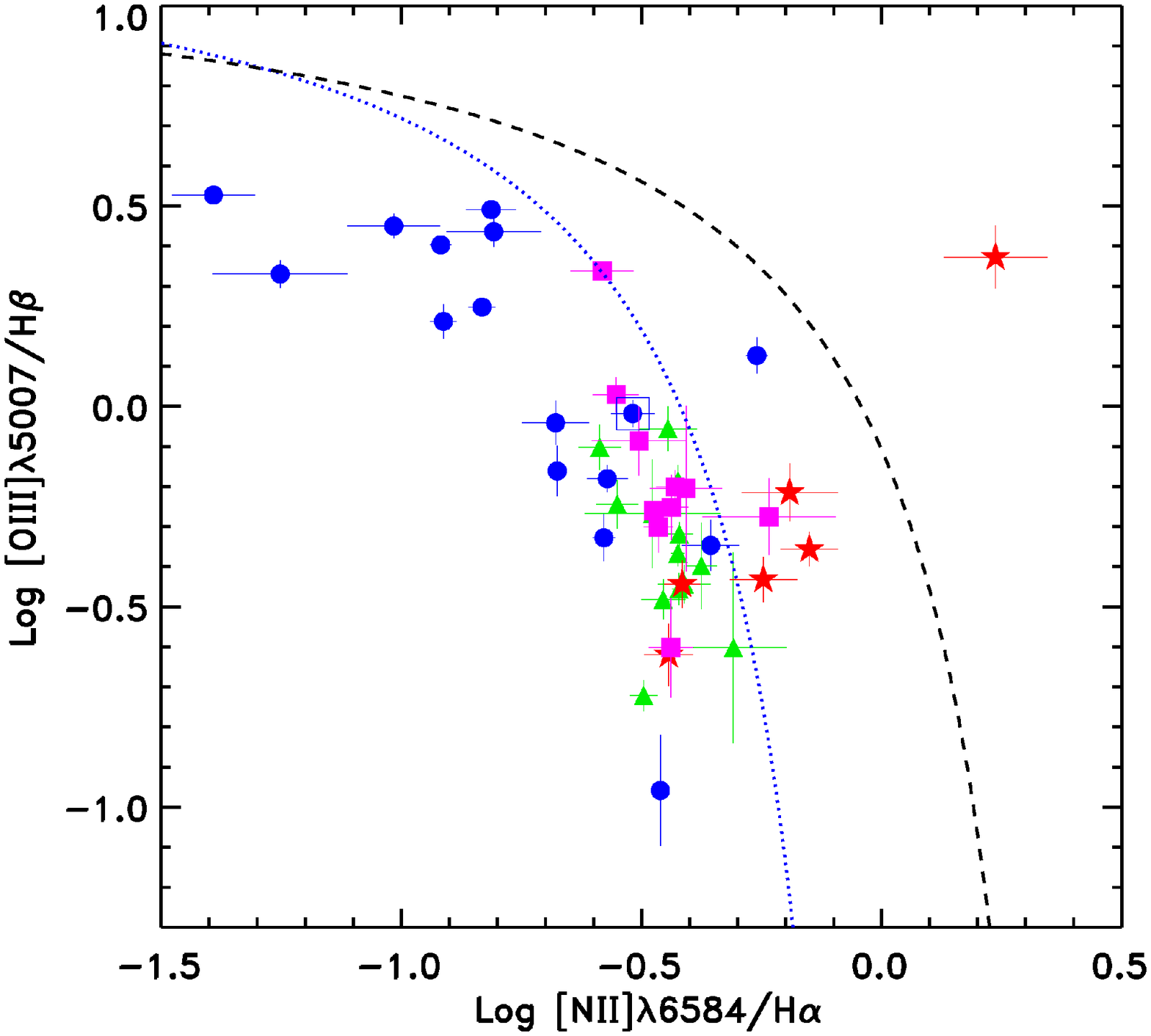}
\caption{BPT diagram and \citet{Kauffmann2003} (continuous blue) and \citet{Kewley2001} (dashed black) separation curves between objects with ionized gas produced by photoionization from massive stars and from non-stellar ionizing sources. 
We use four distinctive colors/symbols for our galaxies: blue filled circles for dwarf/irregular galaxies ($\mathrm{M_B} > -19$), magenta squares for luminous ($\mathrm{M_B} \leq -19$) but integrated galaxies, red stars for the nuclei of 6 spirals that we divide into different parts and green triangles for their corresponding disk components. The two parts of PGC057077 are plotted in magenta filled squares; similarly, the two parts of PGC057064 are plotted in blue filled circles.  We mark with a square the galaxy SDSS J160531.84+174826.1 for which which it has been claimed \citep{Dong2007} to host a Seyfert 1 AGN. 
\label{BPT}}
\end{figure}

Prior to the determination of the gas-phase chemical abundances, the excitation conditions of the gas have to be explored. In order to do that we use the BPT diagrams \citep{BPT81}. The derivation of gas-phase metal abundances is constrained to nebular emission-line spectra generated by photoionization from massive stars. Non-stellar ionizing sources, such as active galactic nuclei (AGN), produce generally distinctive emission-line ratios compared to ordinary \HII regions. In Fig.~\ref{BPT} we present the BPT diagram with \citet{Kauffmann2003} empirical demarcation curve with continuous blue line, and \citet{Kewley2001} theoretical upper limit for starburst galaxies with dashed black line, in order to identify non-stellar ionizing sources.
We use four distinctive colors/symbols for our sample of galaxies. With blue filled circles we plot the 16 integrated dwarf/irregular galaxies ($\mathrm{M_B} > -19$) of our sample\footnote{In figures where the ploted quantities involve measurement of \NII and \SII lines, the galaxy LEDA3085054 is not included}. With red stars we plot the nuclei of 6 spirals and with green triangles their corresponding disk regions. Magenta squares are used for luminous ($\mathrm{M_B} \leq -19$) galaxies for which we integrate the 2D long-slit spectrum into one 1D spectrum (PGC057077 is ploted with magenta squares and PGC057064 with blue filled circles, although the spectra of these peculiar galaxies have been divided into two parts each, for different reasons, see Appendix for details). Although we have divided the spectrum of the irregular galaxy LEDA084703 into three parts, from now on we consider the integrated values, because it shows homogeneous chemical composition (see Appendix for details on this galaxy).   In Fig.~\ref{BPT} we see that  the nuclei of NGC6045, NGC6050A, KUG1602+175A, galaxy  CGCG108+149 and the southern-east part of the peculiar object PGC057064 (see Appendix for special notes on this object) belong to the transition zone between the two separation curves. IC1182 lies on the separation curve of \citet{Kauffmann2003} and there is a controversy in the literature whether there is an AGN hosted in the nucleus of this merger (see \citet{Radovich2005} and references therein), but our long-slit spectrum is slightly off-set from the optical center. A detailed investigation of the nature of this peculiar object is out of the scope of the present paper and it will be presented in a forthcoming work (Petropoulou et al, in prep).  The nucleus of  IC1173 is clearly lying in the AGN region and we do not include it in our abundance analysis. We mark with a square the galaxy SDSS J160531.84+174826.1. It has been claimed  \citep{Dong2007} that this galaxy hosts a Seyfert 1 AGN; the position of this galaxy in the BPT diagram, set by the ratios of the lines in our emission line spectrum, is consistent with the area of the plot populated by SF objects.  

\begin{figure*}
\center
\includegraphics[height=7cm]{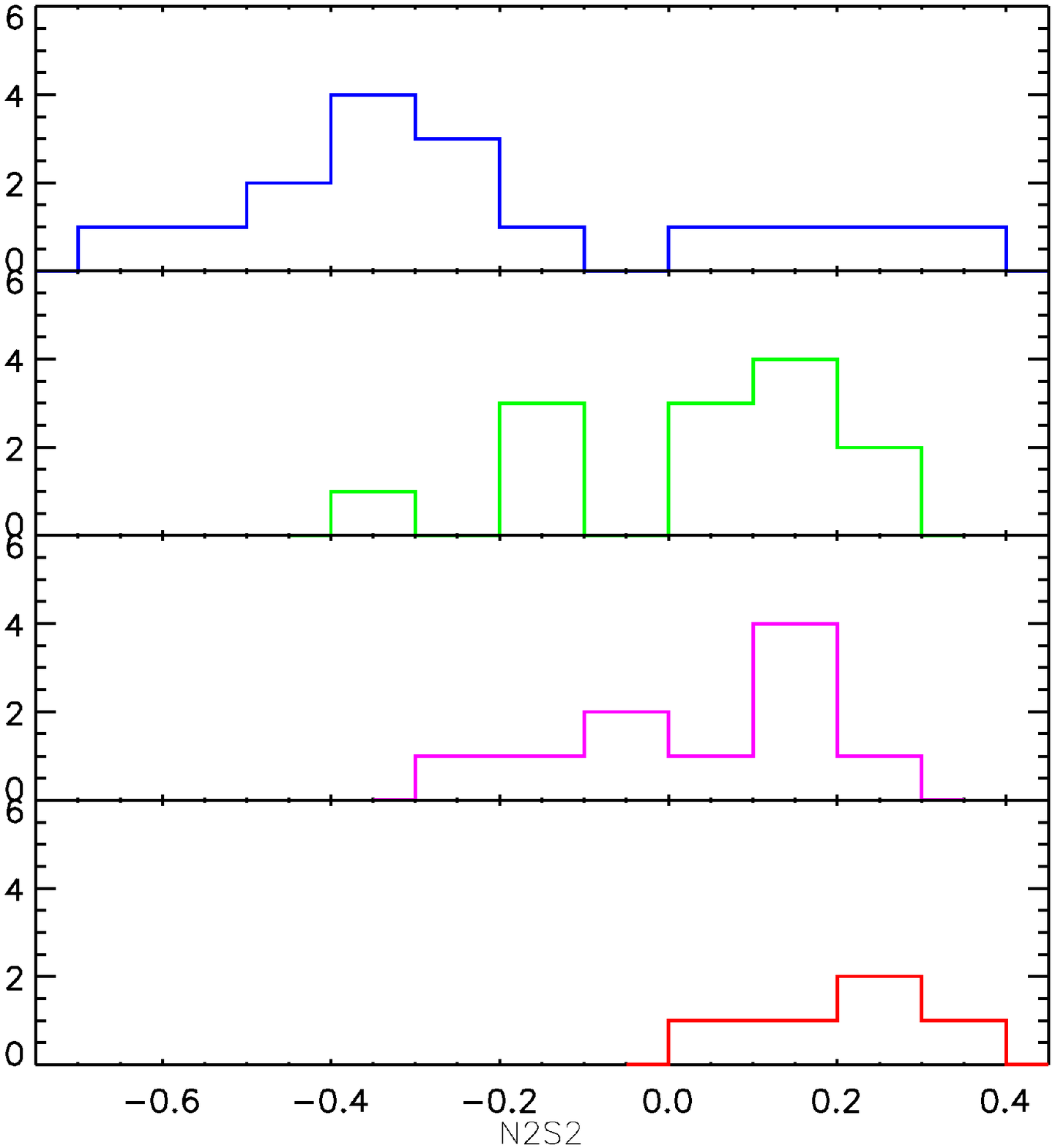}
\includegraphics[height=7cm,angle=0]{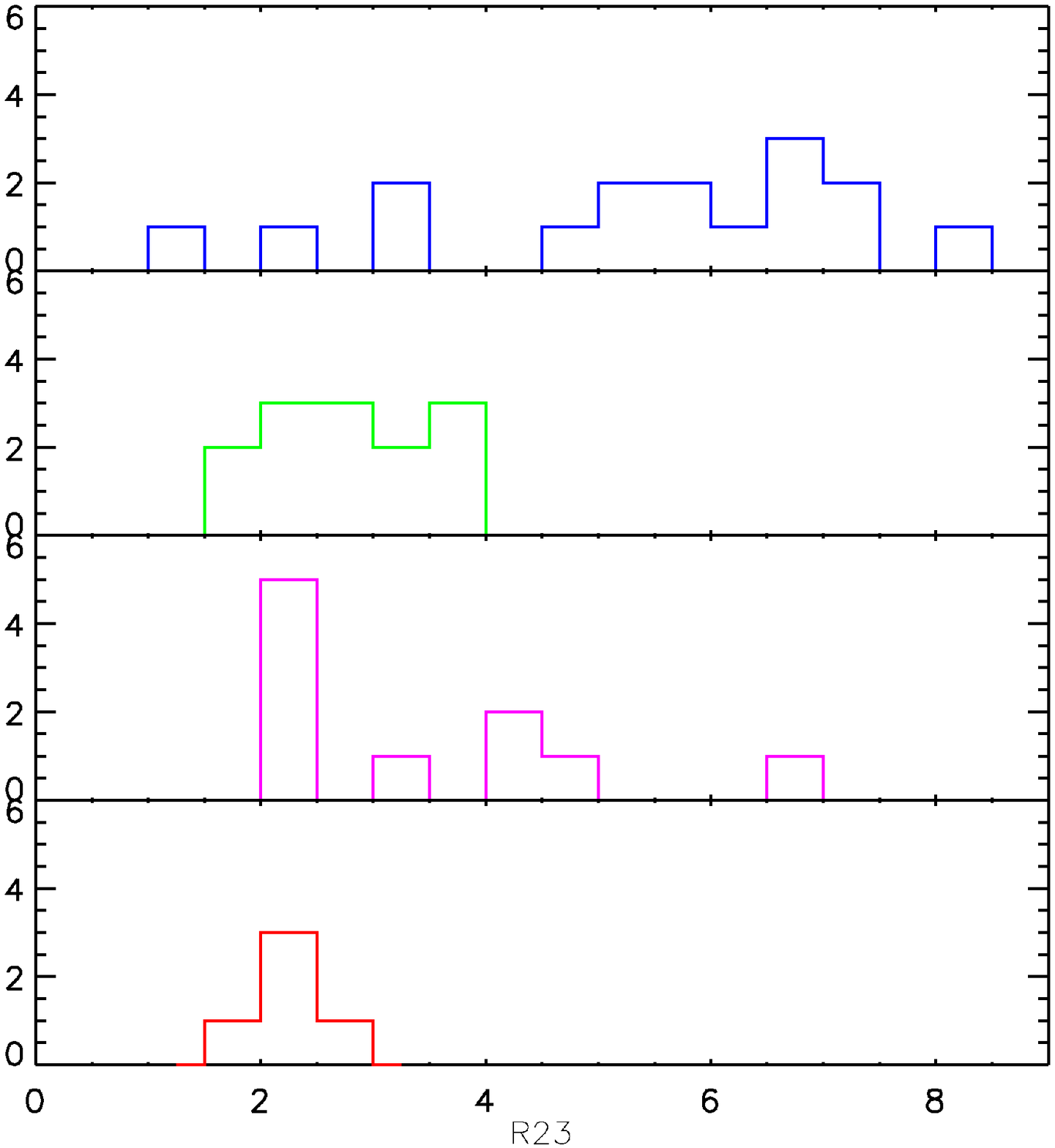}
\caption{The histograms of the distribution of the N2S2 (left) and R23 (right) parameters for our sample of galaxies, using the same colors to separate out our sample in four categories as in Fig.~\ref{BPT}. The histograms in both left and right panel  correspond to top to bottom: dwarf/irregulars, disks of spirals, integrated spirals, nuclei of spirals. \label{SEP}}
\end{figure*}

In Fig.~\ref{SEP} we show the histograms of the distribution of the N2S2\footnote{N2S2=$\mathrm{\log(I_{[NII]\lambda6584}/I_{[SII]\lambda\lambda6717,6731})}$\\ \citep[see][]{PM2009}} and R23\footnote{R23=R2+R3,\\ where R2=$\mathrm{I_{[OII]\lambda3727+\lambda3729}/I_{H\beta}}$\\ and R3=$\mathrm{I_{[OIII]\lambda4959+\lambda5007}/I_{H\beta}}$.\\ When  \OIII$\lambda4959$ is not measured, we assume the theoretical ratio: \OIII$\lambda4959$/\OIII$\lambda5007$=0.33} parameters for our sample of galaxies, using the same colors to separate out our sample in four categories as in Fig.~\ref{BPT}. The histograms in both left and right panel starting from the top correspond to: dwarf/irregulars, disks of spirals, integrated spirals, nuclei of spirals.  The most populated N2S2 bin for the dwarf/irregular galaxies of our sample (left panel, first histogram from the top, with blue line) lies around N2S2 = -0.4, although there are few objects sorting out to higher values. These high values, as we will see later in the discusion, seem consistent with the properties of these galaxies. Spiral galaxies are separated into 3 categories. The galaxies for which we only have integrated spectrum (magenta line, third from the top) show N2S2 values extending up to 0.4 and the most populated range is around 0.1-0.2. The other two categories correspond to massive galaxies divided into nuclei (red line, forth from the top) and disks (green, second from the top). Although both distributions show a common range of N2S2 values from 0.0 to 0.3, there is a tendency for the spectra of the central parts to present higher N2S2 than the disks which can reach values as low as -0.3. In the same line for the R23 distribution, the dwarfs (right panel, first histogram from the top, with blue line) present the larger values of R23 up to 8 with few objects presenting values as low as R23 = 1-3 (as we will show later this is a hint on the  nature of these objects). However, as a possible consequence of the bi-valuated nature of the O/H vs R23 function, the separation among the bright galaxies is not as clear as in the left panel. Nonetheless it can be seen that, for integrated bright galaxies (magenta, third from the top), R23 distribution shows a large range with values from 2 to 7 with the most populated values close to 2; for the spatially resolved galaxies, the range of disks reaches higher values (up to 4) compared to nuclei (up to 3). 
This analysis highlights the fact that using SDSS spectroscopy, which most probably means sampling only galaxy centers (bottom histogram, with red line), could  introduce a bias in the abundance results.  

There are two commonly used methods to determine oxygen abundances in H II regions. The direct method is founded
on a direct measurement of the electron temperature, applicable when the collisionally excited lines such as [O III]$\lambda$4363, [NII]$\lambda$5755, [SIII]$\lambda$6312  are measured. We do not detect any of the temperature diagnostic lines in our spectra. In addition, considering Hercules distance and the resolution of our observations, we are constrained to integrate galaxy spectra over large spatial scales, in some cases containing several HII regions, with potentially different ionization conditions. In these cases, the use of the direct method to derive abundances could result misleading  \citep[e.g][]{Kobulnicky1999}. 

The other way to determine oxygen abundances in HII regions is to use semi-empirical calibrations. There are two different types of calibrations: the empirical ones, based on fits to objects for which an accurate direct derivation of O/H is available \citep[e.g.][]{Pettini2004,Nagao2006,PM2009,Pilyugin2005,Pilyugin2010} and the ones based on predictions of theoretical photoionization models \citep[e.g.][]{McGaugh1991,Kewley2002,Tremonti2004}. For an extensive review see  \citet{Lopez2010} and \citet{Kewley2008}.

It has been shown recently \citep{Bresolin2009} that model calibrations show a positive shift of 0.3 dex compared to oxygen abundances derived with the direct method, whereas the latter agree very well with oxygen abundances derived for the massive young stars. For our sample we also find a 0.3 dex positive shift between the abundances predicted using \citet{McGaugh1991} model calibration  \citep[formulated by][]{Kobulnicky1999} and the empirical calibration  calculated by \citet{Pilyugin2010}. Taking this into account, in this analysis we have chosen using empirical calibrations. 

Our sample spans from very low metallicity dwarfs to evolved disks and galaxy centers, with an important fraction lying in the turn-over region of the R23 versus oxygen abundance relation (see Fig.~\ref{SEP}). Among the empirical calibrations we use the N2-calibration given by \citet{PM2009} (from now on PMC09) and the recent calibration by \citet{Pilyugin2010} (from now on P10), because they have the advantage of being valid along the whole metallicity range of our sample. The N2\footnote{Defined by PMC09 as N2=$\mathrm{\log\left(\mathrm{I_{[N II]\lambda6584}/I_{H\alpha}}\right)}$} PMC09 parameter shows a monotonic relation with oxygen abundance, thus avoiding the degeneracy problem of R23; though this calibration involves a large rms error of up to $\sim0.3$ dex. In order to circumvent this problem  P10 has provided with a new improved calibration which uses a multiparametric function of P, R2, R3, S2, N2\footnote{Defined in P10 as $\mathrm{N2=I_{[NII]\lambda6548+\lambda6584}/I_{H\beta}}$,\\ $\mathrm{S2=I_{[SII]\lambda6717+\lambda6731}/I_{H\beta}}$,  and P=R3/R23.\\ The theoretical ratio \NII$\lambda6548$/\NII$\lambda6584$=0.3 is assumed when necessary.} parameters and produces a very small rms error $\sim 0.07$ dex. In Table~\ref{ABU} we give the oxygen abundance derived using both abundance calibrations. We are aware that using the nitrogen as an abundance indicator to derive oxygen abundances can be potentially  dangerous in case of objects with particular evolution, for example for nitrogen enriched galaxies (see PMC09). Comparing both abundance calibrations, we interpret PMC09 abundances as an upper estimate while P10 as a lower estimate and, being conservative,  we have adopted the mean of the two oxygen abundances. For each galaxy of our sample the adopted O/H value is consistent with the prediction of both calibrations within the statistical errors. For each O/H determination we have estimated the corresponding error as the  rms of a gaussian error distribution produced by a random sampling simulation taking into account the errors of the P, R23, R2, R3, S2, and N2 parameters. This error estimation has been adopted in all cases except in those cases for which it is less than the statistical error provided by the P10 calibration ($\sim 0.07$ dex), which was finally adopted. As can be noted in Table~\ref{ABU}, the error adopted is very close to the half of the difference in O/H provided by both calibrations.  

\begin{deluxetable}{lccccc}
\tablecolumns{6}
\tabletypesize{\scriptsize}
\tablecaption{Chemical Abundances\label{ABU}}
\tablewidth{0pt}
\tablehead{
\colhead{GALAXY} 
& \colhead{12+log(O/H)} 
& \colhead{12+log(O/H)} 
& \colhead{12+log(O/H)} 
& \colhead{error} 
& \colhead{log(N/O)} \\
\colhead{} 
& \colhead{PMC09} 
& \colhead{P10} 
& \colhead{adopted} 
& \colhead{adopted} 
& \colhead{PMC09} 
}
\startdata
PGC057185a &           8.64   &    8.48   &    8.56   &    0.07  &    -1.04     \\ 
PGC057185b &           8.74   &    8.53   &    8.63   &    0.07  &    -0.72     \\
PGC057185c &           8.69   &    8.50   &    8.60   &    0.12  &    -0.86     \\
IC1173a &              8.83   &    8.56   &    8.69   &    0.26  &    -0.63     \\
IC1173e &              8.72   &    8.57   &    8.65   &    0.07  &    -1.05     \\
KUG1603+179Aa &        8.73   &    8.53   &    8.63   &    0.07  &    -0.74     \\
KUG1603+179Ab &        8.88   &    8.57   &    8.72   &    0.07  &    -0.54     \\
KUG1603+179Ac &        8.74   &    8.53   &    8.64   &    0.07  &    -0.69     \\
NGC6050Aa &            8.92   &    8.53   &    8.72   &    0.07  &    -0.55     \\
NGC6050Ab &            8.71   &    8.59   &    8.65   &    0.07  &    -0.69     \\
NGC6050B &             8.63   &    8.52   &    8.58   &    0.07  &    -1.02     \\
LEDA1543586 &          8.43   &    8.34   &    8.38   &    0.07  &    -1.15     \\
NGC60545a &            8.77   &    8.49   &    8.63   &    0.09  &    -0.71     \\
NGC6045b &             8.74   &    8.54   &    8.64   &    0.07  &    -0.57     \\
NGC6045c &             8.95   &    8.54   &    8.75   &    0.07  &    -0.36     \\
NGC6045d &             8.68   &    8.61   &    8.65   &    0.07  &    -0.53     \\
NGC6045e &             8.74   &    8.46   &    8.60   &    0.07  &    -0.74     \\
KUG1602+174A &         8.72   &    8.58   &    8.65   &    0.12  &    -0.81     \\
LEDA084719 &           8.75   &    8.46   &    8.61   &    0.26  &    -0.69     \\
PGC057077a &           8.73   &    8.46   &    8.60   &    0.07  &    -0.64     \\
PGC057077b &           8.70   &    8.54   &    8.62   &    0.07  &    -0.63     \\
UGC10190a &            8.75   &    8.53   &    8.64   &    0.07  &    -1.01     \\
UGC10190b &            8.72   &    8.58   &    8.65   &    0.07  &    -0.86     \\
UGC10190c &            8.61   &    8.50   &    8.55   &    0.07  &    -1.27     \\
LEDA140568 &           8.08   &    8.05   &    8.07   &    0.11  &    -1.60     \\
$[$D97$]$ce-200 &          8.35   &    8.19   &    8.27   &    0.07  &    -1.49     \\
PGC57064a &            8.79   &    8.50   &    8.64   &    0.07  &    -0.58     \\
PGC057064b &           8.86   &    8.35   &    8.61   &    0.07  &    -0.81     \\
LEDA084703int &        8.41   &    8.28   &    8.35   &    0.07  &    -1.29     \\
LEDA084703a &          8.40   &    8.24   &    8.32   &    0.07  &    -1.29     \\
LEDA084703b &          8.42   &    8.29   &    8.35   &    0.07  &    -1.25     \\
LEDA084703c &          8.46   &    8.26   &    8.36   &    0.07  &    -1.40     \\
KUG1602+175 &          8.72   &    8.54   &    8.63   &    0.07  &    -0.63     \\
LEDA084710 &           8.67   &    8.44   &    8.56   &    0.27  &    -0.92     \\
CGCG108-149 &          8.88   &    8.53   &    8.71   &    0.12  &    -0.61     \\
KUG1602+174B &         8.70   &    8.54   &    8.62   &    0.07  &    -0.87     \\
LEDA084724 &           8.53   &    8.42   &    8.48   &    0.07  &    -1.24     \\
SDSSJ160556.98+174304.1&          8.27   &    8.21   &    8.24   &    0.11  &    -1.36     \\
$[$DKP87$]$160310.21+175956.7&    8.61   &    8.50   &    8.56   &    0.07  &    -1.19     \\
SDSSJ160524.27+175329.3&          7.97   &    8.01   &    7.99   &    0.07  &    -1.67     \\
SDSSJ160523.66+174832.3&          8.54   &    8.28   &    8.41   &    0.07  &    -1.30     \\
              IC1182   &          8.61   &    8.41   &    8.51   &    0.07  &    -1.19     \\
  IC1182:[S72]d        &          8.34   &    8.27   &    8.31   &    0.07  &    -1.29     \\
SDSSJ160531.84+174826.1&          8.66   &    8.64   &    8.65   &    0.07  &    -0.45     \\
SDSSJ160304.20+171126.7&          8.62   &    8.44   &    8.53   &    0.07  &    -1.10     \\
SDSSJ160520.58+175210.6&          8.71   &    8.68   &    8.69   &    0.15  &    -0.64     \\
$[$D97$]$ce-143          &            8.43   &    8.26   &    8.35   &    0.07  &    -1.40     \\
LEDA3085054 & \nodata &7.43\tablenotemark{a} & 7.58 & 0.07 & -0.86\tablenotemark{a} \\
\enddata  
\tablecomments{The oxygen abundance derived using \citet{PM2009} and \citet{Pilyugin2010} calibrations. N/O is derived using N2S2 calibration by \citet{PM2009}.}
\tablenotetext{a}{These values are given by the used calibration assuming upper limits for the \NII and \SII lines. The adopted O/H value is the one provided by \citet{Iglesias2003} calculated using the P-method; we see that is in very good agreement with the $O/H$ value that we calculate with the P10 method.}
\end{deluxetable}


Nitrogen is also an important element to investigate galaxy evolution. Nitrogen abundance deserves a more  thorough determination; in particular the importance of possible self-enrichment \citep{LS2010IV, Monreal2010a}  and the differential chemical evolution of N versus O could add some degree of uncertainty. These effects have been extensively considered by PMC09, thus in this work we have adopted their calibration for nitrogen given by the N2S2 parameter which has a typical rms error of $\sim 0.3$ dex. In Table~\ref{ABU} the N/O values are also presented.

\begin{figure*}
\includegraphics[height=9cm,angle=0]{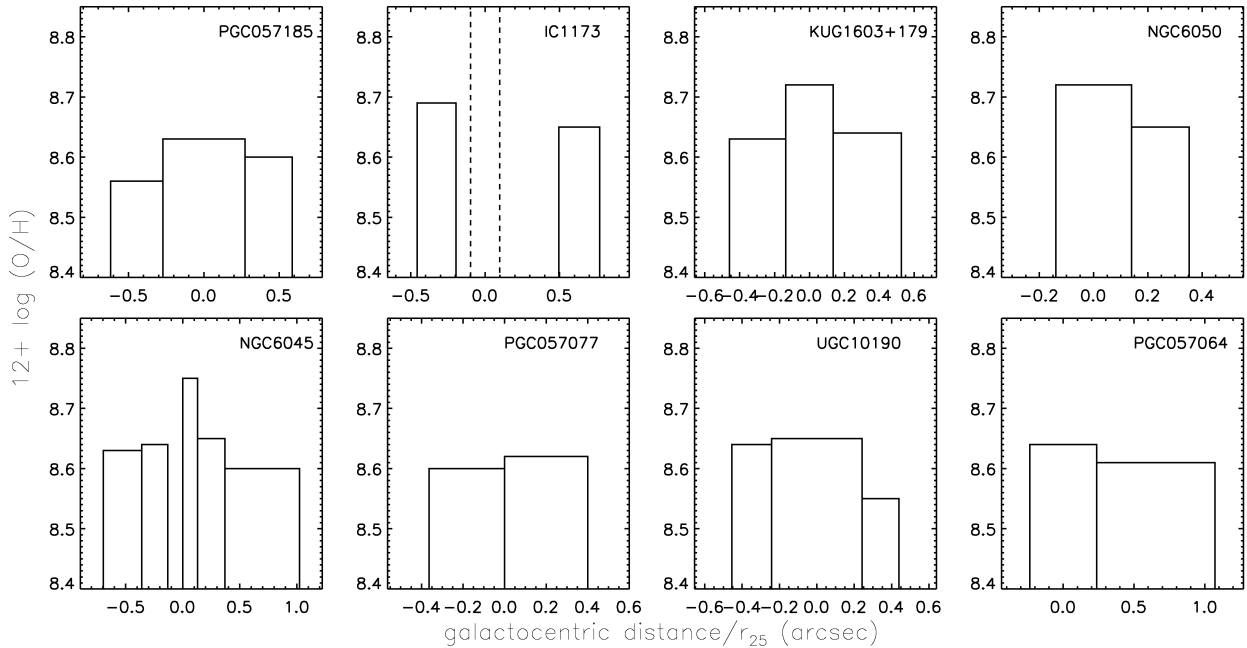}
\caption{Oxygen abundances of the galaxies with spatially resolved spectroscopy in radial bins of galactocentric distance in arcsec normalized to  $r_{25}$. For IC1173, dashed lines indicate the radial bin of the AGN extracted spectrum.\label{GRA1}}
\end{figure*}

\begin{figure*}
\includegraphics[height=9cm,angle=0]{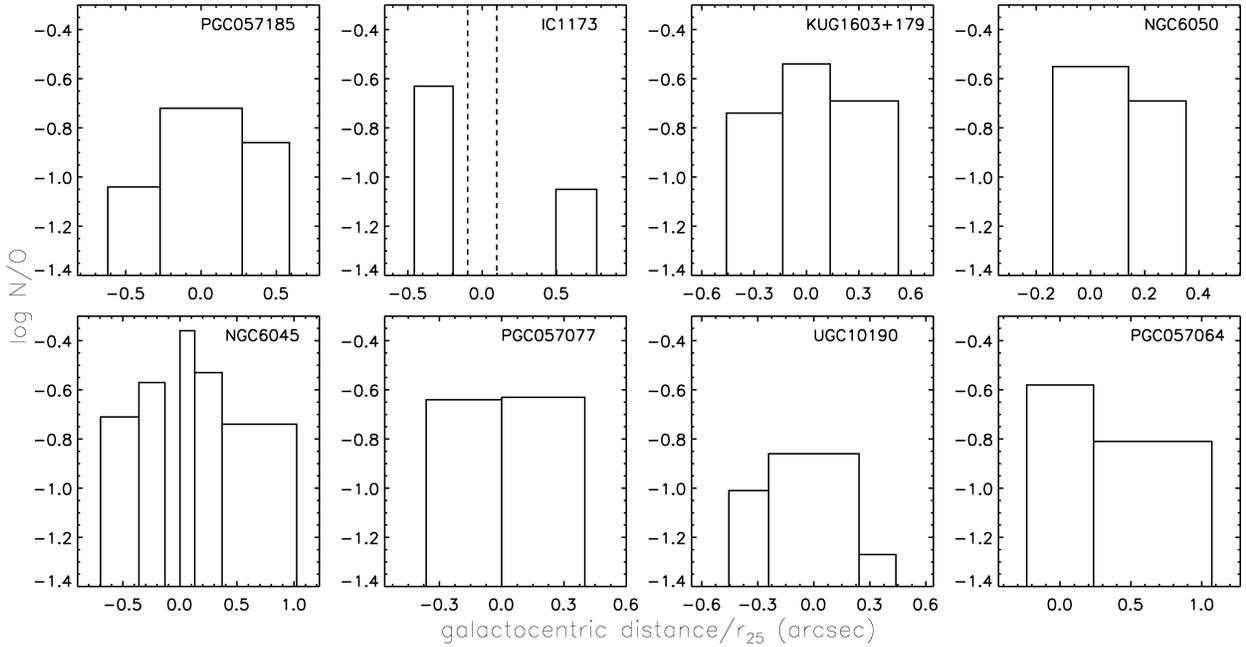}
\caption{The same as in Fig.~\ref{GRA1} for N/O.\label{GRA2}}
\end{figure*}

For the spatially resolved spirals of our sample we have explored the abundance gradients. Figs.~\ref{GRA1},~\ref{GRA2} show the O/H and N/O abundance profiles, respectively, for the 6 spiral galaxies of our sample for which individual spectra, corresponding to different spatial regions, have been extracted.  We plot the chemical abundances derived for each region in bins of galactocentric distance in arcsec, normalized to the $r_{25}$ radius\footnote{Radius $r_{25}$ is the radius of the galaxies at a surface brightness of 25 mag/arcsec$^2$.} extracted from SDSS r' band images (see Table~\ref{PHY}). The plots show mild or flat O/H abundance gradients across a limited radial extension of the disk 

\subsection{Physical Properties}\label{PP}

We derive stellar mass for each galaxy using the kcorrect\_v4 algorithm (version 4\_2) of \citet{Blanton2007}. This IDL code corrects implicitly for dust and assumes a universal IMF of \citet{Chabrier2003} form. The SEDs used to best fit the observed SDSS ugriz photometry are based on \citet{Bruzual2003} stellar evolution synthesis codes. The best-fitting SED provides an estimate of the stellar mass-to-light ratio\footnote{The $\chi^{2}$ of the fit is less than 5 (1) for 73\% (27\%) of our sample}. \citet{Drory2004} argue that broadband color mass estimators, can yield fairly reliable stellar masses for galaxies within $\sim 0.2$ dex over almost 4 decades in mass. \citet{Li2009} demonstrate that, once all estimates are adapted to assume the same IMF, the  \citet{Blanton2007} masses agree quite well with those obtained from the single-colour estimator of \citet{Bell2003} and also with those derived by \citet{Kauffmann2003} from a combination of SDSS photometry and spectroscopy. More recently, \citet{Zahid2010} also state that optical bands are sufficient to constrain the SED fit for determining masses of SF galaxies. The galaxy stellar masses given from the algorithm were converted to H$_0$=73 km s$^{-1}$ Mpc$^{-1}$ adopted in this work and are presented in Table~\ref{PHY}.
 
We also calculate the two-dimensional local galaxy density $\Sigma$, using two methods.
We use the density estimator based on an average of the projected distances (d, in Mpc) to the fourth and fifth nearest neighbor as given by \citet{Mouhcine2007} and used by \citet{Ellison2009}
\begin{equation}
\log \Sigma_{4,5}=\frac{1}{2} \log\left(\frac{4}{\pi d^2_4}\right)+\frac{1}{2} \log\left(\frac{5}{\pi d^2_5}\right)\label{MOUH} 
\end{equation} 
considering all galaxies with SDSS spectra in the Hercules region  at the corresponding redshift range, and presenting a difference in radial velocity smaller than 750 km/s, a value approximately equal to $\mathrm{\sigma_V}$ of the cluster. 

To account for possible fiber collisions which may lead to an under-estimate of the local density in regions rich in projected galaxies as cluster centers, we calculate the local galaxy number density to the 10$^{th}$ nearest neighbor $\Sigma_{10}$, using a magnitude limited sample of galaxies taken out from SDSS database without velocity constrain. 
Counts of galaxies as a proxy of local environment has the main weakness that is missing the luminosity information of the counted galaxies. To account for this problem, we limit to luminous galaxies $M_{r'} = -19, m_{r'}=17$ mag when corrected for galactic extinction.
To  correct for background and foreground galaxies, we use  the field galaxy counts as given by \citet{Yasuda2001}. The local galaxy number density is calculated to the 10$^{th}$ neighbor in order to improve statistics.
As expected, the corresponding densities were  higher, but they do not affect the conclusions of the MZ relation analysis as will be discused in \S\ref{DIS}. 
The density estimates $\log \Sigma_{4,5}$ and $\log \Sigma_{10}$ are quoted in Table~\ref{PHY}.

We then estimate the projected distance of our galaxies to the cluster center. As already mentioned in \S\ref{INTRO}, Hercules is a peculiar cluster of galaxies, where the maximum of the cluster galaxy distribution (16$^{\mathrm{h}}$05$^{\mathrm{m}}$15.0$^{\mathrm{s}}$ +17\dd44\mm55\scn,  extracted from  NED) is not coincident with the primary X-ray maximum found by \citet{Huang1996}.  The center of the brightest X-ray component coincides with the brightest cluster galaxy NGC6041A (16\hh04\mm35.8\scn +17\dd43\mm18\scn). In Table~\ref{PHY} we give the projected distance of our sample galaxies to both centers in Mpc, referred to the cosmological corrected distance of Hercules cluster 158.3 Mpc. 

We calculate the SFR of our galaxies, from their \Ha emission given by C09 using the \citet{Kennicutt1998} calibration. We use c(H$\beta$) derived by our optical spectroscopy to correct \Ha emission from extinction, assuming the \citet{miller1972} extinction law with $R_V=3.2$. When spectra of different parts of a galaxy are considered, the c(H$\beta$) used to calculate the global SFR of the galaxy is the mean value of all these c(H$\beta$) derived spectroscopically.  We also correct the \Ha flux from \NII contamination using the empirical correction given by \citet{Reverte2008}: 
\begin{eqnarray}
\log \mbox{EW(H$\alpha$)}&=&(-0.34\pm0.03)+\nonumber\\
(1.13&\!\!\!\!\pm&\!\!\!\!0.02)\log \mbox{EW(H$\alpha$+[N {\scshape{ii}}])}
\end{eqnarray}
This empirical correction was derived using line fluxes integrated for entire galaxies, extracted from the extended spectrophotometric galaxy sample of \citet{Jansen2000}, suitable for this kind of analysis. 
The errors on SFR quoted take into account only the  \Ha flux error.  

Finally we note that the HI survey of \citet{Dickey1997} covers the whole central region of A2151 where all our SF galaxies lie. We have adopted the HI masses provided in that work (converting them to H$_0$ = 73 km s$^{-1}$ Mpc$^{-1}$).

Table~\ref{PHY} summarizes for each galaxy the physical properties calculated in this work and others extracted from the literature: the $\B$ absolute magnitude, the galaxy radius at 25 mag arcsec$^{-2}$ in arcsec (extracted from  the SDSS $iso_{A}$ parameter for the r' band), the local density estimates log$\Sigma_{4,5}$ and log$\Sigma_{10}$, the projected distance to the cluster center and the projected distance to the X-Rays center in Mpc, the stellar mass calculated with kcorrect algorithm, the HI mass from \citet{Dickey1997}, and the SFR in $M_\sun$ yr$^{-1}$.


\begin{deluxetable}{lccccccccc}
\tabletypesize{\scriptsize}
\tablecaption{Physical Properties\label{PHY}}
\tablewidth{0pt}
\tablehead{
\colhead{Galaxy} & \colhead{$\B$}  &\colhead{$iso_A$} & \colhead{$\log \Sigma_{4,5}$} & \colhead{$\log \Sigma_{10}$} & \colhead{$R$} &\colhead{$R_X$} & \colhead{$M_\star$} & \colhead{$M_{HI}$}& \colhead{SFR}\\
\colhead{} & \colhead{mag}  &\colhead{arcsec}  &\colhead{} &\colhead{}&\colhead{Mpc} &\colhead{Mpc} & \colhead{$10^8 M_\sun$} & \colhead{$10^8 M_\sun$} & \colhead{$M_\sun yr^{-1}$}}
\startdata
PGC057185                       &-19.51  & 24.24     &1.4 &   1.5 &  1.06  &  1.47  &    169.8        &  \nodata                     & $ 0.8 \pm  0.1  $     \\
IC1173                          &-20.79  & 30.68     &1.1 &   1.5 &  0.9   &  0.92  &    444.7        &  37.3                        & $ 7.3 \pm  0.7  $     \\
KUG1603+179A                    &-20.05  & 23.54     &2.0 &   2.4 &  0.18  &  0.62  &    275.9        &  128.5                       & $ 3.9 \pm  0.2  $     \\
NGC6050A                        &-20.87  & 30.10     &1.7 &   2.4 &  0.1   &  0.53  &    27.6         &  53.1                        &   \nodata             \\
NGC6050B                        &-19.30  &\nodata    &2.3 &   2.4 &  0.1   &  0.53  &    262.7        &  \nodata                     &   \nodata             \\
LEDA1543586                     &-18.46  & 13.43     &1.8 &   1.9 &  0.3   &  0.53  &    6.3          &  27.6                        & $ 0.19 \pm  0.01$     \\
NGC6045                         &-21.04  & 36.76     &2.0 &   2.4 &  0.08  &  0.37  &    806.9         &  46.8\tablenotemark{\dagger}& $15.2 \pm  1.0  $     \\
KUG1602+174A                    &-19.24  & 13.47     &1.2  &   2.3 &  1.17  &  1.03  &    72.6         &  \nodata                     & $ 0.6 \pm  0.1  $     \\
LEDA084719                      &-19.20  & 10.60     &1.6  &   1.9 &  0.29  &  0.4   &    83.7         &  \nodata                     & $ 1.5 \pm  0.3  $     \\
PGC057077                       &-19.11  & 11.00     &2.0  &   2.2 &  0.22  &  0.65  &    53.3         &  \nodata                     & $ 8.8\tablenotemark{\ddagger}$\\
UGC10190                        &-19.21  & 26.40     &1.9  &   2.1 &  0.19  &  0.56  &    60.8          &  100.8                      & $ 0.4 \pm  0.1      $ \\
LEDA140568                      &-17.42  & 13.65     &1.2  &   1.5 &  0.49  &  0.93  &    2.1          &  36.4                        & $ 0.017 \pm  0.002  $ \\
$[D97]$ce-200                   &-17.74  &  8.54     &2.3  &   2.3 &  0.13  &  0.38  &    3.0          &  9.0                         & $ 0.091 \pm  0.006  $ \\
PGC057064                       &-18.65\tablenotemark{\star} & 18.65\tablenotemark{\star}   & 2.1 &   1.9 &   0.26  &  0.64  &   104.0\tablenotemark{\star} & \nodata & $ 3.9 \pm  0.9$\tablenotemark{\star} \\
LEDA084703                      &-18.79  & 15.94    &1.6 &   1.7 &   2.13  &  1.81  &    37.9         &  26.5                        & $ 0.73 \pm  0.03    $ \\
KUG1602+175                     &-20.57  & 21.92    &1.3 &   1.6 &   0.89  &  0.76  &    191.4         &  99.6                       & $ 4.7 \pm  0.3      $ \\
LEDA084710                      &-19.23  & 18.85    &1.2 &   1.7 &   1.05  &  0.8   &    122.9        &  \nodata                     & $ 3.9 \pm  0.2      $ \\
CGCG108-149                     &-20.14  & 18.87    &1.3 &   1.6 &   0.97  &  1.39  &    311.5        &  15.5\tablenotemark{\dagger} & $ 3.2 \pm  0.3      $ \\
KUG1602+174B                    &-19.70  & 18.65    &1.2 &   1.8 &   1.15  &  1.04  &    96.1         &  49.9                        & $ 1.9 \pm  0.2      $ \\
LEDA084724                      &-18.72  & 14.82    &1.0 &   1.4 &   0.57  &  0.86  &    42.6          &  \nodata                    & $ 1.002 \pm  0.03   $ \\
SDSS J160556.98+174304.1        &-17.09  &  5.54    &1.2 &   1.5 &   0.47  &  0.89  &    3.3           &  \nodata                    & $ 0.210 \pm  0.009  $ \\
$[$DKP87$]$160310.21+175956.7   &-17.41  &  7.59    &2.0 &   1.8 &   0.34  &  0.67  &    5.7          &  \nodata                     & $ 0.06 \pm  0.02    $ \\
SDSS J160524.27+175329.3        &-16.16  &  3.53    &1.3 &   1.8 &   0.41  &  0.71  &    1.1          &  \nodata                     & $ 0.022 \pm  0.002  $ \\
SDSS J160523.66+174832.3        &-16.76  &  5.88    &1.9 &   2.1 &   0.19  &  0.58  &    3.4           &  \nodata                    & $ 0.07 \pm  0.02    $ \\
IC1182                          &-20.98  & 27.24    &2.1 &   2.0 &   0.28  &  0.7   &    654.9         &  \nodata                    & $22.1 \pm  0.6      $ \\
IC1182:[S72]d                   &-17.73  &  9.62    &1.9 &   1.9 &   0.33  &  0.76  &    1.3           &  186.0                      & $ 0.83 \pm  0.02    $ \\
SDSS J150531.84+174826.1        &-17.23  &  5.21    &2.2 &   2.0 &   0.25  &  0.66  &    15.5         &  \nodata                     & $ 0.146 \pm  0.009  $ \\
SDSS J160304.20+171126.7	&-17.42  & 9.05     &1.6 &   1.7 &   2.1   &  1.78  &   24.8         &  \nodata                     & $0.55 \pm  0.03     $ \\
SDSS J160520.58+175210.6	&-18.51  & 13.85    &1.9 &   1.9 &   0.34  &  0.64  &   10.6         &  \nodata                     & $0.20 \pm  0.02     $ \\ 
$[D97]$ce-143			&-17.56  & 8.29     &1.4 &   1.9 &  0.33   &  0.64  &    14.9         &  16.7                     & $0.025 \pm  0.004   $ \\ 
LEDA3085054 			&-16.23  &  4.65    &1.4 &   1.5 &  0.46   &  0.88  &    0.7          &  19.7                        & $0.045 \pm  0.002   $ \\
\enddata                                                                                                       
\tablecomments{Column1: galaxy name from C09; Column 2: $B$ absolute magnitude, calculated from SDSS g' magnitudes, assuming a distance to A2151 of 158.3 Mpc and an average correction of $g'-B=-0.3$ mag \citep{Fukugita1995}. All the quantities of this table are refered to the cosmological corrected distance of Hercules cluster assuming H$_0$=73 km s$^{-1}$ Mpc$^{-1}$; Column 3: SDSS $iso_A$ parameter for the r' band in arcsec;  Column 4: local density estimator $\log \Sigma_{4,5}$; Column 5: local density estimator $\log \Sigma_{10}$; Column 6: projected distance to the cluster center in Mpc; Column 7: projected distance to the X-Rays center in Mpc; Column 8: stellar mass calculated with kcorrect; Column 9: HI mass from \citet{Dickey1997}; Column 10: SFR in $M_\sun$ yr$^{-1}$.}
\tablenotetext{\dagger}{For NGC6045 and CGCG108-149 \citet{Dickey1997} data are incomplete or non existent, then we have used HI masses from \citet{Giovanelli1981}. For NGC6045 HI measurements provided by these two works are consistent, for the galaxy part that was included in \citet{Dickey1997} due to the velocity cut-off of the spectrometer used, see Appendix for details}
\tablenotetext{\ddagger}{The derived SFR is uncertain, as this galaxy shows two parts suffering completely different extinctions. The average extinction has been used.}
\tablenotetext{\star}{These physical properties have been derived on the basis of integrated photometric data of this merger being considered as one galaxy.}
\end{deluxetable}


\section{DISCUSSION}\label{DIS}

\subsection{Metallicity vs Local Density}\label{LD}

In this paper we study the relation between metallicity and environment for our sample of SF galaxies in the Hercules cluster. We examine the potential impact of the environment on the MZ and LZ relations and investigate different evolutionary scenarios of these cluster galaxies. 

The Hercules cluster three dimensional galaxy distribution is extremely clumpy \citep{Bird1995}, rendering it a very interesting vivid environment difficult to be disentangled only in the projected space without taking into account galaxy velocity.   The local density of galaxies has been widely used as an indicator of their environment; in this work we have studied the behavior of metallicity versus local density. The local density $\Sigma_{4,5}$, derived to the 4$^{th}$ and 5$^{th}$ nearest neighbor would describe the local environment 
at the group-scale.

\begin{figure*}
\includegraphics[height=7cm]{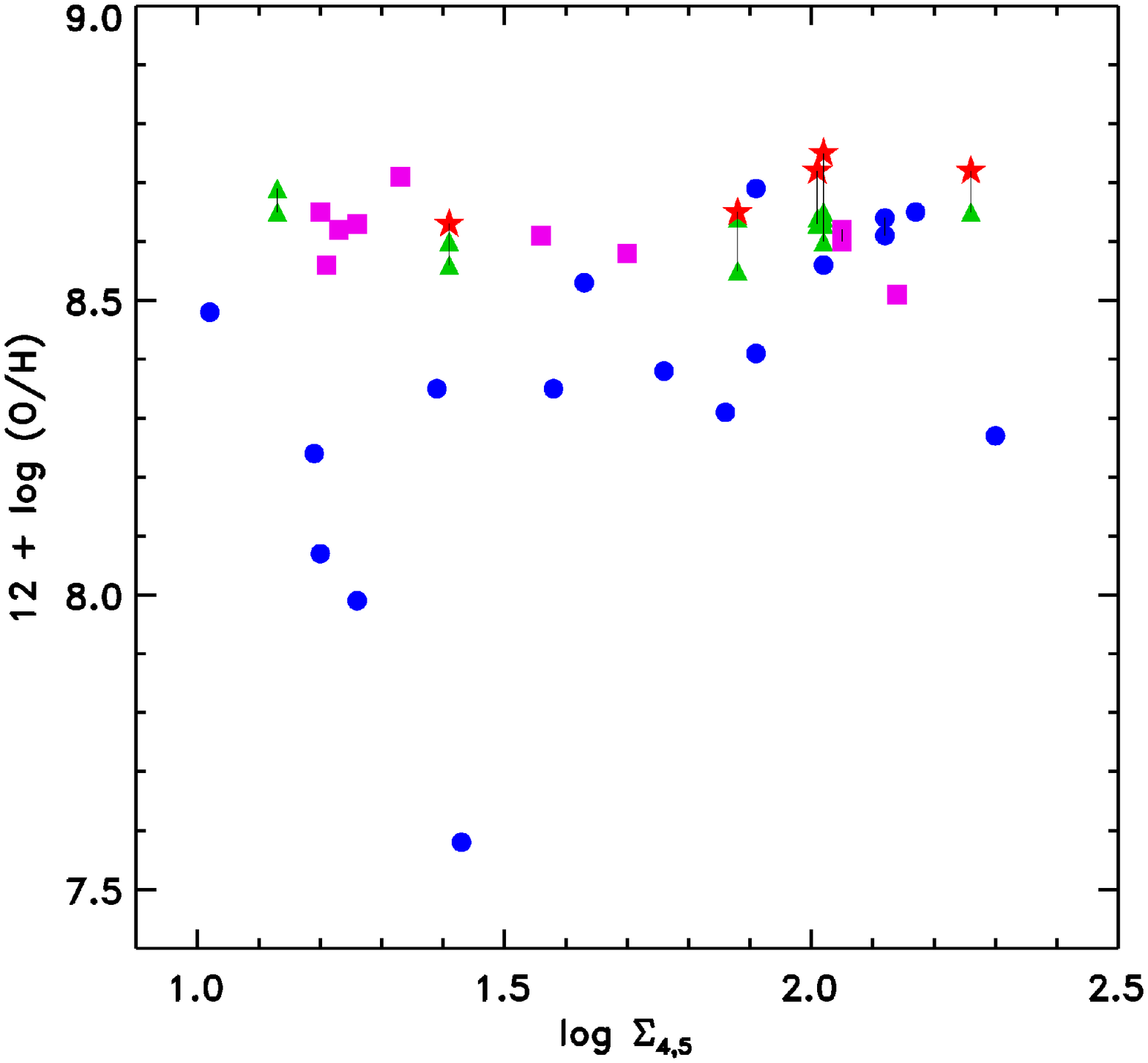}
\includegraphics[height=7cm]{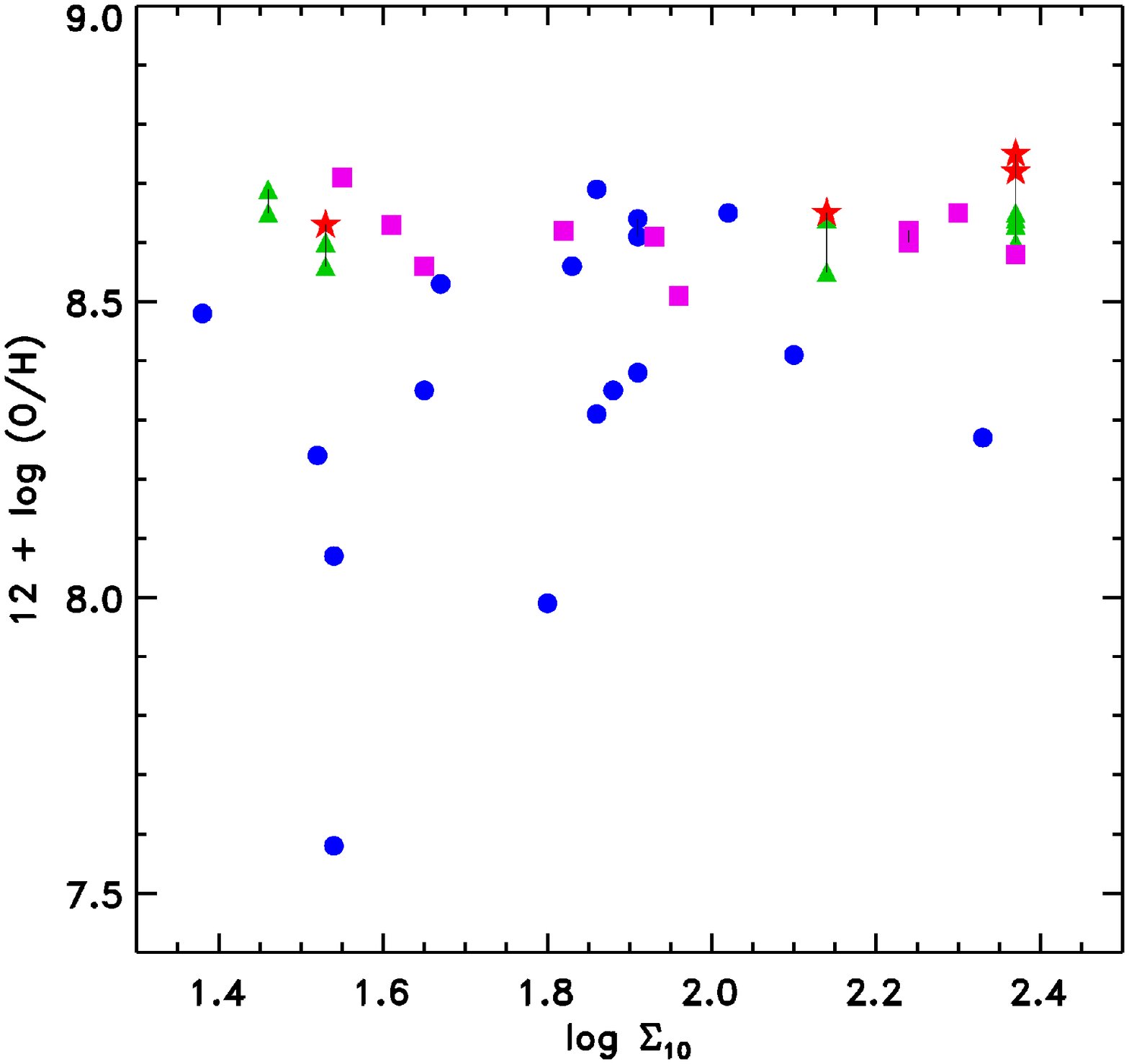}
\caption{Oxygen abundance of galaxies versus their local density. Left: O/H vs $\Sigma_{4,5}$, the local galaxy number density to the average of the projected distances to the fourth and fifth nearest neighbor, where the galaxies considered are secure members of the Hercules cluster, as follows by their SDSS spectroscopic redshift. Right: O/H vs $\Sigma_{10}$, the local galaxy number density to the 10$^{th}$ nearest neighbor, where a magnitude limited sample ($\mathrm{M_{r'}} \leq -19$ mag) of galaxies is used and we remedy for background and foreground galaxies using the field galaxy counts of \citet{Yasuda2001}. Colors and point features as in Fig.~\ref{MZ}: blue filled circles for dwarf/irregular galaxies ($\B > -19$), magenta squares for spirals ($\B \leq -19$)  integrated galaxies, red stars for the nuclei of 6 spirals that we divide into different parts and green triangles for their corresponding disk components (nuclei and disks are connected with lines). Both parts of PGC057077 are plotted in magenta filled squares; similarly, both parts of PGC057064 are plotted in blue filled circles.\label{ZDens}}
\end{figure*}

In Fig.~\ref{ZDens} we plot the galaxy gas-phase oxygen abundances versus the local density estimators $\Sigma_{4,5}$ (left) and $\Sigma_{10}$ (right) calculated in \S\ref{PP}. We keep the same color/symbol distinction as in Fig.~\ref{BPT}. Additionally, the points corresponding to the same galaxy (nuclei and disks), are connected with straight lines. We see that for luminous galaxies, the oxygen abundance does not show any significant dependence on the local galaxy number density. Luminous galaxies of this sample span the whole range of densities and all are found to have nearly solar oxygen abundance. Dwarf galaxies show a noticeable variation: the $\sim80\%$ of the less metallic dwarfs ($12+\log \mathrm{O/H}<8.4$)  are located  at $\Sigma_{4,5}<1.85$, whereas the $\sim70\%$ of the higher metallicity ($12+\log \mathrm{O/H} >8.4$) dwarf galaxies are located at very high local densities $\Sigma_{4,5}>1.85$. A substantial fraction of these more metallic dwarfs have been identified  to be affected by interactions and they are described in detail in the Appendix.  This dual behavior is not so evident when $\Sigma_{10}$ density estimator is used.  The dependency observed in Fig.~\ref{ZDens} of the metallicity of dwarf/irregular galaxies with local density could be interpreted as follows:  at the highest local densities, i.e. approaching the cluster center, only the more ``robust'' --i.e. more massive and  more metallic-- dwarf galaxies can survive. Conversely, the less metallic dwarf galaxies should have been incorporated recently to the cluster.  This ``newcomers'' scenario for dwarfs is additionally supported by the fact that the majority of the low metallicity dwarf galaxies present radial velocities which differ from the radial velocity of the brighter cluster galaxy NGC6041A --located at the X-ray maximum of the cluster-- by more than $\mathrm{\sigma_V}$,  possible evidence of infall (\S\ref{SAMP}).

One point worth of mentioning here is that, in this work, we study the effect of local galaxy density sampling  much denser environments (from $\log \Sigma_{4,5} = 1.0$ to $2.5$) than in previous works \citep{Mateus2007, Mouhcine2007, Cooper2008, Ellison2009} typically  reaching $\log \Sigma \sim 1.5$. Additionally, the works mentioned above use the SDSS database and include few dwarf galaxies, mainly due to the magnitude limit of SDSS in combination with galaxy-size limits or redshift constraints applied in order to minimize possible aperture effects. This study, therefore, is complementary and goes beyond these previous works, dealing also with the dwarf galaxy population. In this sense, the Hercules cluster, being an ideal laboratory to study the environmental effects on SF galaxies,  was not included in the SDSS-DR4 used in all these previous studies. 

\subsection{Mass and Luminosity vs Metallicity}\label{MASS}

\begin{figure*}
\includegraphics[height=7cm,angle=0]{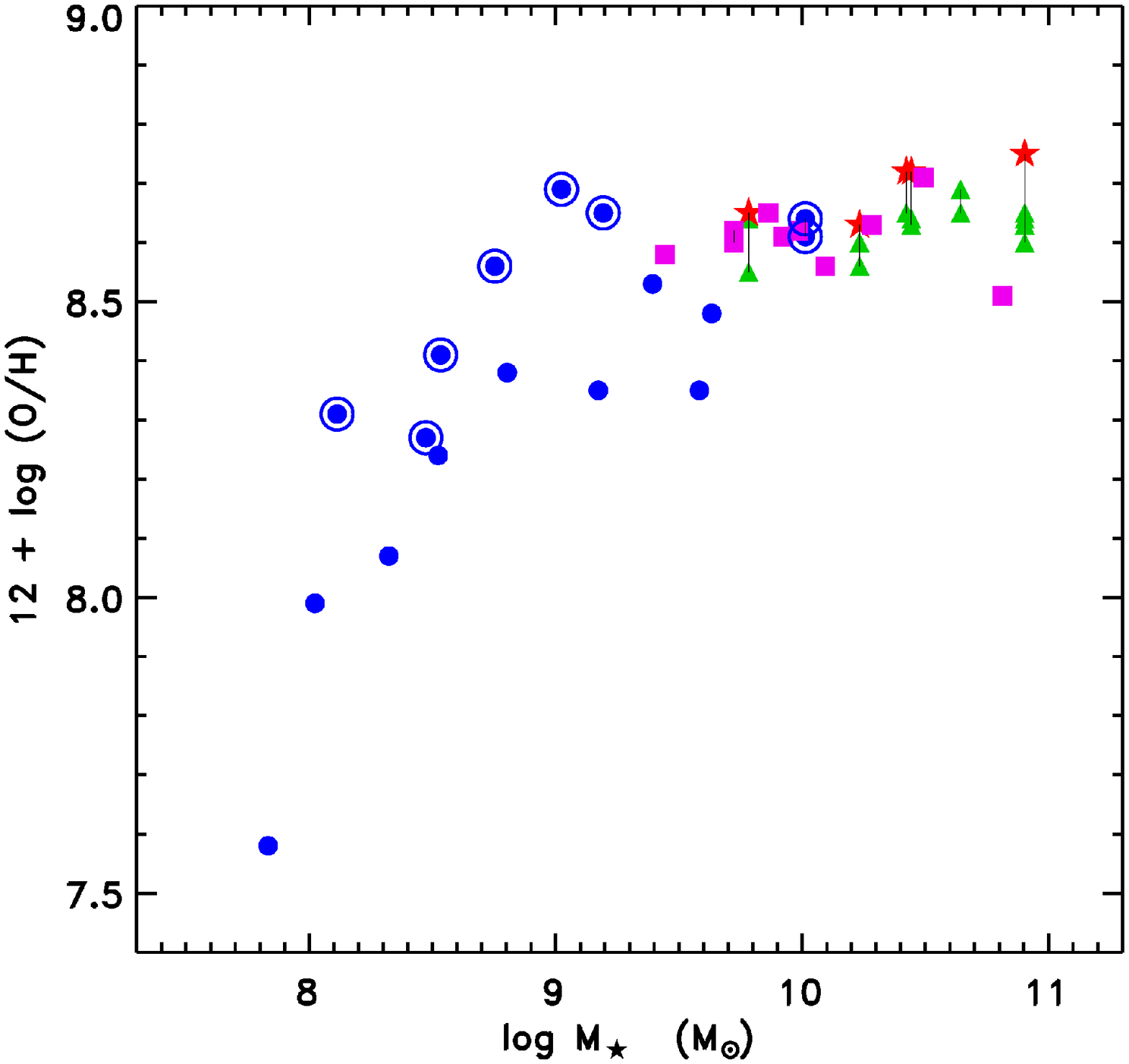}
\includegraphics[height=7cm,angle=0]{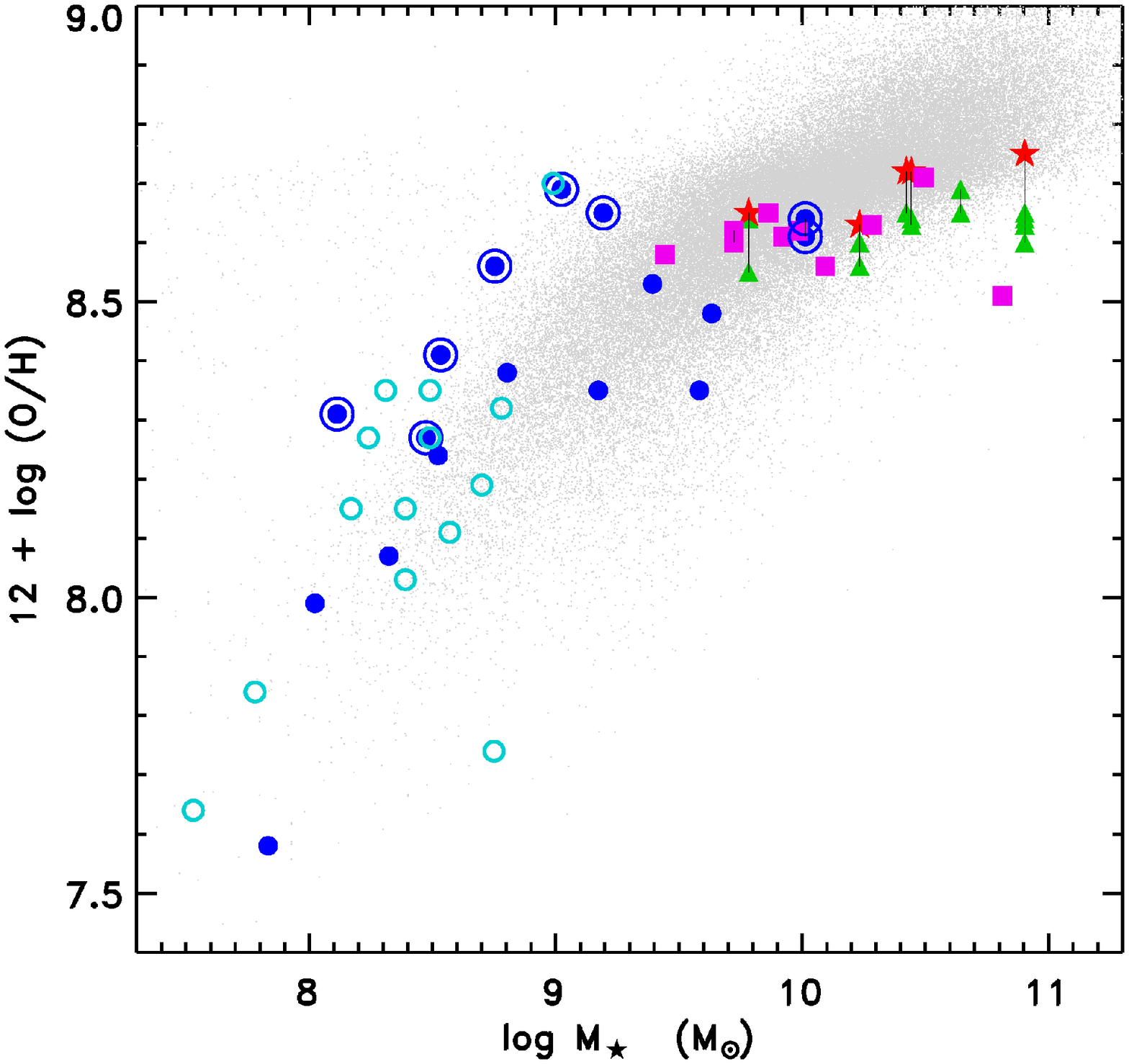}
\caption{Left: Gas-phase oxygen abundance vs galaxy stellar mass for the Hercules galaxies. We keep the same color distinction as in Fig.~\ref{ZDens}. We mark with circles the galaxies located at densities $\log\Sigma_{4,5} > 1.85$ which appear preferentially located above the general relation. Right: On the same plot we add the Virgo dIs and BCDs from  \citet{Vaduvescu2007} (light blue open circles) and the MPA/JHU SF galaxies (gray points) used as a reference sample by \citet{Amorin2010}.  \label{MZ}}
\end{figure*}

Fig.~\ref{MZ} (left) shows the gas-phase oxygen abundance versus galaxy stellar mass for our sample of galaxies  (colors and point features as in Fig.~\ref{ZDens}). We see that Hercules SF galaxies follow a well defined sequence on this plot, which reaches a saturation value  $\sim Z_{\sun}$ for galaxies with $\sim 10^{10} M_{\sun}$. We observe that the set of  dwarfs/irregulars populating the higher local densities (i.e. $\log \Sigma_{4,5}>1.85$, see \S\ref{LD}), marked here with circles, appear shifted towards higher metallicities for their mass. This fact suggests a different evolution for these galaxies in the environment of the cluster,  thus providing a physical reason for the dispersion in the MZ relation. These findings are in the line of the results of \citet{Cooper2008} who attribute $\sim 15 \%$ of the measured scatter of the MZ relation to the environment.  

For the sake of clarity, in Fig.~\ref{MZ} (right) we added the 15 Virgo dIs and BCDs (light blue open circles) from \citet{Vaduvescu2007} and references therein (excluding VCC641 because of its uncertain oxygen abundance). For consistency with our data, when a direct abundance estimation is not available, we recalculate their abundances using the P10 method. We have recalculated the stellar mass for the Virgo dwarf galaxy sample using the kcorrect code (as for the Hercules sample) and we found a good agreement with the mass estimate given by \citet{Vaduvescu2007} calculated via K-band photometry (we use the latter values in our Fig.~\ref{MZ}). We see that Virgo dwarf galaxies couple nicely with our Hercules data on the MZ plot. We also compare with the extensive sample of SF galaxies used by \citet{Amorin2010}, comprised by all the emission-line galaxies listed in the Max Planck Institute for Astrophysics/Johns Hopkings University (MPA/JHU) Data catalog of the SDSS DR 7\footnote{Available at http://www.mpa-garching.mpg.de/SDSS/} which covers a redshift range $0.03\leq z\leq0.37$. Oxygen abundances were calculated using the N2 calibration of PM09 and mass estimates were driven from the MPA/JHU catalog. On this plot Hercules SF galaxies lie within the same range as the emission-line galaxies of SDSS DR 7, except in the higher mass range, where our data show a shift towards lower metallicities by $\sim 0.15$ dex as compared to the \citet{Amorin2010} sample. We tentatively attribute this observed shift to the different oxygen abundance calibration used in each case; in addition, another important effect is the different spatial coverage of SDSS spectra versus the spatially resolved spectra used in this work.   

\begin{figure*}
\includegraphics[height=7cm,angle=0]{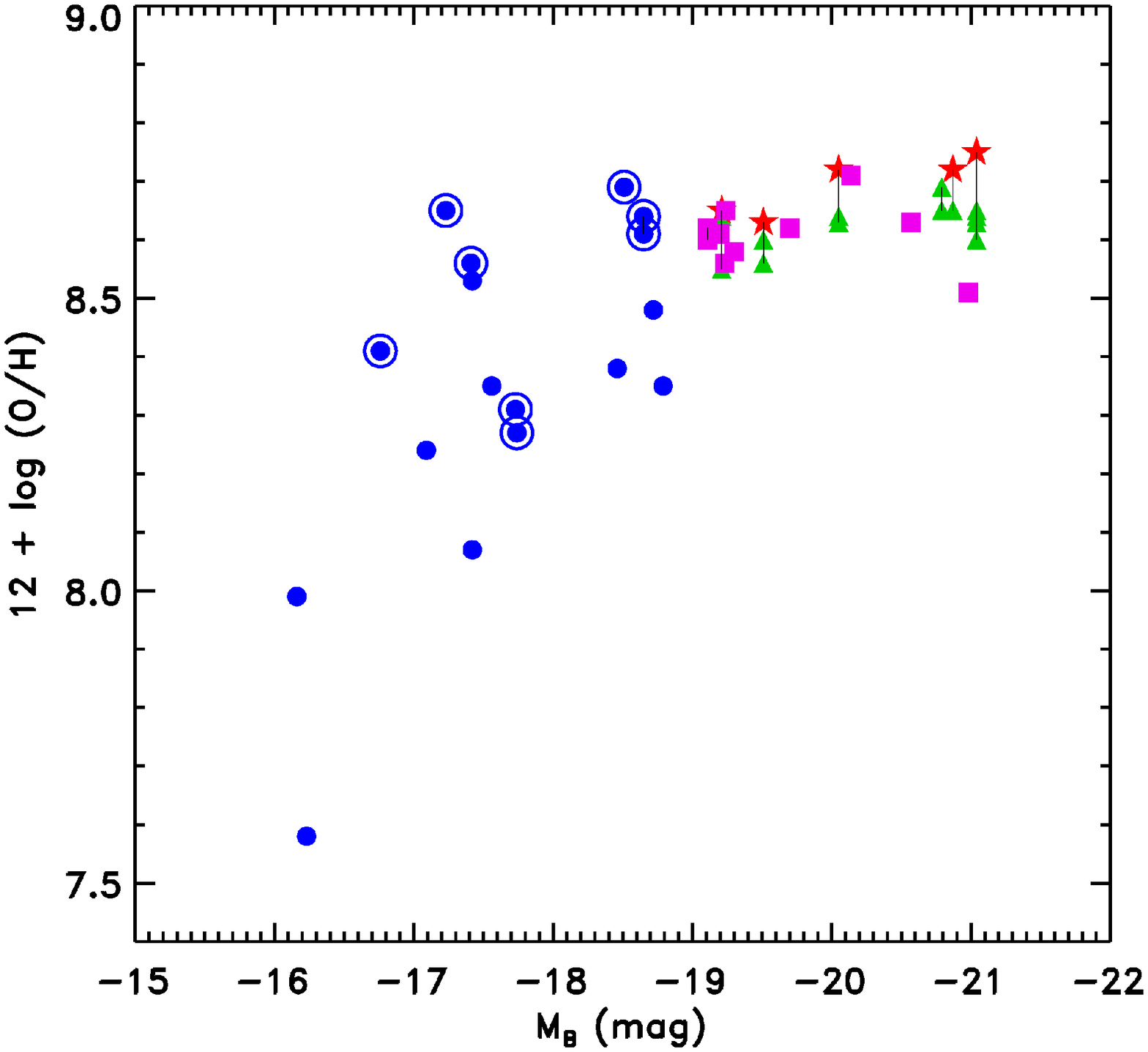}
\includegraphics[height=7cm,angle=0]{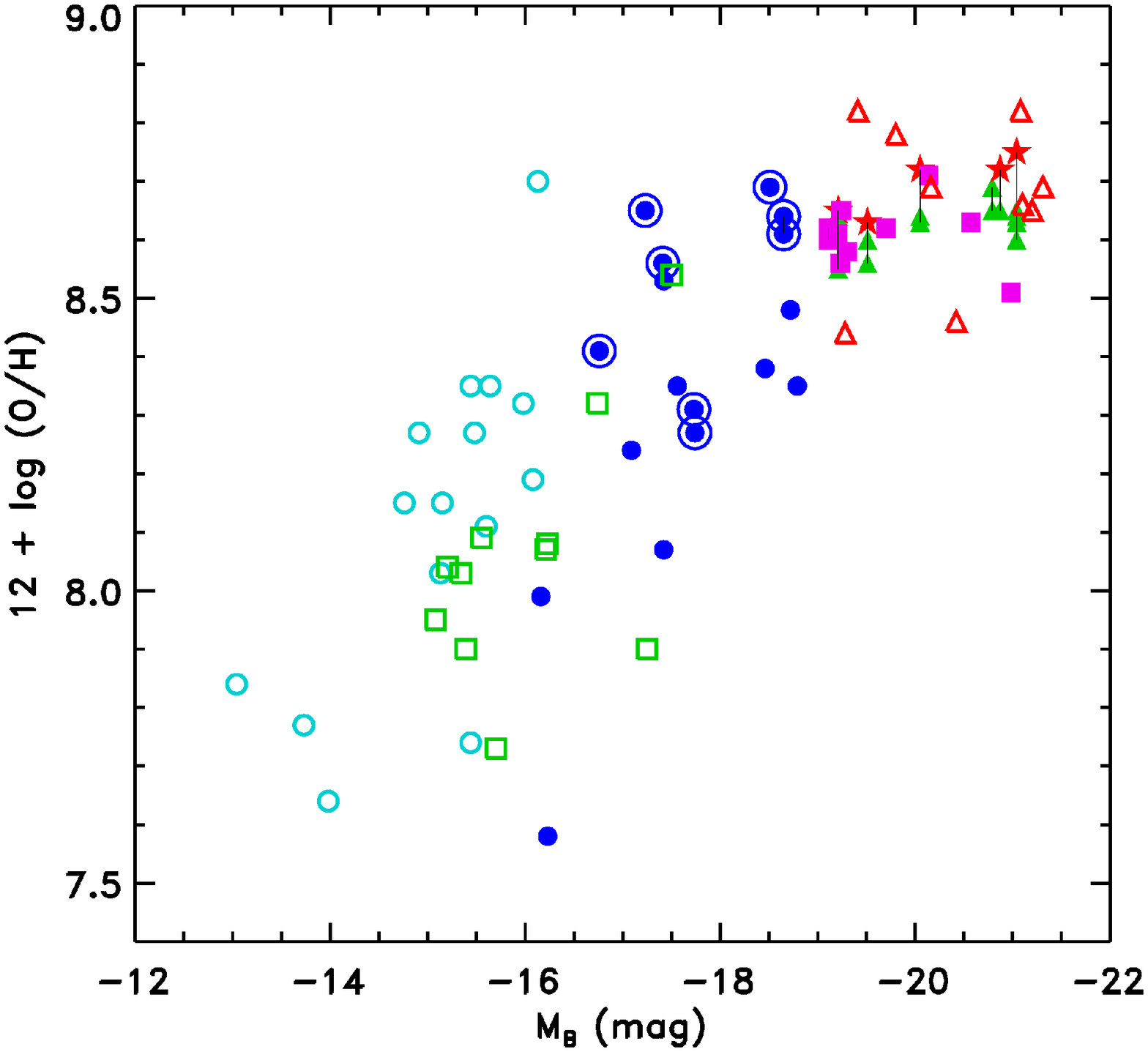}
\caption{Left: Gas-phase oxygen abundance  vs galaxy $\B$ absolute magnitude  for the Hercules sample (colors and symbols as in Fig.~\ref{MZ}). As in the MZ relation, galaxies (marked with circles) at high density ($\log\Sigma_{4,5} > 1.85$) appear shifted towards higher O/H. Right: We add the Virgo dI and BCDs  (light blue circles)  as given by \citet{Vilchez2003} and \citet{Vaduvescu2007}, the Virgo spirals (central abundance; red open triangles) from \citet{Pilyugin2002} and references therein, as well as the Hydra dwarfs (green open squares) by \citet{Duc2001,Duc1999}.\label{LZ}}
\end{figure*}

In Fig.~\ref{LZ} (left) we plot the gas-phase oxygen abundance versus galaxy $\B$ absolute magnitude. Our galaxy sample follows a LZ relation, where the same differential behavior identified in the MZ relation is clear for the dwarf/irregular galaxies at high local densities (i.e. $\log \Sigma_{4,5}>1.85$). On the plot to the right we show again the Virgo dI and BCDs (light blue open circles) \citep{Vilchez2003, Vaduvescu2007}, and the (centers of) Virgo spirals (red open triangles) from \citet{Pilyugin2002} and references therein. We also add the Hydra dwarfs (green open diamonds) from \citet{Duc2001,Duc1999}. All the abundances were recalculated  using P10 method and $\B$ of the Virgo spirals was derived by the $L_B$ given by \citet{Skillman1996}. We see that  SF galaxies of Hydra, Virgo, and Hercules clusters appear to follow the overall LZ relation, though is apparent that  the scatter has increased.

\begin{figure*}
\includegraphics[height=7cm,angle=0]{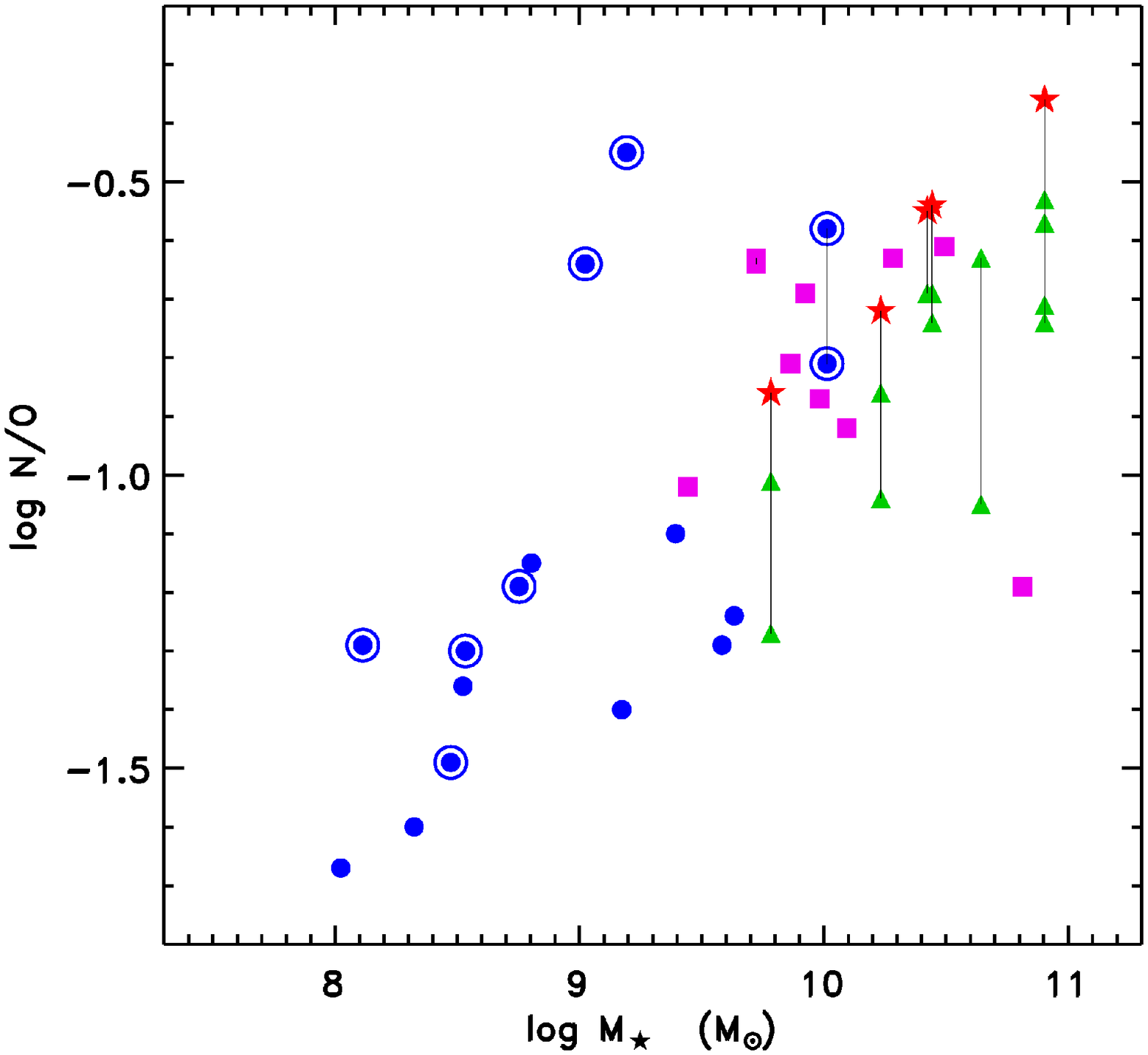}
\includegraphics[height=7cm,angle=0]{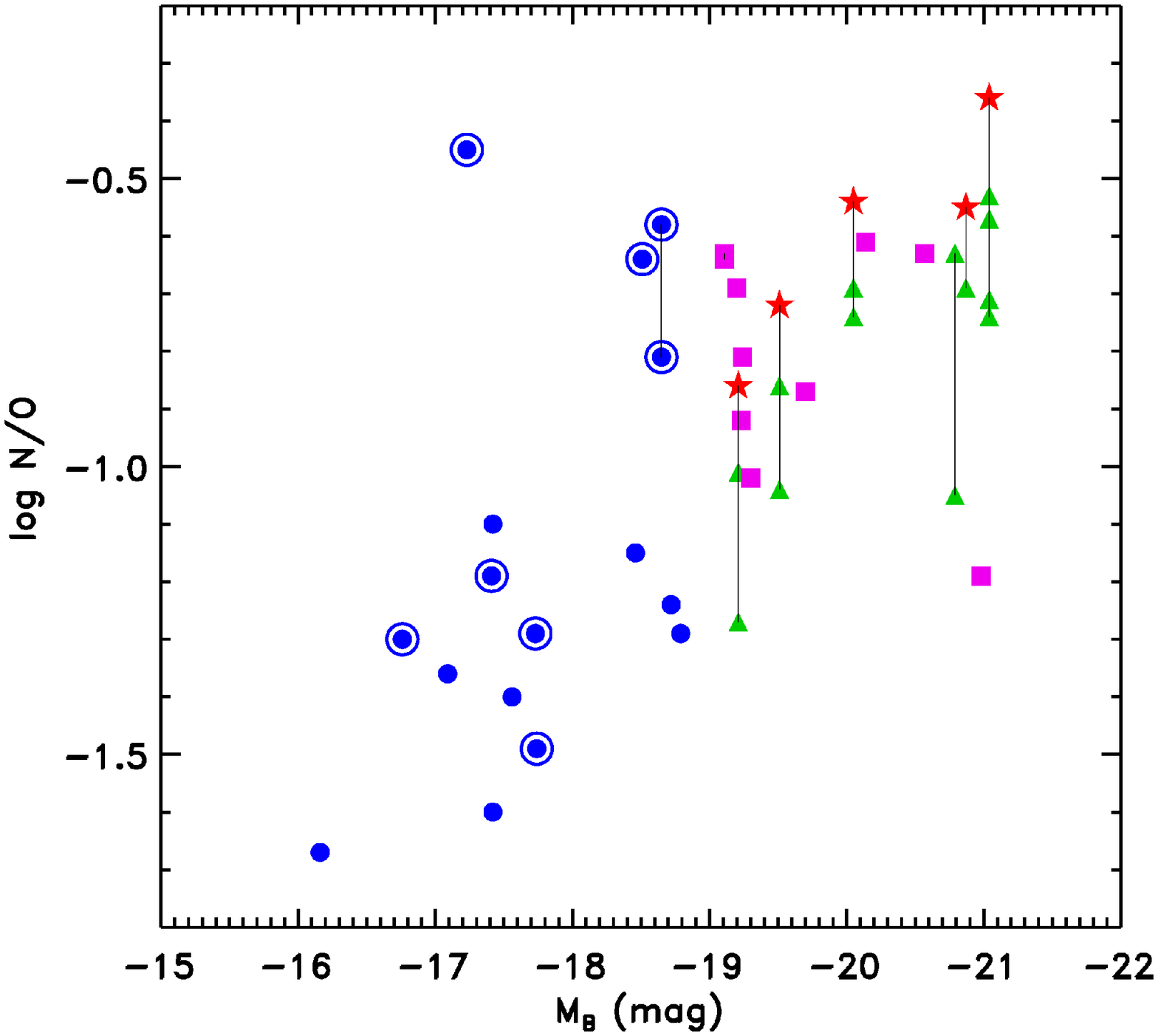}
\caption{N/O vs galaxy stellar mass (left) and $\B$ absolute magnitude (right) for the Hercules galaxies (colors and symbols as in Fig.~\ref{MZ}). A significant N/O abundance difference can be seen between galaxy nuclei and disks (points connected with straight lines).\label{MLNO}}
\end{figure*}

Moreover, Hercules SF galaxies also follow a well defined sequence on the plots of the N/O ratio versus galaxy stellar mass and galaxy $\B$ absolute magnitude  shown in Fig.~\ref{MLNO}. In this plot we can see how some of the dwarf galaxies shifted towards higher O values in the MZ relation  appear overabundant in N/O (see also \S\ref{STAR}). Additionally, the N/O ratio makes evident a significant abundance difference between nuclei and disks (points connected with straight lines), that will be discussed in more detail in \S\ref{ALL}.

Very recent works \citep{Mannucci2010, Lara2010} discuss the existence of a more general relation between stellar mass, gas-phase metallicity and star formation rate in the local universe. We have checked that, for the galaxy mass range we have in common, Hercules SF galaxies lie in the same range as \citet{Lara2010} sample in the 3D plot which combines stellar mass, gas-phase metallicity and star formation rate. However, in this work we prefer not extending this discussion until our spectroscopic study of several nearby clusters is finished (Petropoulou et al. in prep), where metallicities, stellar masses, and SFR are computed for a large sample of SF cluster galaxies. With  this  new dataset  a study of the fundamental plane for SF galaxies in clusters could be performed.  

\subsection{Environment and chemical enrichment in Hercules galaxies}\label{EVO}

\subsubsection{Comparison with the closed-box model}\label{CBOX}

In order to study possible environmental effects on the (gas-phase) chemical enrichment in galaxies, it is useful to compare with the predictions of the so-called closed-box model \citep{Edmunds1990}. According to this model, a galaxy consists initially of gas with no stars and no metals. The stellar IMF is assumed to be constant on time and the products of stellar nucleosynthesis are assumed to  enrich the ISM instantaneously. Throughout its life, the metal content of a galaxy is  neither diluted by infalling pristine gas nor lost via outflow of enriched gas. Hence, the  
metallicity at any given time is only determined by the fraction of baryons which remains in gaseous form.
The model equation can be written as 
\begin{eqnarray}
\label{yield} {Z_o} = y_{o}\ln(1/\mu)
\end{eqnarray}
where $Z_o$ is the oxygen mass fraction, $y_o$ is the yield by mass and $\mu$ is the ratio of the gas mass to the baryonic mass,  
$\mu=M_{gas}/M_{bar}$. The gas mass corresponds to the hydrogen atomic gas with a correction for neutral helium ($M_{gas}=1.32 M_{HI}$;  molecular gas mass was not taken into account here) and $M_{bar}=M_{gas}+M_{\star}$ where $M_{\star}$ is the mass in stars. 

\begin{figure}[h]
\includegraphics[height=7cm,angle=0]{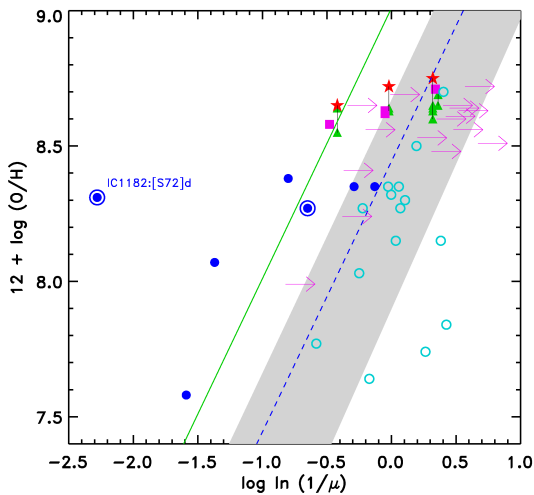}
\caption{Oxygen abundances for our sample galaxies  vs the ratio of the gas mass to the baryonic mass,  
$\mu=M_{gas}/M_{bar}$ (colors and symbols as in Fig.~\ref{MZ}). The green continuous line indicates the closed-box model with $y_o=0.0074$ and the blue dashed line corresponds to  $y_o=0.002$. The gray strip gives the best fit of \citet{Lee2003} to their field dIs. The cyan circles correspond to the Virgo dI and BCDs from \citet{Vaduvescu2007}. Arrows mark objects with upper limit in HI. \label{YIE}}
\end{figure}

Fig.~\ref{YIE} compares the observed oxygen abundances  for our sample galaxies  to the prediction of the closed-box model, plotted with lines with  
different $y_o$. The green continuous line indicates the model with $y_o=0.0074$, that is the theoretical yield of oxygen expected for a Salpeter IMF and constant star formation rate, for  stars with rotation following \citet{MM2002} models \citep{vZH2006}. Both, infall and outflow of well-mixed material will result in effective yields that are less than the true yields, as the enriched material is either diluted (infall) or lost from the system (outflow).  A fraction of the sample of isolated dwarf irregular (dI) galaxies of \citet{vZH2006} was found to follow the theoretical yield, and the rest appears consistent with a lower yield $y_o=0.002$, almost 1/4 of the model prediction  (blue dashed line in  Fig.~\ref{YIE}).  The gray strip indicates the relationship given by \citet{Lee2003} between oxygen abundance and the baryonic gas fraction for a sample of local Universe dIs.  The cyan open circles correspond to the Virgo dI and BCDs from \citet{Vaduvescu2007}. Most of Virgo dwarfs appear consistent with the  \citet{Lee2003} locus for local dwarf galaxies, though some of them still present lower gas fractions.

We can see in Fig.~\ref{YIE} that four of our low mass galaxies for which we have HI measurements (Leda1543586, LEDA140568, $[D97]$ce-200, and LEDA3085054), are in agreement with the closed-box model predictions with $y_o=0.0074$ (green line). This fact suggests that these galaxies are  falling into the central region of the cluster and encountering now the dense ICM gas for the first time; as a consequence, gas removal by ram-pressure stripping might not yet be observable. These four galaxies can be considered as prototypes of the ``newcomers'' to the cluster introduced in \S\ref{LD}. This  result is additionally supported by their disturbed H$\alpha$ characteristic morphologies (as seen in the  H$\alpha$ maps in C09) as described below. The \Ha emission map of Leda1543586  reveals a strong episode of star formation concentrated into an asymmetric arc located on the side of the galaxy facing the cluster center and almost no emission in the opposite side. This typical ``bow'' morphology  has been observed in other cluster galaxies \citep[e.g.][]{Gavazzi2001} and is very suggestive of a ram-pressure event. After  closer  inspection of the C09 \Ha maps we identify that the galaxy LEDA140568 also show a  ``bow-shock'' morphology  on the side of the galaxy facing the cluster center. Additionally, the other 2 galaxies, $[D97]$ce-200, and LEDA3085054 present strongly asymmetric \Ha emission, one-sided and offset from the galaxy optical center.  These star-bursts could be the signature of pressure-triggered star formation by the ICM within the cluster environment \citep{Treu2003}. Additionally these four galaxies have abs($\Delta \mathrm{V}) > \mathrm{\sigma_V}$ (see \S\ref{SAMP}) and  they have a median projected distance to the center of the X-ray distribution of $\sim700$ kpc; thus they are approaching to the edge of the main X-ray emitting region which extends up to $\mathrm{R_{X}}=678$ kpc \citep{Huang1996}. 

The blue points close to the blue dashed line in  Fig.~\ref{YIE} ($y_o=0.002$) correspond to the dwarf/irregular galaxies [D97]ce-143 and LEDA084703 (see the Appendix for details). [D97]ce-143 is located at high local density (\S\ref{LD}) and \citet{Dickey1997} reports that its correspondent HI cloud shows two elongations and is highly reminiscent of the Magellanic Stream. LEDA084703 shows a long HI plume \citep{Dickey1997} reaching about $1'$ from the optical center to the southeast \citep[also at the east part of this galaxy is located the supernova  quoted by][]{Zwicky1969}. These peculiar HI morphologies suggest that some gas mass-loss effect has taken place, explaining thus their location on Fig.~\ref{YIE}.

A very interesting case is the galaxy IC1182:[S72]d labeled in Fig.~\ref{YIE}. This tidal dwarf candidate  \citep[TDC,][]{Iglesias2003} shows an HI distribution which extends well beyond the galaxy disk \citep{Dickey1997}. This morphology, combined with the information we got on its gas-phase chemical content (this galaxy lies above the MZ and mass-N/O relations), together with its old stellar population (see \S\ref{STAR}), indicate a particular formation scenario for this galaxy. This galaxy has probably  been the result of  a ``block'' produced during the  merger IC1182 and seems dominated by an old stellar population. This stellar ``block''  could have acquired a large mass of gas from the late type galaxy of the merger IC1182. This formation scenario can explain why IC1182:[S72]d shows a much higher gas fraction than the value expected according to the closed-box model. Thus, the active environment of the cluster can provide an explanation for the location of a galaxy in such  forbidden region  on Fig.~\ref{YIE} \citep{Edmunds1990, vZH2006}. There exist examples of dwarf galaxies formed from the gas lost in a merger, e.g. the old TDG VCC 2062 in the Virgo cluster studied by \citet{Duc2007}; according to these authors this galaxy has probably been formed out of the gas clouds lost by a gas-rich galaxy involved in a merger. 

For the galaxies of our sample that were not detected in \citet{Dickey1997}  we assume an HI mass upper limit corresponding to the detection threshold of this survey  ($\leq 2.6 \times 10^8 M_\sun$). In Fig.~\ref{YIE} we add those galaxies with right pointing arrows (representative of the central oxygen abundance) to indicate the upper limit for their HI mass. Excluding the galaxies already discussed, the rest of the galaxies in Fig.~\ref{YIE} (including upper limits) on average suggest effective yields below the closed box model and are consistent with the field sample of \citet{Lee2003}. Finally, some points (few Virgo dwarfs and one Hercules upper limit for the IC1182 merger) still appear in Fig.\ref{YIE} displaced towards even lower values of gas fractions, as it would be expected if these  cluster galaxies suffered important environmentally induced gas removal (e.g. from ram-pressure stripping).

\subsubsection{Gas-phase metallicity vs properties of the underlying stellar component}\label{STAR}

In order to search for possible environmental footprints on the chemical history of our sample galaxies, we compare the stellar population properties, such as the  mass-weighted stellar metallicity $Z_{\star,M}$ and  mass-weighted stellar age $\tau_{\star,M}$, brought forth by the STARLIGHT model fitting (\S\ref{STL}), with the gas-phase  abundances derived in this work.   The fact that these properties (for stars and gas) have been obtained following a completely different methodology should render our analysis more robust. 

\begin{figure*}
\includegraphics[height=7cm,angle=0]{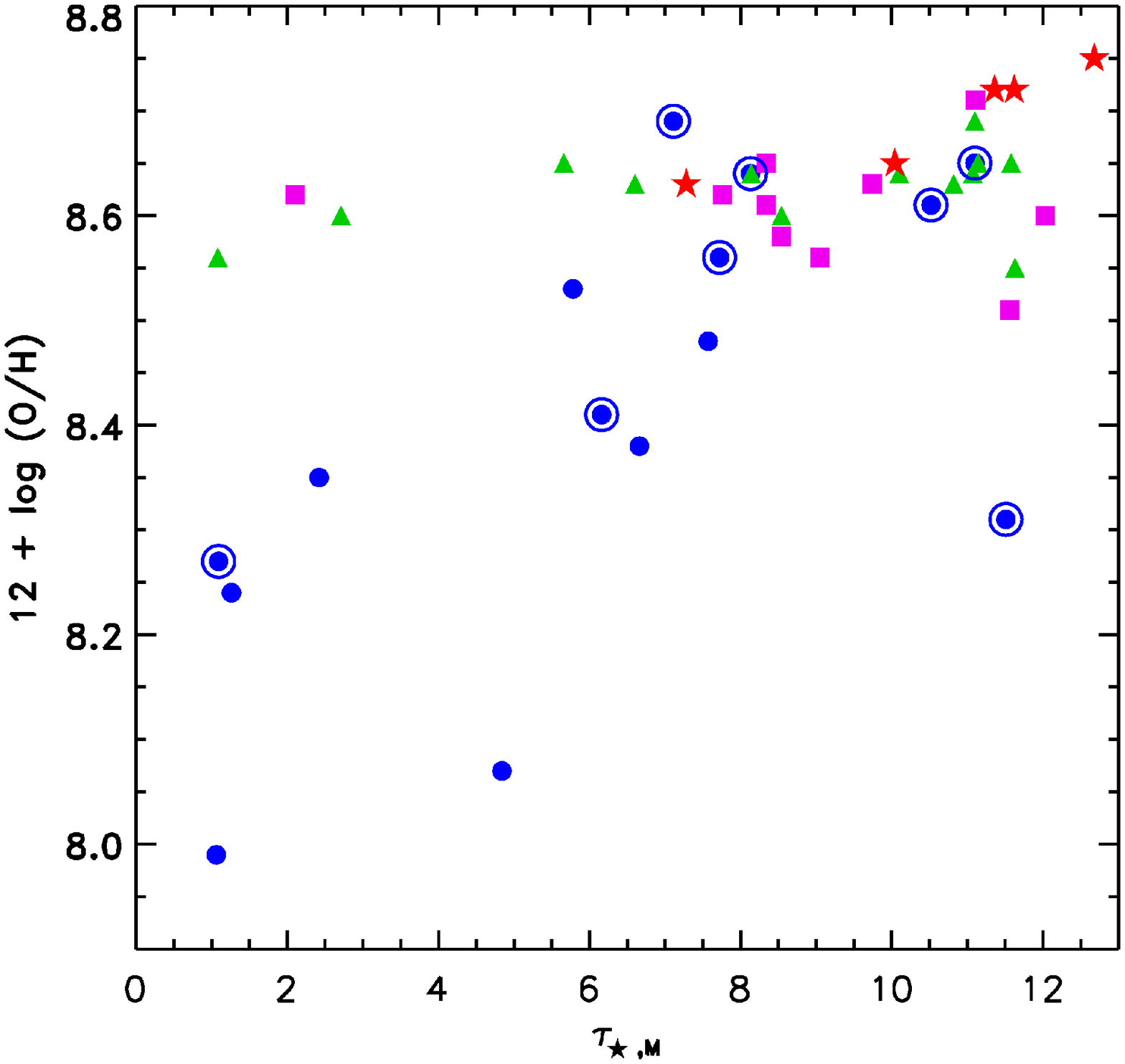}
\includegraphics[height=7cm,angle=0]{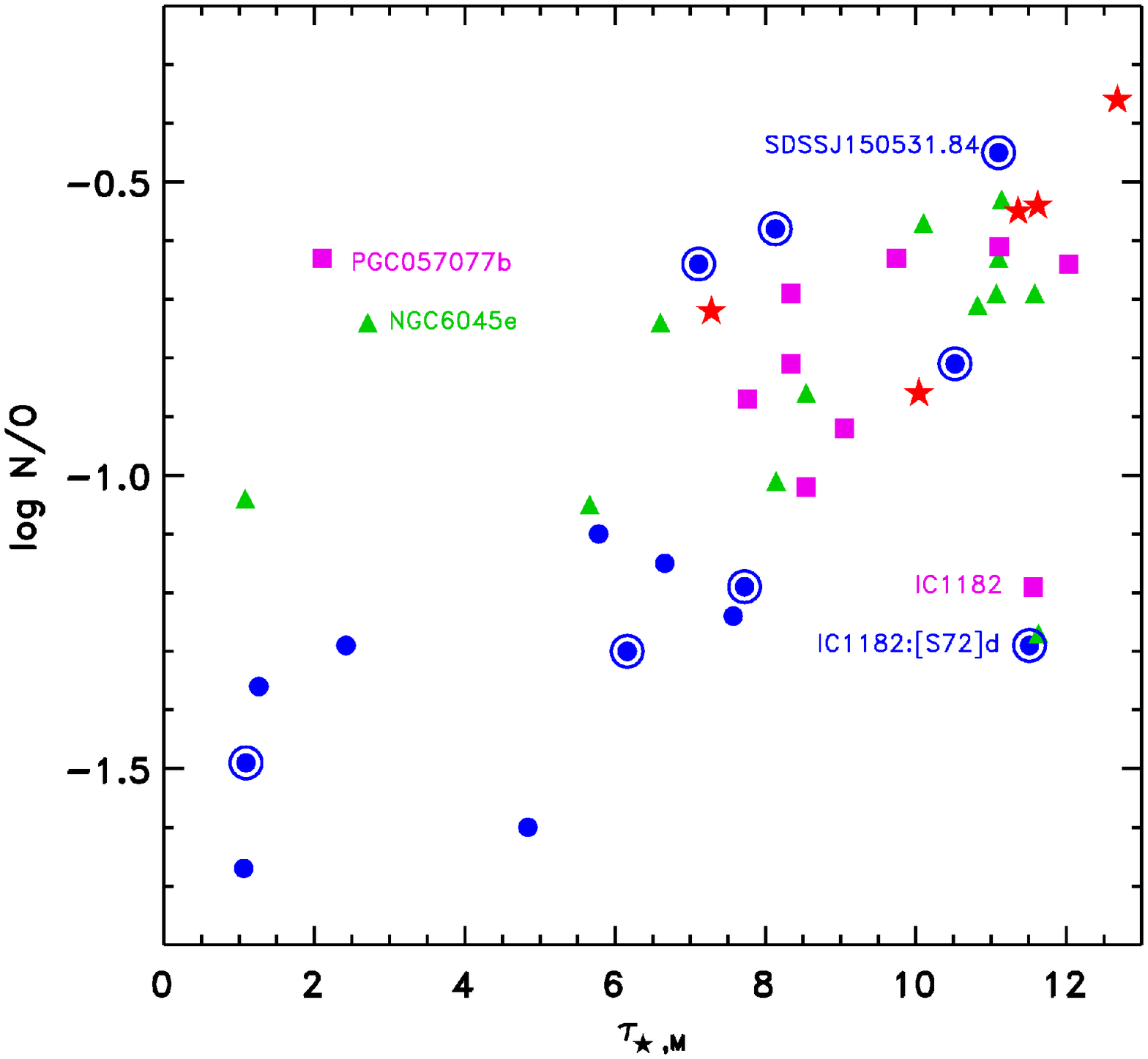}
\caption{The gas-phase oxygen abundance (left) and N/O ratio (right) versus the mass-weighted stellar age $\tau_{\star,M}$ as given by STARLIGHT model fitting (colors and symbols as in Fig.~\ref{MZ}). An overall positive trend  can be seen (see the text for details).\label{AGE}}
\end{figure*}

Fig.~\ref{AGE} shows the gas-phase oxygen abundance (left) and the N/O ratio (right) versus the mass-weighted stellar age $\tau_{\star,M}$. We see an overall positive trend, more prominent for  N/O, which should  reflect the different time-scales for the delivery of these two elements to the ISM. Oxygen, produced in Type II supernovae, is released after $\sim 10$ Myr, while nitrogen is produced and released over a substantially longer period, $\gtrsim 250$ Myr. Overall, we can see how nitrogen abundance seems to correlate better with the mass-weighted stellar age. Additionally, the dispersion in N/O becomes smaller at larger age, possibly reflecting the averaging effects of many SF episodes, while the larger dispersion seen at small ages could reflect the stochastic effects of few episodes of SF or other possible environmental effects such as e.g. gas inflows.

In Fig.~\ref{AGE} we can see that $\gtrsim85\%$ of the dwarf galaxies residing at high local density  environments ($\log\Sigma_{4,5}> 1.85$) present old stellar populations of mass weighted age $\geqslant6$ Gyr. Conversely, $\sim 70\%$ of the dwarfs located at density  $\log\Sigma_{4,5}< 1.85$ present mass weighted ages below this value; a hint suggesting that only the more robust galaxies --e.g. more massive, evolved and more metallic-- could have survived in the environment of highest galaxy density. The outliers to this general correlation also can give us important clues on possible environmental effects on their  chemical histories. On the right plot of Fig.~\ref{AGE} we can identify the galaxy merger IC1182 and the two dwarfs associated  to it (SDSS J150531.84+174826.1 and IC1182:[S72]d; see Appendix). Interestingly enough, all three objects present almost the same (very old) stellar age. Moreover, the galaxy SDSS J150531.84+174826.1 appears to be more chemically enriched with respect to galaxies of similar mass (see also Figs.~\ref{MZ} and \ref{MLNO}). Taken altogether, these properties of SDSS J150531.84+174826.1 remind those of the central part of a massive galaxy after having lost its outer parts (e.g. during a past interaction with the neighbor merger IC1182). The TDC galaxy IC1182:[S72]d, despite hosting an old stellar population, does not seem to be that chemically evolved (especially in N/O) in accordance with the scenario already proposed for it: the accretion of a large mass of gas from the late type galaxy of the merger IC1182.  For IC1182, our spectrum sampled the slightly off-center starburst giving a gaseous oxygen abundance slightly lower than the average observed trend in the MZ relation. This fact, together with its low N/O ratio, suggest that this gas should have an origin external to the host underlying galaxy. A similar scenario has been invoked before by \citet{Moles2004} \citep[see also][]{Radovich2005} who claimed  that this gas was provided by the late type galaxy of the merger. A detailed study of this complex system is out of the scope of this work and will be presented in a forthcoming paper (Petropoulou et al, in prep.). 

Two other objects can be seen above the correlation in Fig.~\ref{AGE}: the eastern part of the disk of NGC6045 (NGC6045e) and  PGC057077b, both galaxies appear  affected by interactions and  show intense SF (the highest SFR values measured for our sample after the merger IC1182). Though it has been shown that high SFR can be  a key ingredient to enhance N/O \citep{Molla2006}, what moves these two objects out of the general age-N/O relation seems to be related to their derived ages: the spectra of these objects are sampling just the starbursts, hence the derivation of $\tau_{\star,M}$ should then be dominated by the contribution of the young starburst; in contrast, the results derived for other parts of these two galaxies follow the general trend of N/O vs $\tau_{\star,M}$.

\begin{figure}
\includegraphics[height=7cm,angle=0]{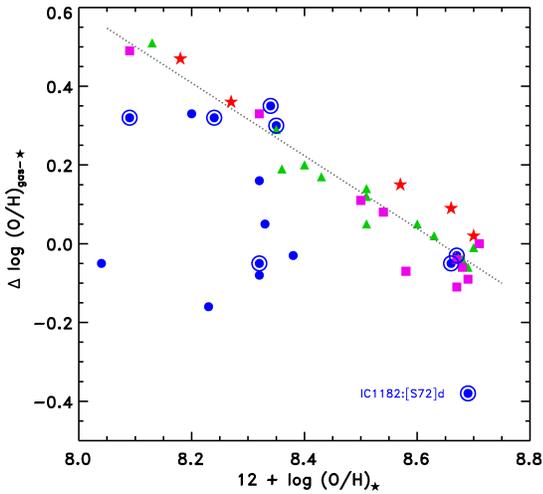}
\caption{The difference between gas-phase and stellar oxygen abundances versus stellar oxygen abundance $12+\log (O/H)_{\star}$  derived  from the STARLIGHT mass-weighted stellar metallicity $Z_{\star,m}$   (colors and symbols as in Fig.~\ref{MZ}). \label{STGAS}}
\end{figure}

In Fig.~\ref{STGAS}  we compare the gas-phase and the stellar oxygen abundances derived for the sample of galaxies. We compute the latter abundance using the mass-weighted metallicity,  $Z_{\star,M}$,  given by STARLIGHT, assuming 12+log(O/H)$_{\sun}=8.69$ \citep{Asplund2009}. We plot the difference between gas-phase and stellar oxygen abundance $\Delta \mathrm{log(O/H)}_{gas-\star}$ versus the stellar oxygen abundance 12+log(O/H)$_{\star}$. In this plot we can see two main behaviors.  For luminous galaxies ($\B \leq -19$), we can see the following correlation: 
\begin{equation}
 y=(7.99\pm0.47)-(0.925\pm0.055)x
\end{equation}
where $y=\Delta log(O/H)_{gas-\star}$ and $x=12+\log (O/H)_{\star}$ ($\chi^2=0.074$). 
Assuming our model fitting hypotheses, this ``upper bound'' correlation would mean that for a chemically evolved stellar population, of order $\sim Z_{\sun}$, gas and stars present the same metallicity. However, even when the stellar metal content of these galaxies goes down, we can see the observed gas abundance remains close to $\sim Z_{\sun}$ \citep[see also][]{Gallazzi2005,Asari2007}. This behavior could reflect the fact that, for our more massive galaxies, gradients of stellar metallicity can be more conspicuous than for gas, consistent with  model predictions \citep{Ferrini1994} suggesting that abundance gradients should flatten with age.

For the lower luminosity galaxies we can see that the two groups of dwarfs (referred to in \S\ref{MASS},\S\ref{CBOX}) split up in Fig.\ref{STGAS} too. The ``newcomers''  show similar oxygen abundances for gas and stars whereas the more chemically evolved dwarfs found at high local densities (marked with circles), hosting older stellar populations ($\geqslant6$ Gyr), show that the gas oxygen abundance is higher than the abundance of the stars by up to $0.3$ dex. An outlier to this relation, given its peculiar formation  already unraveled, is the galaxy IC1182:[S72]d, for which the derived stellar metallicity is higher than the gaseous one.
Whether these two  behaviors, shown in Fig.~\ref{STGAS}, result from different chemical evolutionary paths or rather reflect the environmental impact of the cluster remains to be disentangled.

\subsection{Searching for the cluster influence in Hercules SF galaxies}\label{ALL}

A considerable amount of work has been done in order to constrain the physical mechanisms that drive the SFH of galaxies, these mechanisms being either internal to the galaxy or related to the environment \citep{Haines2007,Bretherton2010,Weisz2011}. 
In the cluster environment it has been found that quenching mechanisms can even suppress the SF of galaxies with respect to their field counterparts and morphological changes can operate converting spirals into anemic \citep{Balogh2004}.  
Various processes have been proposed to describe the environmental actions on cluster galaxies, these can be classified into three broad categories: 1) galaxy-ICM interactions, 2) galaxy-cluster gravitational interactions, 3) galaxy-galaxy interactions \citep[for a review see e.g.][]{Boselli2006, Treu2003}. 

In this work we have searched for possible observable imprints of the cluster environment on the metallicity and the chemical history of a sample of SF galaxies in the Hercules cluster. Most of our sample galaxies are located within the central region of the Hercules cluster ($<R_{200}$) where the cluster potential is expected to be steep, the ICM is measurable and most of the physical mechanisms proposed above are effectively at work. We have examined the metal content of our sample galaxies as a function of local density, stellar mass, luminosity and chemical evolution. On the basis of this analysis, three main subgroups of objects have emerged: 1) the more massive  SF galaxies (including all spirals), 2) the  set of dwarf/irregulars labeled  ``newcomers'' in \S\ref{CBOX} and 3) a group of more chemically evolved dwarfs residing at the highest local densities observed in our sample. In addition, three dwarf/irregular galaxies SDSS J160524.27+175329.3, SDSS J160556.98+174304.1 and LEDA084724, could not be fully adscribed to any of these groups (see the Appendix).

Massive galaxies ($\B \leq -19$)  follow the global  MZ/LZ relations, though they present evidences of being affected by the cluster environment, judging from their H$\alpha$ structure/morphology and their abundance gradients. Most spirals in Hercules show  \Ha emission (see H$\alpha$ maps in C09) less extended than their optical disks and sometimes offset from the center of the stellar continuum. This can be an evidence for truncation of ionized gas in the disks  of these cluster spirals, since we observe star-formation occurring mostly in the inner parts. Gas removal from the outer parts of cluster spirals is believed to be a consequence of the ISM-ICM interaction. This effect has been observed e.g. in the Virgo cluster spirals by \citet{Koopmann2004}, where over half of their Virgo sample show truncation of the star-formation in the disks. In fact, \emph{ram-pressure} gas stripping could be effective on our Hercules spirals since all are located within $R_{200}$.  

The oxygen abundances derived for massive galaxies are close to solar, and for the more \Ha extended spirals, mild or flat O/H abundance gradients have been obtained, a result in the line of previous findings  by \citet{Skillman1996} for Virgo spirals. In contrast, for the N/O ratio even oversolar values have been measured for the central part of some galaxies, showing prominent N/O spatial variations; this picture could result from the effect of gas infall in the center of these galaxies, suffering the action of the ICM \citep{Vollmer2001}. Such infall would dilute the abundance at the central parts of the galaxies flattening  the O/H gradient, while the N/O ratio is not expected to be affected. Overall, the question which remains to be explored is whether these spirals are chemically evolved because they reside in such high density environments or we are just observing an effect of the morphology density relation \citep{Dressler1980}. Further observations are needed to answer this question.

Regarding the dwarf/irregular galaxies, they appear to form two main groups with substantial differences as it has emerged in the previous sections. Overall, all the dwarfs present similar levels of SFR as derived from their H$\alpha$ luminosities. The group of ``newcomers'' are metal poor dwarfs, they present a young stellar population (Fig.~\ref{AGE}), and their stellar and gas metallicities are similar (Fig.~\ref{STGAS}). They avoid the highest local densities (\S\ref{LD}), appear located close to the boundaries of the X-ray cluster core and show bow-shock/offset H$\alpha$ morphologies. These structures host intense bursts of  SF, possibly tracing the contact discontinuity of the ISM of the galaxy with the  X-ray emitting ICM, an observable signature of \emph{pressure-triggered star formation}. Though, the ``newcomers'' follow the closed-box model predictions, suggesting that ram-pressure stripping has not yet substantially reduced their HI gas. Ram-pressure stripping is expected to act on a time-scale $\sim5 \times 10^7$ yr \citep{Abadi1999}, while typical \HII regions lifetimes are $\sim5-10 \times 10^6$ yr. Based on this scheme we should conclude that the ``newcomers'' are those dwarfs observed right on time when they are set to fire by their first encounter with the ICM; before removal of the galactic gas is accomplished by the ICM and prior to subsequent quenching of SF rendering them undetectable in \Ha. 

The majority of the dwarf/irregulars populating the highest local densities of our sample ($\log \Sigma_{4,5}>1.85$) show higher metallicities for their mass and luminosity and they appear located above the overall MZ and LZ relations (see Figs.~\ref{MZ},\ref{LZ}). Their gaseous and stellar abundances differ by up to $\sim 0.3$ dex, pointing towards a dominant old stellar population. Indeed, as seen in Figs.~\ref{AGE} their stellar population presents ages exceeding $\sim6$ Gyr. These  galaxies have been  also found close to the  X-ray cluster core, but their H$\alpha$ morphologies do not suggest ISM-ICM interaction. However we can not discard they could have been possibly affected by ram-pressure stripping. Although HI masses are not available for the vast majority of these objects, if we assume  an HI mass upper limit for them as in \S\ref{CBOX}, many of these galaxies would be shifted in Fig.~\ref{YIE} out of the canonical closed-box model, towards the zone of HI deficiency. 

We have seen that high local density is a key parameter that separates this group of chemically evolved dwarfs from the rest of the dwarfs. At these high density environments \emph{preprocessing} has been claimed to operate under the combined action of tidal forces among group members and the ram-pressure by the ICM \citep{Cortese2006}. We suggest that these dwarf galaxies --over-metallic for their mass--  could originate from enriched material stripped by tidal forces among group members, as it has been suggested for the group of galaxies falling into Abell 1367 by \citet{Cortese2006}. Indeed we have identified a good number  of these more metallic dwarfs affected by tidal interactions.

\citet{Mahajan2011} have found two sets of blue dwarf galaxies with different H$\alpha$ emission properties in the Coma supercluster, presenting  strong environmental dependence. These authors suggest that the more evolved dwarf population could be the progenitors of passive dwarf galaxies seen in the centers of clusters. The two dwarf galaxy groups we have identified in Hercules could  match this scheme. 

In this work we have studied SF galaxies located in the cluster central region ($< R_{200}$); as a consequence the discussion on the physical mechanisms affecting galaxies in the cluster environment is restricted to SF galaxies and it could not have been exhaustive by no means. It would be necessary a larger sample covering a more extended area of the cluster in order to explore the full action of other processes able to suppress SF, like starvation of the gaseous component, harassment, or interactions with the global cluster potential.  An extended H$\alpha$ survey reaching up to the Hercules cluster periphery is in progress. This will enable us to study in depth the interesting general environment of the Hercules cluster. \citet{Bird1995} using optical and X-ray data suggested the presence of at least three distinct subclusters in Hercules cluster, the central A2151C, eastern A2151E and northern A2151N, see also Fig.~\ref{XRAY}.  These authors suggested that the A2151E and A2151N subcluster have recently undergone a merger event. Additionally, the velocity distribution of A2151C points towards the existence of two subgroups, one possibly originating from A2151N via infall. The X-ray emission is associated with the two galaxy groups in the central subcluster (Fig.~\ref{XRAY}). All this information supports the idea that Hercules is at a relatively early stage of development.

\section{SUMMARY}\label{SUM}

The Hercules cluster is one of the most exciting nearby dense environments, showing abundant sub-structures unraveled in X-ray emission and broadband imaging.  This cluster constitutes an ideal laboratory to explore the effects of the environment on galaxy evolution. We have studied the environmental effects on the metallicity and the chemical evolution of 31 SF cluster galaxies.

Spatially resolved spectroscopy has been obtained for a sample of SF galaxies and spectral synthesis model fitting has been performed for all the spectra analyzed in order to provide an effective correction of the underlying stellar absorption on emission line spectra, as well as to derive the characteristic properties of the galaxy stellar populations. Line fluxes and chemical abundances of O/H and N/O have been obtained for all the galaxies, and whenever possible for different part of galaxies, of the sample.

The main conclusions of this work can be summarized as follows:

1) From the study of the metallicity vs. galaxy local density we have seen a dual behavior separating the dwarfs from the more luminous galaxies. The luminous galaxies have metallicities $\sim Z_{\sun}$ and reside at all densities studied in this work. The set of dwarfs found at higher local densities ($\log \Sigma_{4,5}> 1.8)$ are found to be  more metallic ($12+\log\mathrm{(O/H)}>8.4$) while the
observed less metallic dwarfs ($12+\log\mathrm{(O/H)}<8.4$ are found preferentially at lower densities and some of them seem to be ``newcomers" to the cluster.

2) We have found that our sample of Hercules SF galaxies shows well defined sequences of blue luminosity vs. metallicity and stellar mass vs. metallicity (using both O/H and also the N/O ratio), following the general behavior found for SF galaxies. Besides this global behavior, we have found that dwarf/irregular galaxies populating the densest regions seem to crowd the upper part of the global sequences, thus providing a source of the dispersion observed in these relations. These more metallic dwarfs could be parts of more massive galaxies, fragmented by tidal interactions among group members.

3) Most of the luminous galaxies are chemically evolved spirals with oxygen abundance close to solar and truncated disks of ionized gas, possibly by the action of ram-pressure stripping. From our spatially resolved spectroscopy we have found that the \Ha extended spiral galaxies present shallow oxygen abundance gradients, an expected result of possible gas infall at their centers. For the N/O ratio,  even oversolar values have been obtained for the central parts of some galaxies and a significant spatial variation has been observed.

4) A detailed study of the chemical history of the sample galaxies has been performed, combining information on their gas-phase abundances, HI content and stellar mass. Most of the dwarf galaxies with available HI mass seem to be  ``newcomers'' to the cluster and appear consistent with the predictions of the closed box model. This fact agree with the scenario that these galaxies experience a pressure-triggered starburst, right before the ram-pressure stripping privies them from their gas component.  The rest of the galaxies with HI measurement on average show lower values of gas fractions, though most of them are still consistent with the loci defined by samples of field galaxies. 

5) The properties of the underlying stellar population, such as stellar age and stellar metallicity, have been explored and compared with the gas-phase metallicity. The ``newcomers'' dwarfs present a young stellar population and their stellar and gas metallicities are similar.  The more metallic dwarf galaxies host an old stellar population, resembling to the evolved blue dwarfs refered by \citet{Mahajan2011}. An overall positive trend has been found in the gas-phase oxygen abundance versus the mass-weighted stellar age $\tau_{\star,M}$, which becomes more prominent in the case of the N/O ratio. 

We have learned that the local environment of a galaxy is one of the  key parameters in order to understand its chemical history.  In the variety of the Hercules cluster ecosystem we have seen  galaxy-galaxy interactions, galaxy ISM-ICM interactions and candidates of tidal dwarfs galaxies. Further observations are needed to disentangle the role of all these environmental effects from the expected intrinsic galaxy evolution.

\acknowledgments

V.P. would like to thank Enrique Perez Montero, Ricardo Amorin, Ovidiu Vaduvescu, Ana Monreal-Ibero and Josep Maria Solanes for their help and suggestions. We thank the anonymous referee for the useful suggestions which helped to improve the paper.
V.P also thanks the Osservatorio di Arcetri, the Centro de Astrof\'isica de La Palma and the Department of Astronomy in the University of Athens for hospitality during the stays to carry out part of this work. Special thanks are owed to M. Kontizas and E. Kontizas for their support. 

V.P. acknowledge financial support from the Spanish Ministerio de Ciencia e Innovaci\'on under grant FPU AP2006-04622. 
We also acknowledge financial support by the Spanish PNAYA, projects
ESTALLIDOS (grant AYA2007-67965-C03-02)
and CSD2006-00070 “1st Science with
GTC” from the CONSOLIDER 2010 programme of the Spanish MICINN. We thank the Spanish CSIC for financial support (Proyecto Intramural 200850I018). P.P. is supported by a Ciencia 2008 contract, funded by FCT/MCTES (Portugal) and POPH/FSE (EC). 

The INT and WHT are operated on the island of La Palma by the Isaac Newton Group in the Spanish Observatorio del Roque de los Muchachos of the Instituto de Astrof\'isica de Canarias. We acknowledge CAT for the allocation of telescope time to this project. We thank the support astronomers of WHT (ING) for  Service Time observations on June 26-27, 2009 and the directors of IAC and ING for the allocation of DDT on July 19, 2009. 

This research has made use of the NASA/IPAC Extragalactic Database (NED), which is operated by the Jet Propulsion Laboratory, California Institute of Technology, under contract with the National Aeronautics and Space Administration. We also have benefitted from Sloan Digital Sky Survey (SDSS) database. Funding for the SDSS and SDSS-II was provided by the Alfred P. Sloan Foundation, the Participating Institutions, the National Science Foundation, the U.S. Department of Energy, the National Aeronautics and Space Administration, the Japanese Monbukagakusho, the Max Planck Society, and the Higher Education Funding Council for England. The SDSS was managed by the Astrophysical Research Consortium for the Participating Institutions. IRAF is distributed by the National Optical Astronomical Observatory, which is operated by the Associated Universities for Research in Astronomy, Inc., under cooperative agreement with the National Science Foundation. 

{\it Facilities:} \facility{ING:Newton}, \facility{ING:Herschel}.

\appendix

\section{NOTES ON INDIVIDUAL GALAXIES}\label{NOTES}

\subsection{Interacting galaxies}

\emph{IC1182, IC1182:[S72]d and SDSS J160531.84+174826.1}:

IC1182 is a late-merger that undergoes vigorous nuclear starburst activity with a star formation rate, estimated from its infrared and extinction-corrected H$\alpha$ luminosity, to be between 11 and $\sim$90 M$_{\odot}$/yr \citep{Radovich2005, Moles2004}. IC1182 shows an extended ($\sim$60 kpc) stellar tail to the east, where a tidal dwarf candidate IC1182:[S72]d is discernible (identified by \citet{Iglesias2003}; named ce-061 after \citet{Dickey1997}). To the north of the system and to a comparable distance there are also distinguished faint tidal features and plumes. In addition, to the west of the merging system, at a projected distance  of $\sim$60 kpc, there is the dwarf galaxy SDSS J160531.84+174826.1 which has been claimed \citep{Dong2007} to be a Seyfert 1 AGN (but see \S\ref{RES}). 
\citet{Dickey1997} reports a long HI structure extending northwest of IC1182, creating a bridge of HI clumps towards SDSSJ150531.84+174826.1 (see Fig.~3.a of IP03), although the galaxy SDSSJ150531.84+174826.1 itself  was not detected as a consequence of the velocity cuttoff (9820 km/s) of the spectrometer used by \citet{Dickey1997}. These lines of evidence point towards a physical connection among the three galaxies. A forthcoming work will discuss the global picture of this system (Petropoulou et al., in prep).

\emph{NGC6050}:

It has long been considered as a collision between two spiral galaxies (NGC6050A and NGC6050B). From our long-slit spectrum we found that these two galaxies show important velocity difference ($\sim 1600$ km s$^{-1}$) suggesting that they might be just a chance galaxy alignment. On the other hand, to the north of the system we have identified another galaxy, on the basis of its underlying stellar component (clearly seen in the 2MASS imaging) and its SDSS spectrum. This galaxy is probably in interaction with NGC6050A.
\citet{Dickey1997} quotes an HI mass for the NGC6050 system. We attributed the HI mass exclusively to NGC6050B because NGC6050A is out of the velocity cutoff of the spectrometer used by \citet{Dickey1997}. The HI distribution is centered on NGC6050B. 

\emph{NGC6045}:

This edge-on spiral could  be interacting with a smaller spiral companion located at the east of it. On its H$\alpha$ map (C09) we can see intense star-burst activity on the east part of the galaxy disk, probably triggered by this interaction.  \citet{Huang1996} suggested a possible interaction of NGC6045 with the radio galaxy NGC6047 which is about $1'.6$ ($\sim 74$ kpc) to the south. \citet{Dickey1997} reports that in HI only half of the disk is visible. From our long-slit spectroscopy we find a large large velocity gradient along the disk of the galaxy and half of the disk shows velocity out of the cut-off of the spectrometer used by \citet{Dickey1997}.

\emph{PGC057064}: 

This peculiar galaxy,  until now considered as one galaxy, has turned out to be a merger of two galaxies, which on our long-slit spectrum show different rotation velocities. We have considered the corresponding 2D spectrum parts separately. The 1D spectrum part (a) corresponds to the NW  galaxy and (b) to the SE.  The two dwarf galaxies comprising this merger do not show any significant chemical variation.

\subsection{Peculiar galaxies}

\emph{PGC057077}:
This is a peculiar galaxy previously cataloged as an elliptical in LEDA. The C09 H$\alpha$ survey showed that PGC057077 is a very intense and compact  H$\alpha$  emitter with $f_{H_\alpha}= 87 \times 10^{-15}$ $erg$ $cm^{-2}$ $s^{-1}$ and  $EW=148$\AA. Although the galaxy is very compact and does not show any structural properties, we noted on its $g-i$ map, illustrated in Fig.~\ref{SLITS}, a conspicuous color gradient. In our long-slit spectrum we were able to identify two regions with important difference of the continuum emission. In order to check for velocity gradient and differences in the chemical content between these two regions, we separated the 2D spectrum in two corresponding 1D spectra. The important continuum emission difference of the 1D spectra can be contemplated in Fig.~\ref{SPE}. From our emission-line analysis we do not find any significant velocity variation between the two different galaxy parts, neither any difference in the chemical abundances. The difference in the continuum emission is attributed then to the high extinction suffered by both the stellar population and the gas in the NW part of the galaxy which correspond to our (a) spectrum part. This peculiar galaxy is situated close to the edge of the principal X-Ray emitting region, at 650 kpc from the X-Ray center. In the NE of the galaxy there is discernible a low surface blue plume, that could point toward the existence of a faint companion. The existing data are not conclusive  whether it could have been the ICM-ISM interaction or a galaxy merger that could have triggered the unusual starburst of this galaxy. Further mid-infrared observations would be of interest there.

\emph{SDSSJ160520.58+175210.6} appears to be embedded into the HI structure assigned to the galaxy \emph{[D97]ce-200} (SDSSJ160520.64+ 175201.5); \citet{Dickey1997} reports that this system looks like the Magellanic Clouds and the Magellanic Stream. However these two galaxies have a velocity difference of 1300 km/s a fact that renders unlikely an actual interaction of them. The velocity of the HI cloud is very similar to the velocity of the galaxy [D97]ce-200.

\emph{LEDA084703}: 
This is one of the two galaxies of this sample situated in the substructure almost 1 Abell radius SW from the cluster center. It is a quite blue galaxy, classified Sd/Irr and it is situated almost one Abell radius away from the cluster center. There has been a SN explosion reported by \citet{Zwicky1969} in the East part of this irregular galaxy. There is a long plume extending towards southeast detected in the optical images as well as in HI by \citet{Dickey1997}. The west part of the disk hosts a very intense star-burst activity as can be seen in its H$\alpha$ EW map in Fig.~\ref{SLITS}.

\emph{SDSSJ160524.27+175329.3}:  
It has typical H$\alpha$ morphology and colors of a blue compact galaxy. This compact starburst, located close to the X-ray maximum in the center of the Hercules cluster could have resulted from the compression of the interstellar gas of a dwarf galaxy when entering the cluster core. Similar compact starburst galaxies have been found in the cores of other nearby clusters by \citet{Reverte2007}.

\emph{SDSS J160556.98+174304.1} The \Ha map of this dwarf galaxy shows an off-center starburst. Additionally this galaxy is located close to the edge of the primary X-ray maximum in the cluster center and presents a large velocity difference with NGC6041A. These evidences could suggest that the starburst has been triggered by the effect of the hot ICM. \citep{Dickey1997} have not detected molecular gas for this galaxy, suggesting that the ram-pressure stripping could have already taken off its molecular gas.  

\emph{LEDA084724} presents an extended knotty \Ha structure NW, in the direction towards the X-ray cluster center. This morphology could be suggestive of a starburst triggered by the ICM. \citep{Dickey1997} have not detected molecular gas for this galaxy.

\emph{SDSSJ160304.20+171126.7} is one of the two SF galaxies of this spectroscopic sample situated in the substructure almost 1 Abell radius SW from the cluster center. This galaxy has a close companion, the galaxy SDSS J160305.24+171136.1. The disturbed H$\alpha$ structure of both could suggest a possible interaction. 
 
\emph{SDSSJ160523.66+174832.3} is located very close to the faint galaxy SDSSJ160523.67+174828.8 (\#22 and \#23 in C09). The H$\alpha$ map as well as the $g-i$ color map of SDSSJ160523.66+174832.3 seem to reveal a disturbed morphology. There could be the possibility that the two galaxies are interacting, although SDSSJ 160523.67+174828.8 does not have spectroscopic data and SDSS provide a photo-z estimate of $z=0.15\pm 0.02$, nominally not consistent with the velocity range of Hercules cluster.

\emph{[DKP87]160310.21+175956.7} is projected close to a faint galaxy located southern-east. The H$\alpha$ EW map of [DKP87]160310.21+175956.7 shows an intense star-burst event towards that direction. However the close companion galaxy has no spectroscopic data and the photo-z estimate provided by SDSS $z=0.25\pm 0.06$ is not consistent with the velocity range of Hercules cluster.

\emph{SDSSJ160547.17+173501.6}:
It has turned out to be a background galaxy with $z=0.12\pm0.01$. The various SDSS photo-z estimates were $0.06\pm0.01$, $0.06\pm0.05$ and $0.10\pm0.04$, thus the criteria adopted by C09 for cluster membership were fulfilled.

\begin{figure*}
\epsscale{1.0}
\plotone{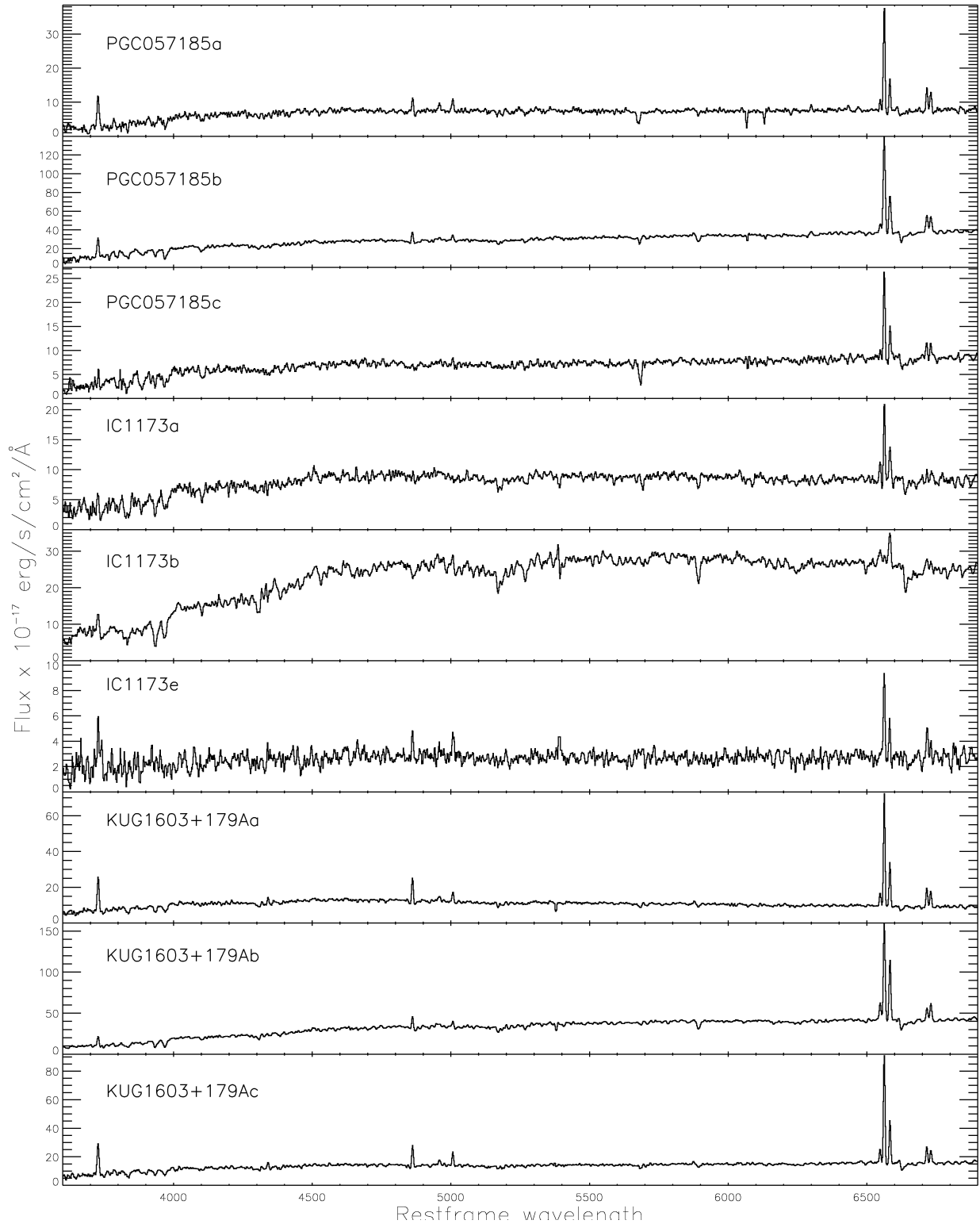}
\caption{Integrated 1D observed spectra for each galaxy or part of a galaxy by INT and WHT.\label{SPE}}
\end{figure*}

\addtocounter{figure}{-1}
\begin{figure*}
\epsscale{1.0}
\plotone{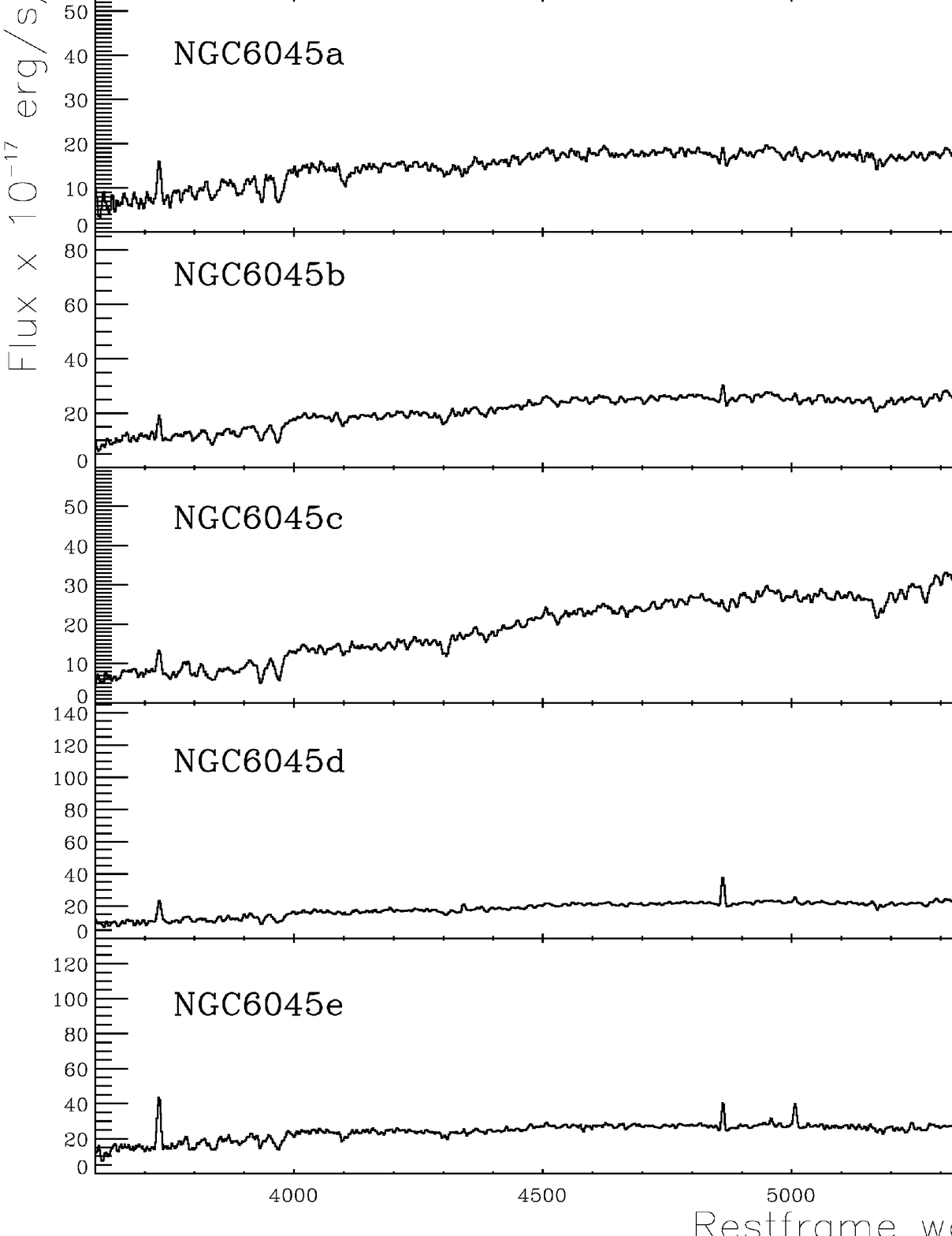}
\caption{Continued.}
\end{figure*}

\addtocounter{figure}{-1}
\begin{figure*}
\epsscale{1.0}
\plotone{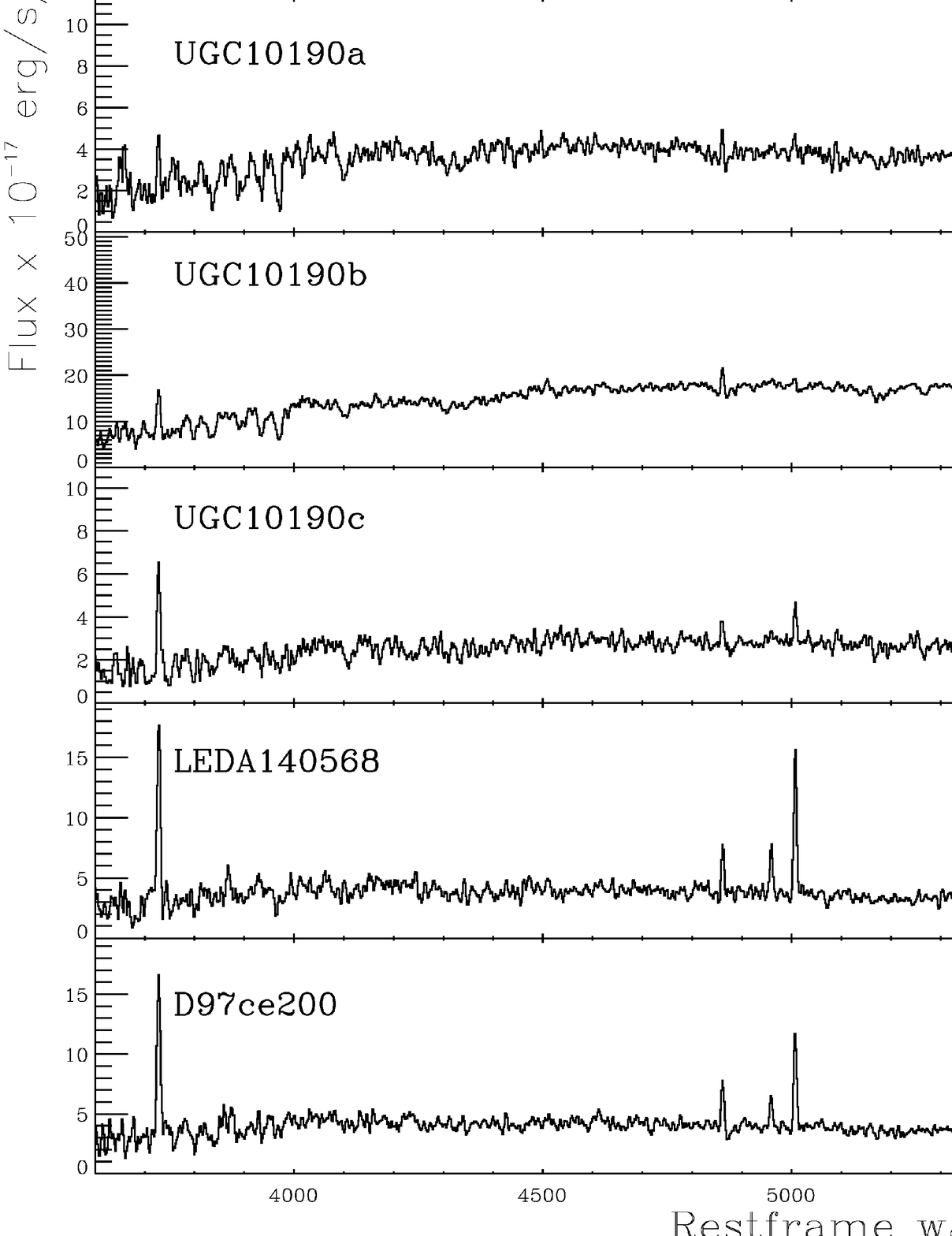}
\caption{Continued.}
\end{figure*}

\addtocounter{figure}{-1}
\begin{figure*}
\epsscale{1.0}
\plotone{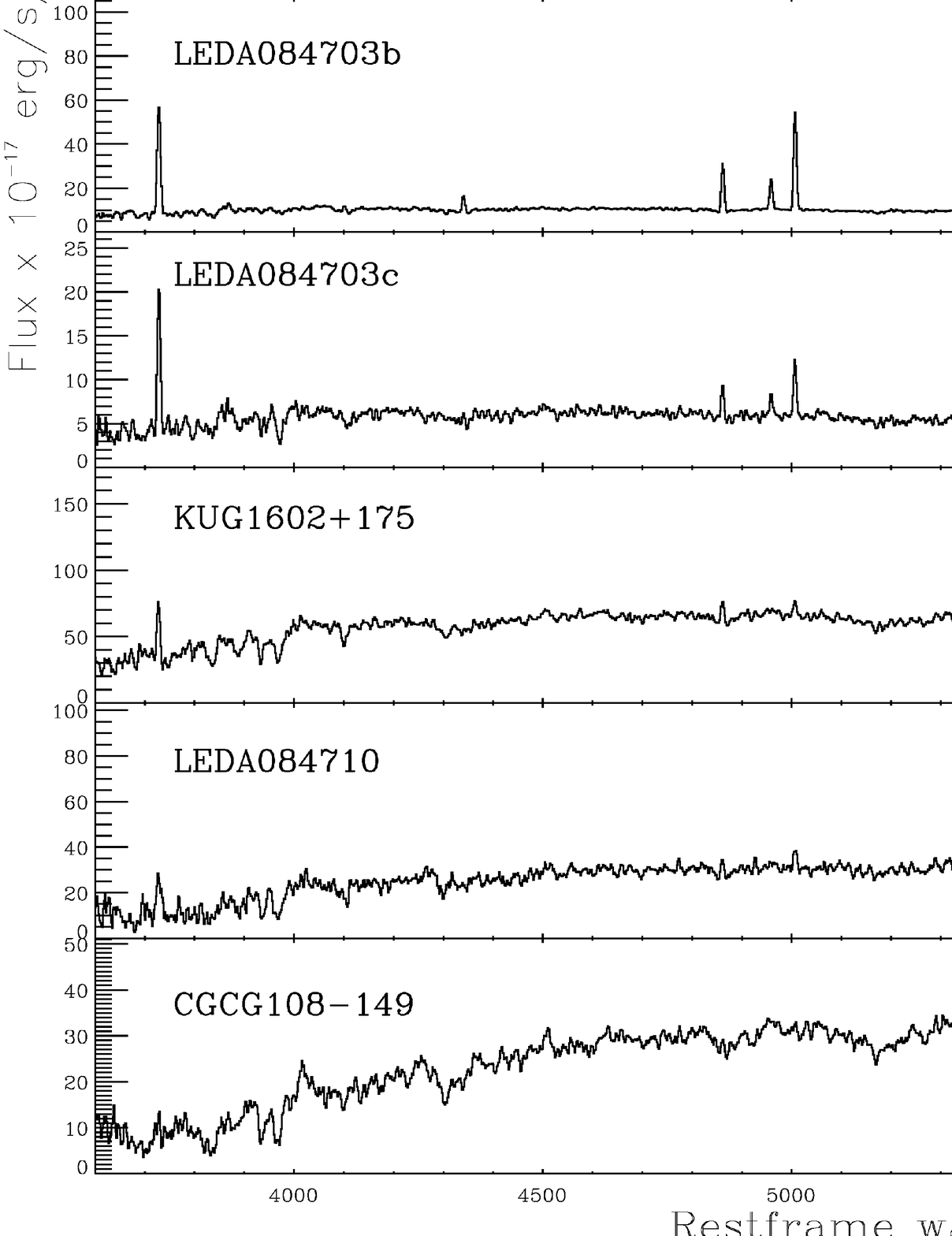}
\caption{Continued.}
\end{figure*}

\addtocounter{figure}{-1}
\begin{figure*}
\epsscale{1.0}
\plotone{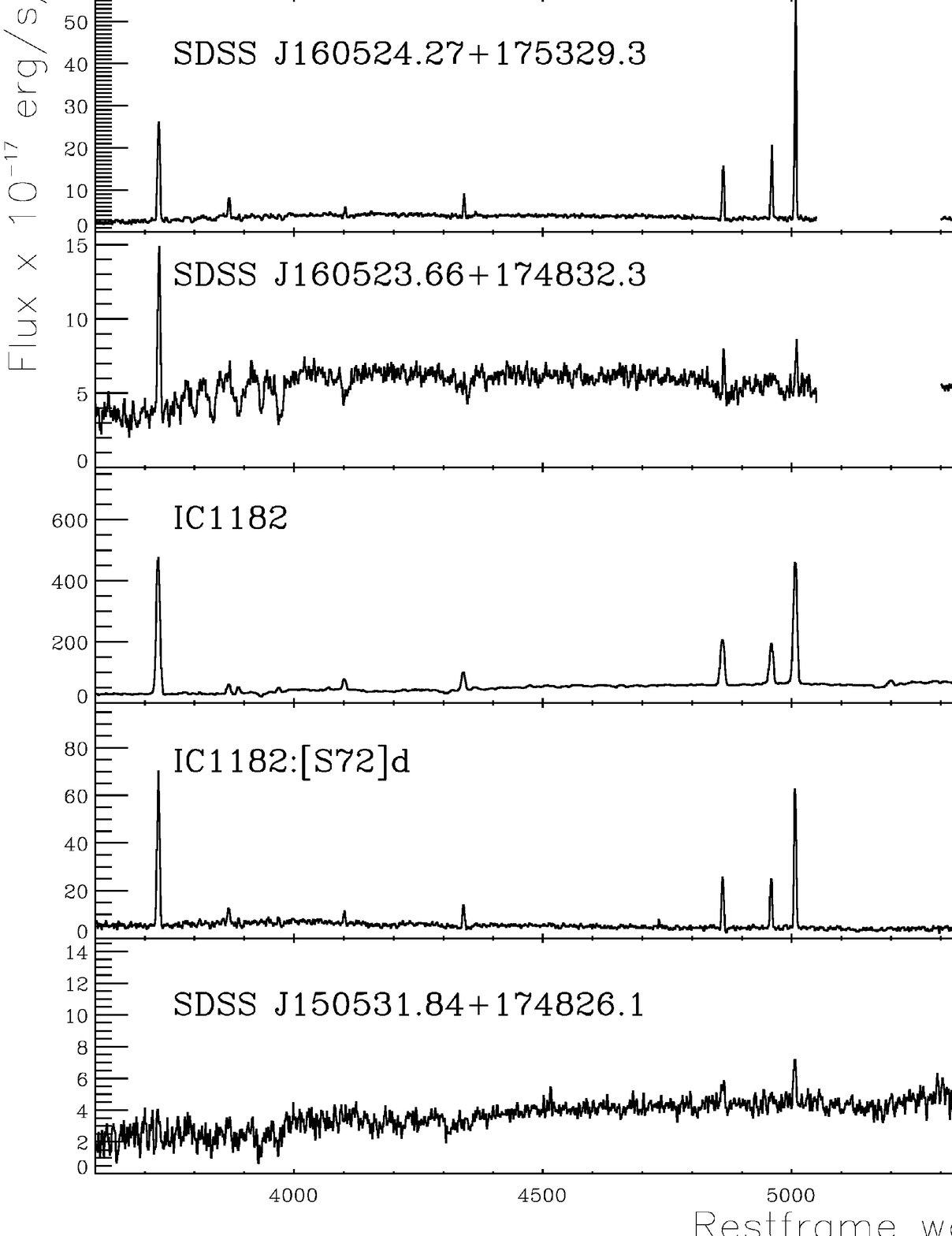}
\caption{Continued.}
\end{figure*}

\begin{figure*}
\includegraphics[height=5cm,width=5cm,angle=0]{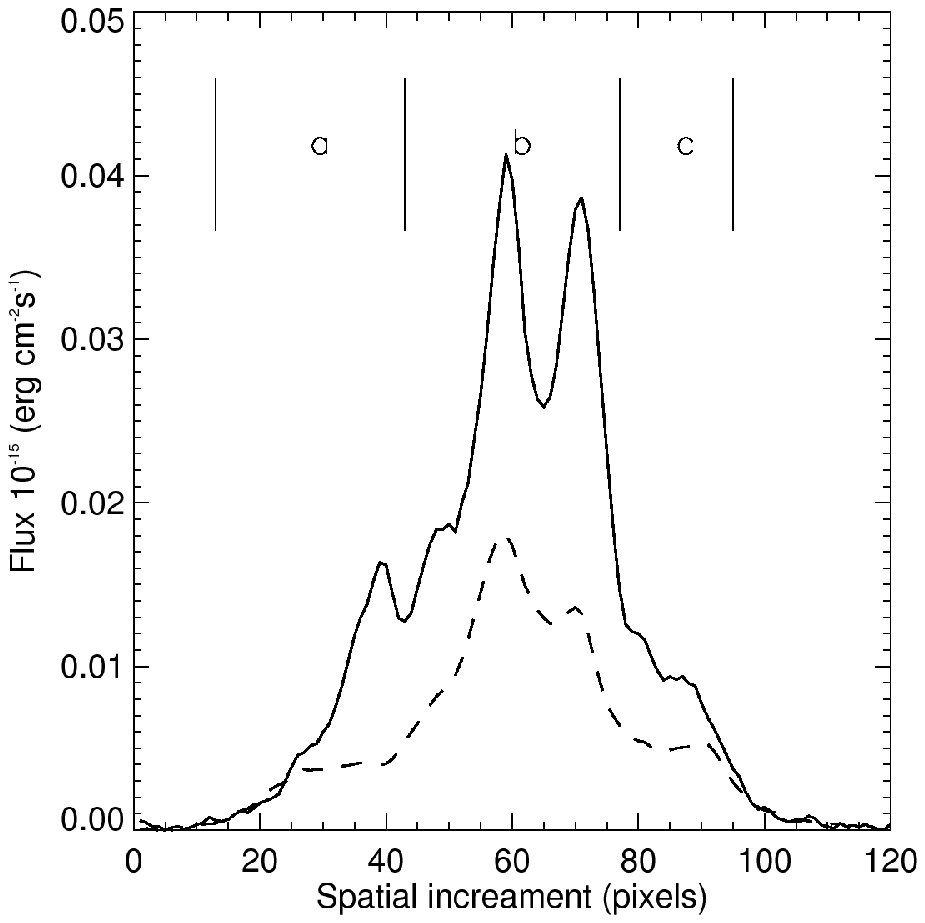}
\includegraphics[height=5cm,angle=0]{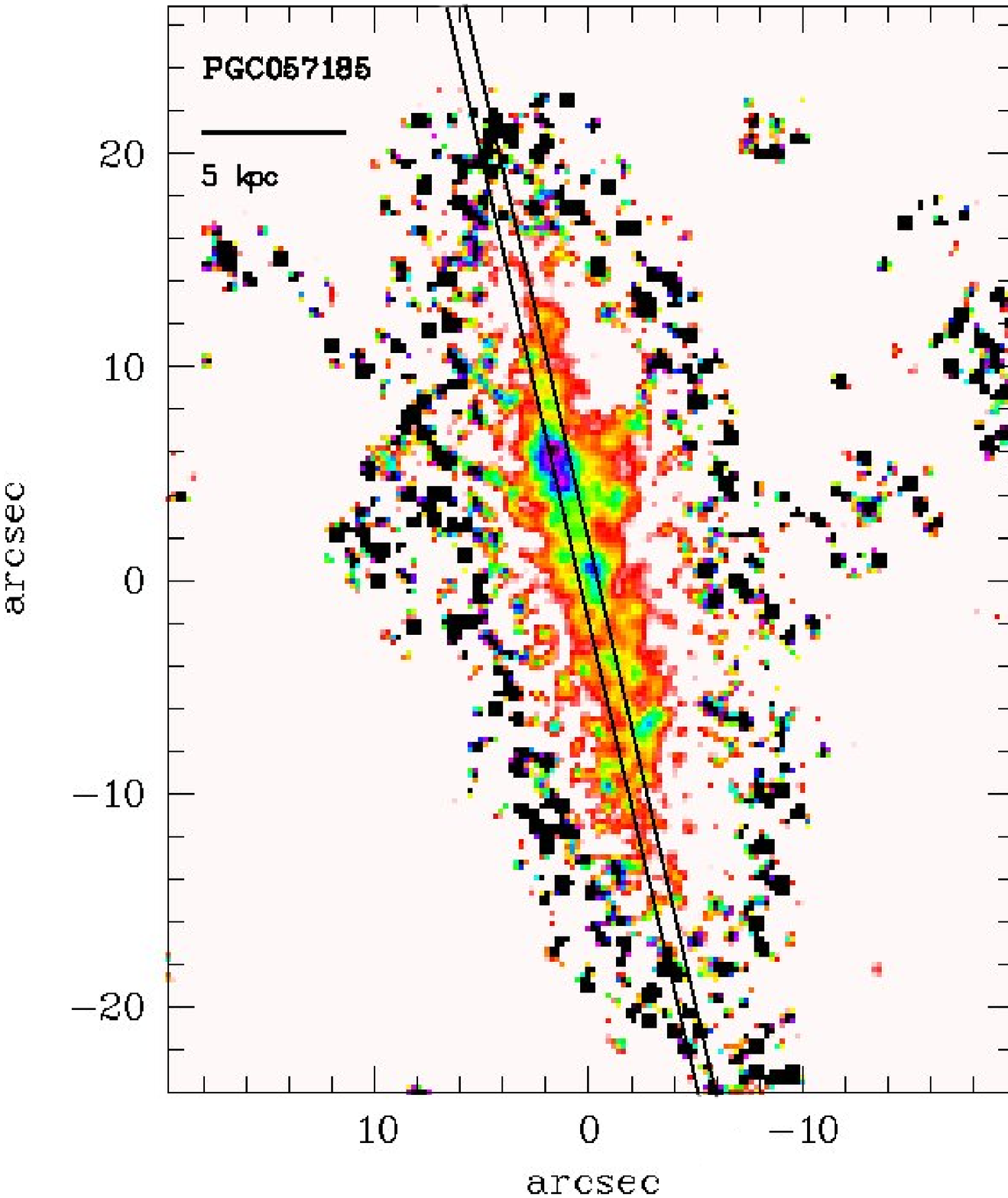}
\includegraphics[height=5cm,angle=0]{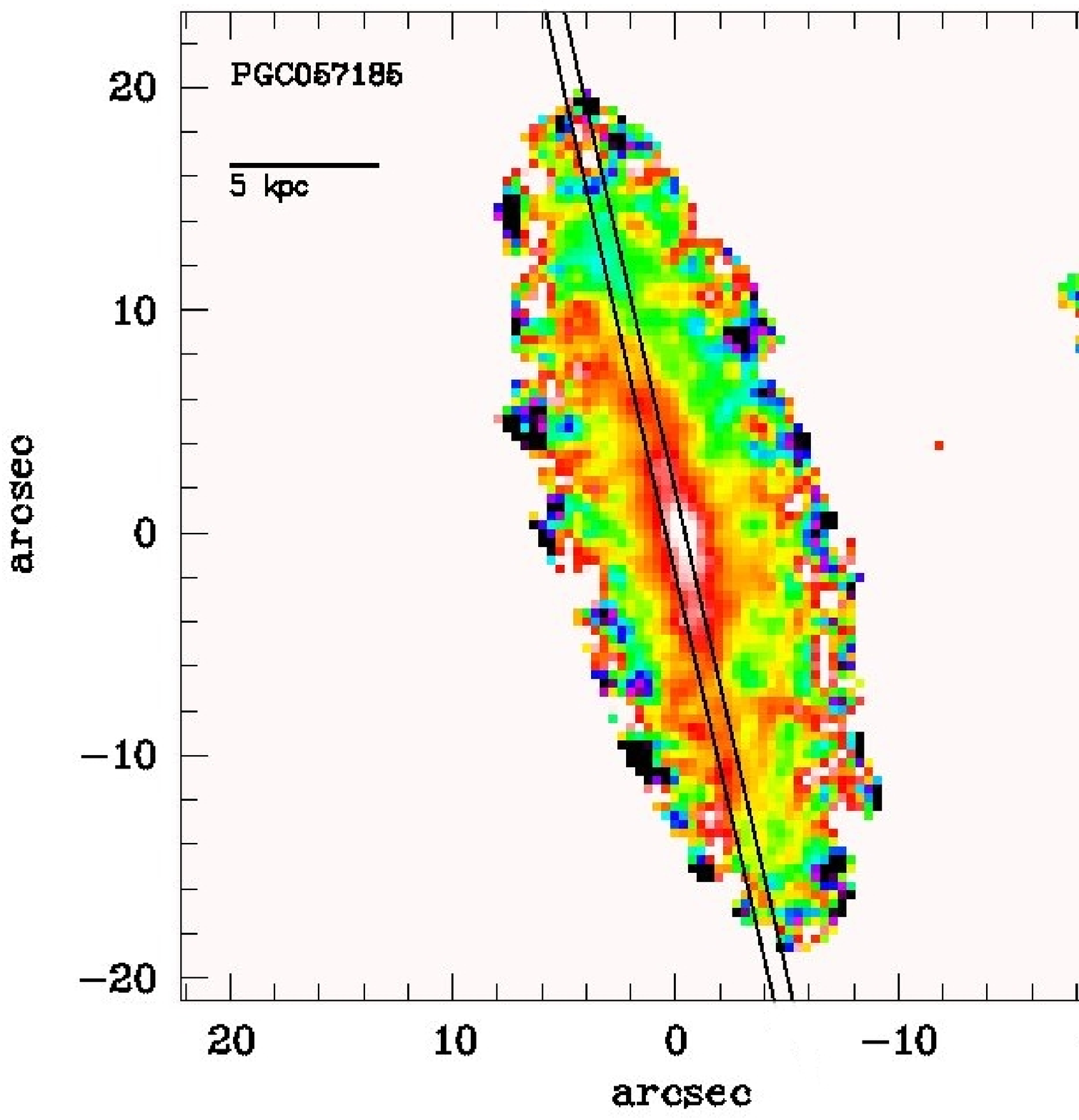}

\includegraphics[height=5cm,angle=0]{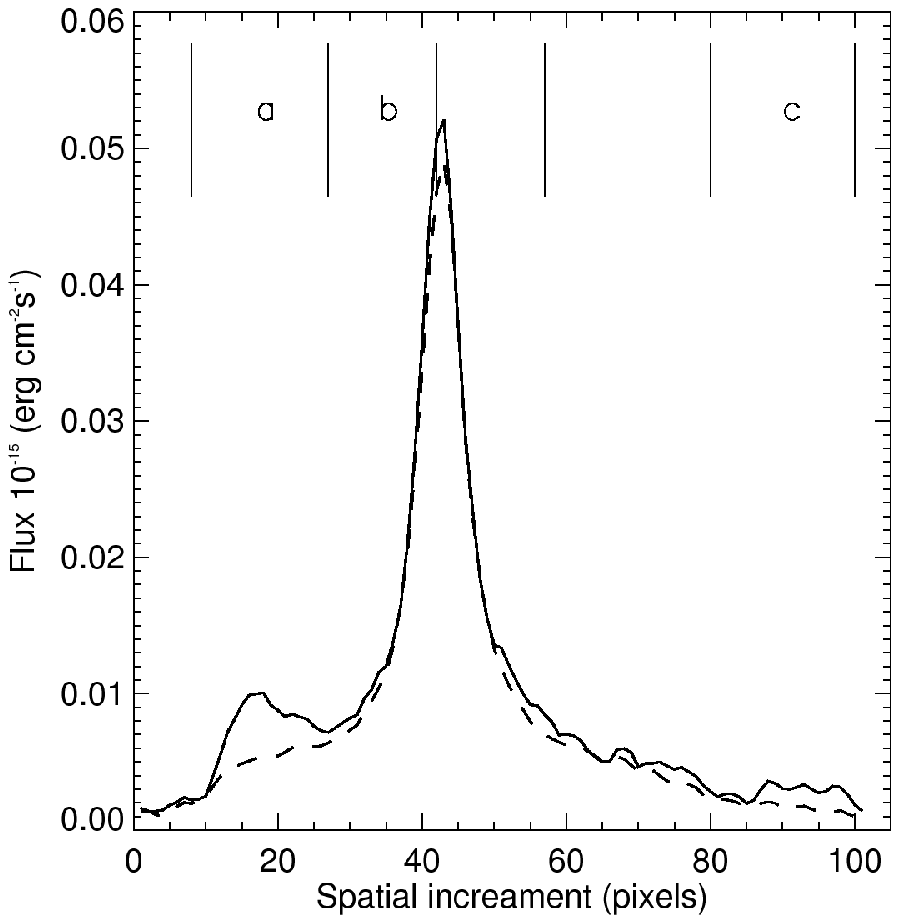}
\includegraphics[height=4.5cm,angle=0]{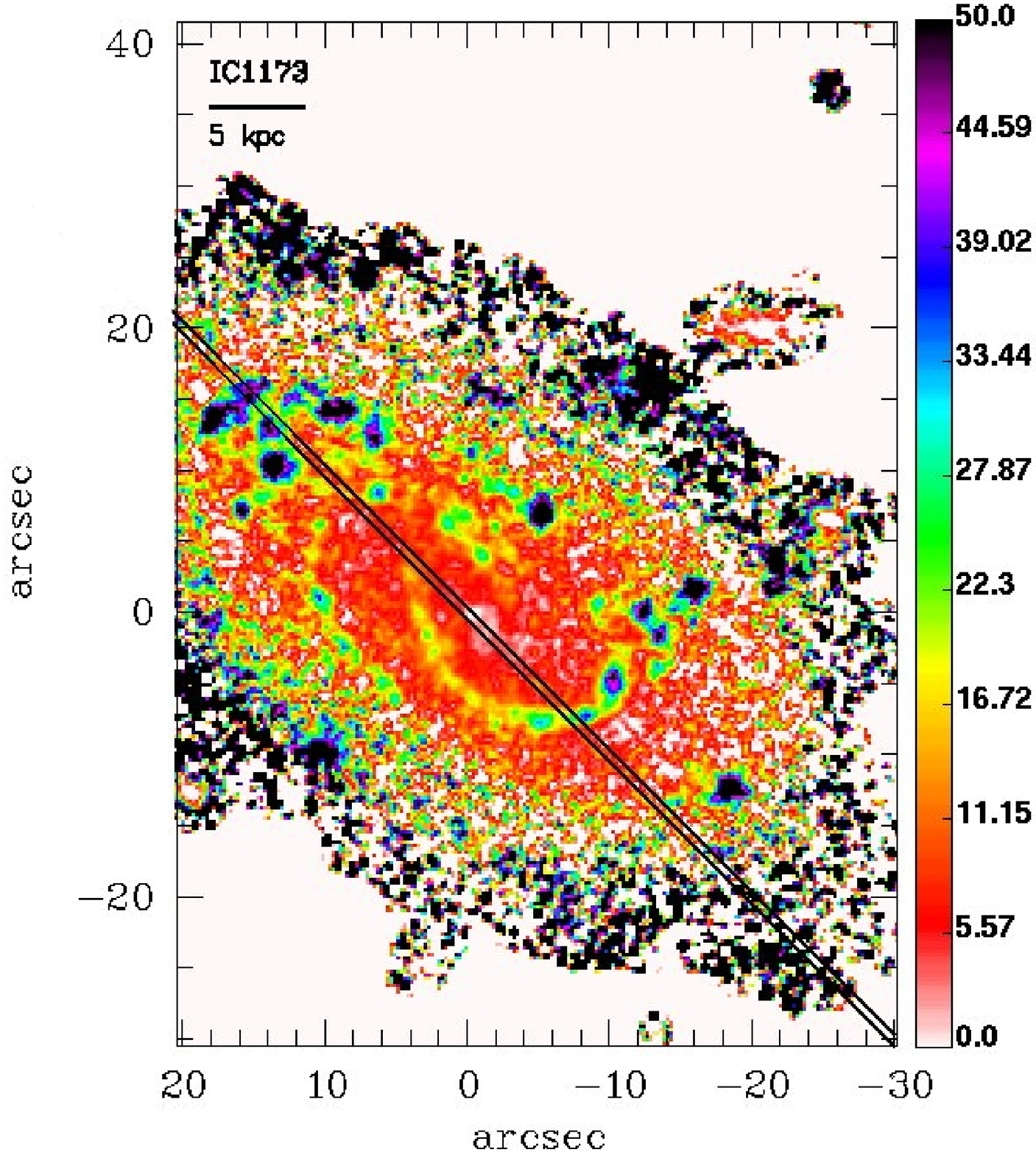}
\includegraphics[height=5cm,angle=0]{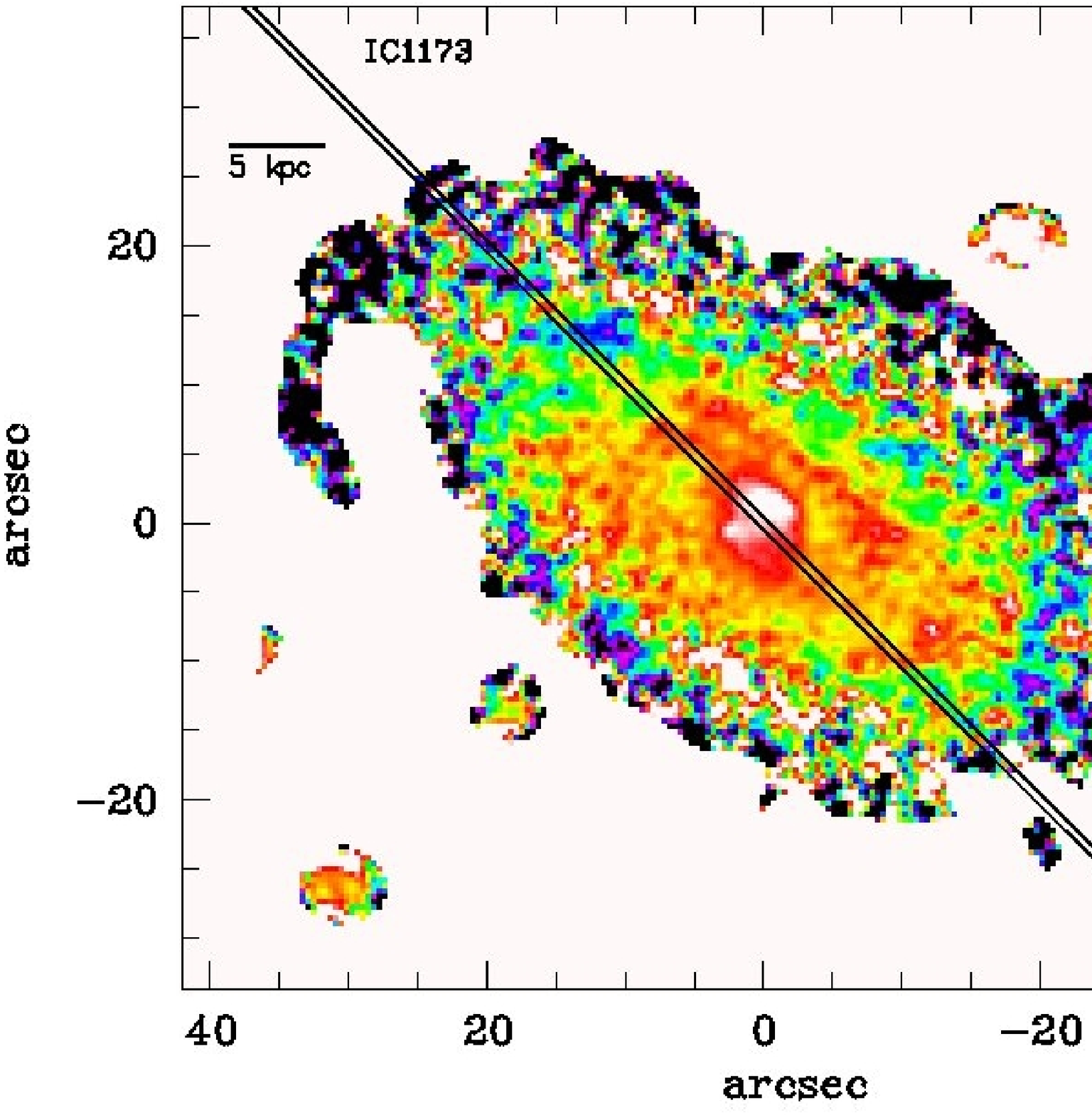}

\includegraphics[height=5cm,angle=0]{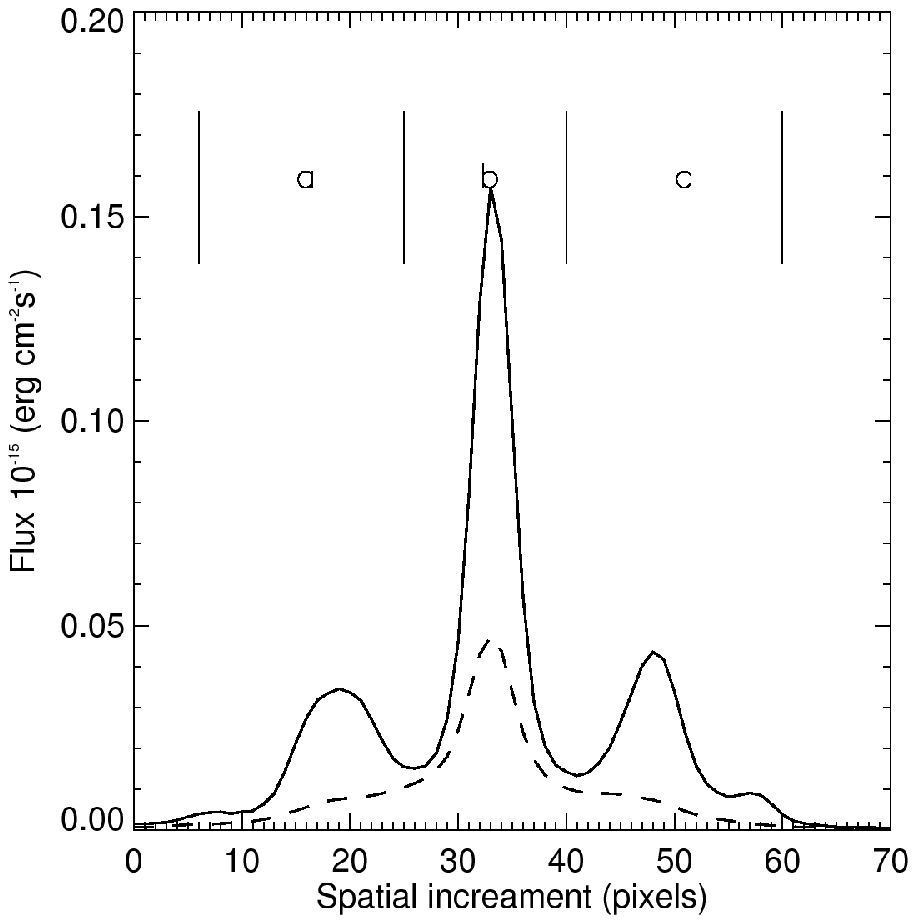}
\includegraphics[height=5cm,angle=0]{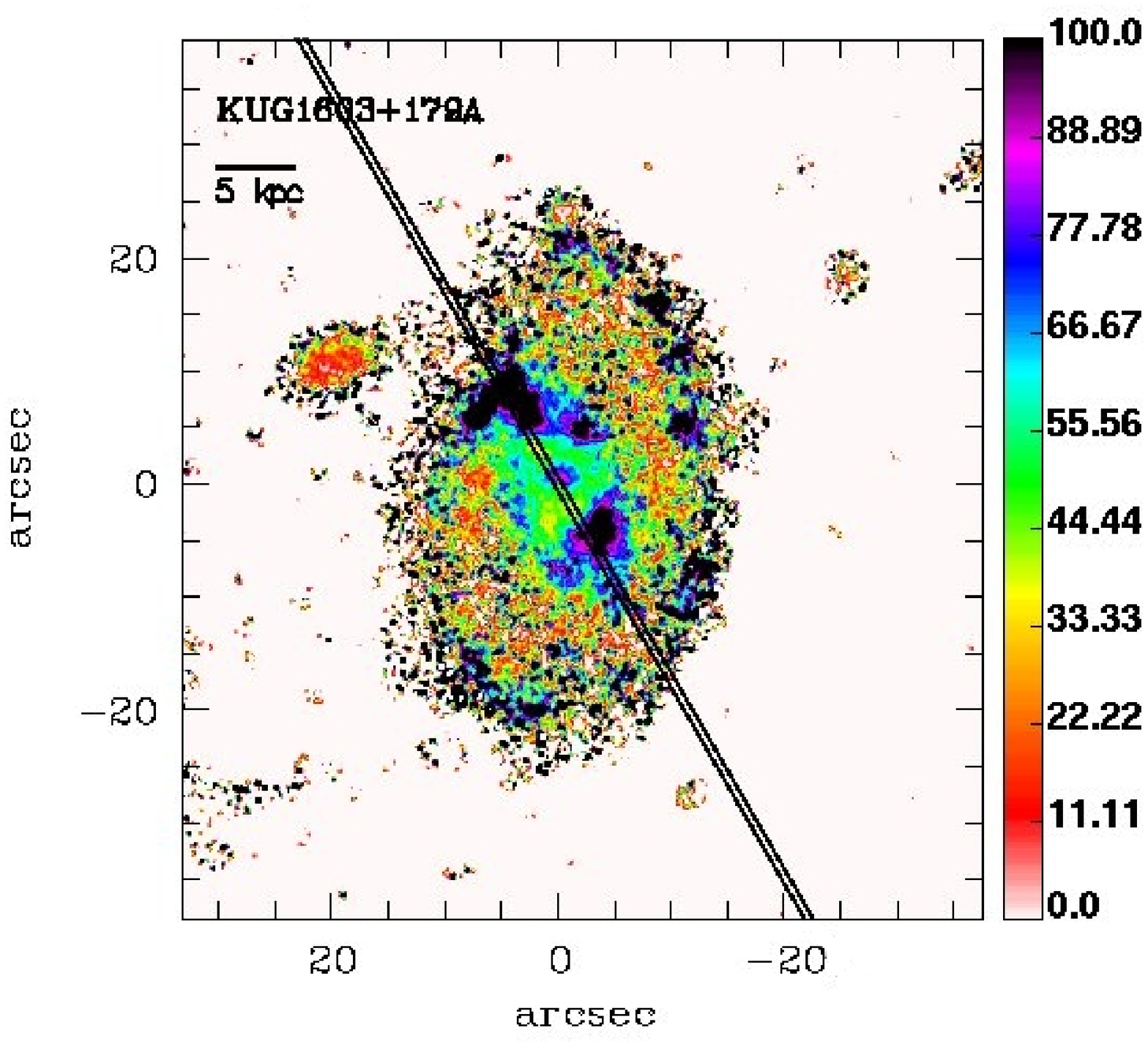}
\includegraphics[height=5cm,angle=0]{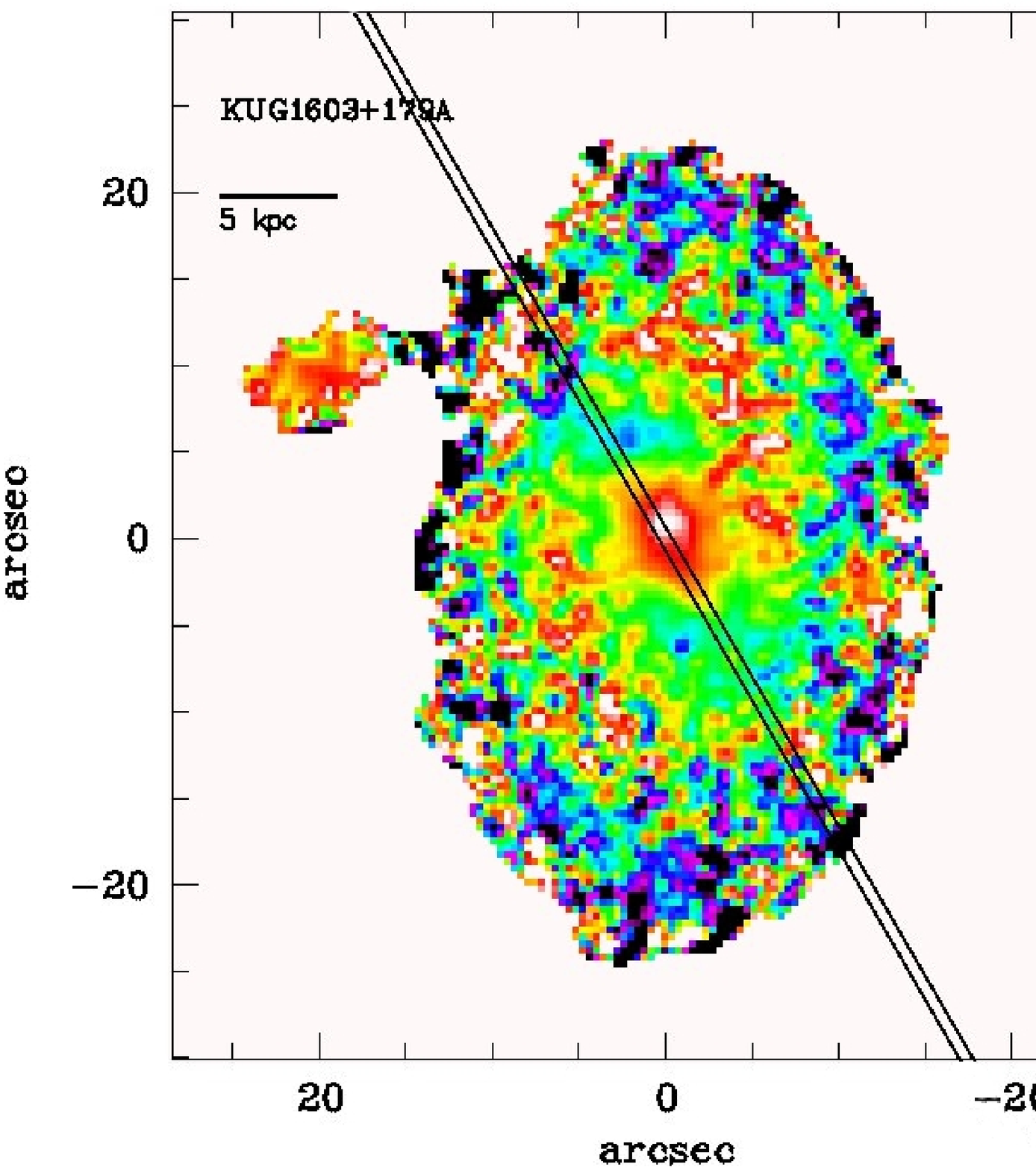}
\caption{The first column shows the spatial profiles of the \Ha line (continuous) and nearby continuum (dashed line) along the slit extracted from the 2D spectra,  for the galaxies showing rich spatial structure. The different sub-regions used to divide their 2D spectra into different 1D spectra are shown. The second and third column show respectively the color (g'-i') maps and the \Ha EW maps, with the slit position overploted.\label{SLITS}}
\end{figure*}

\addtocounter{figure}{-1}
\begin{figure*}
\includegraphics[height=4.9cm,angle=0]{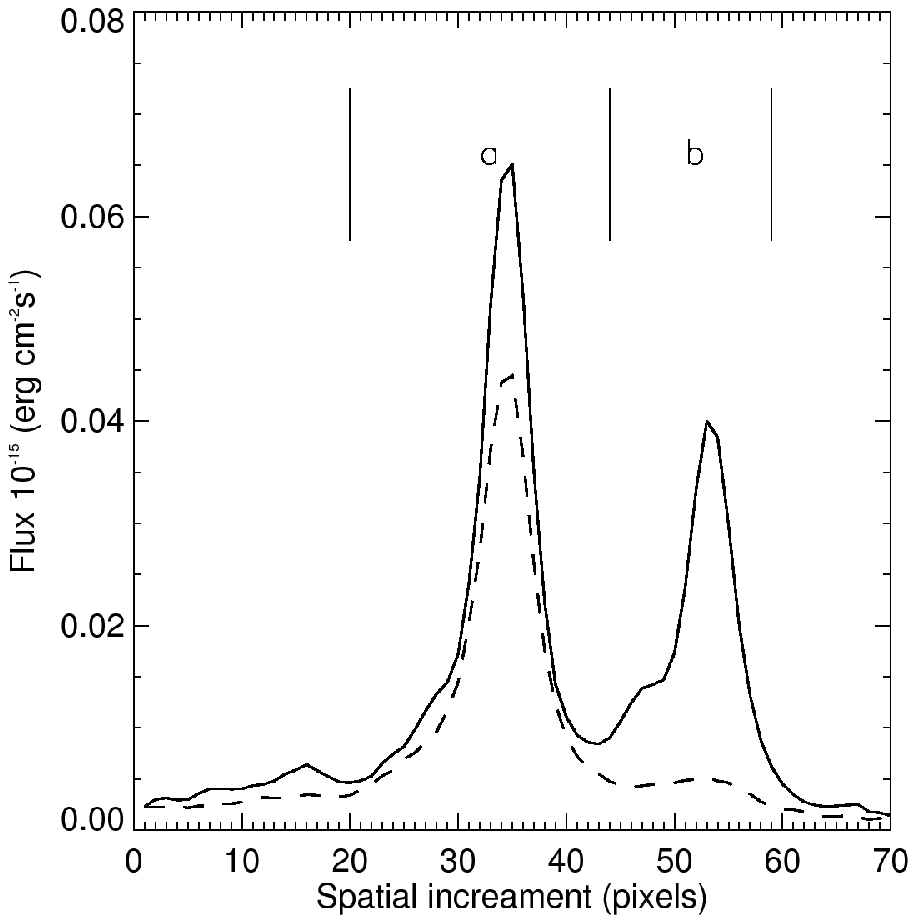}
\includegraphics[height=4.9cm,angle=0]{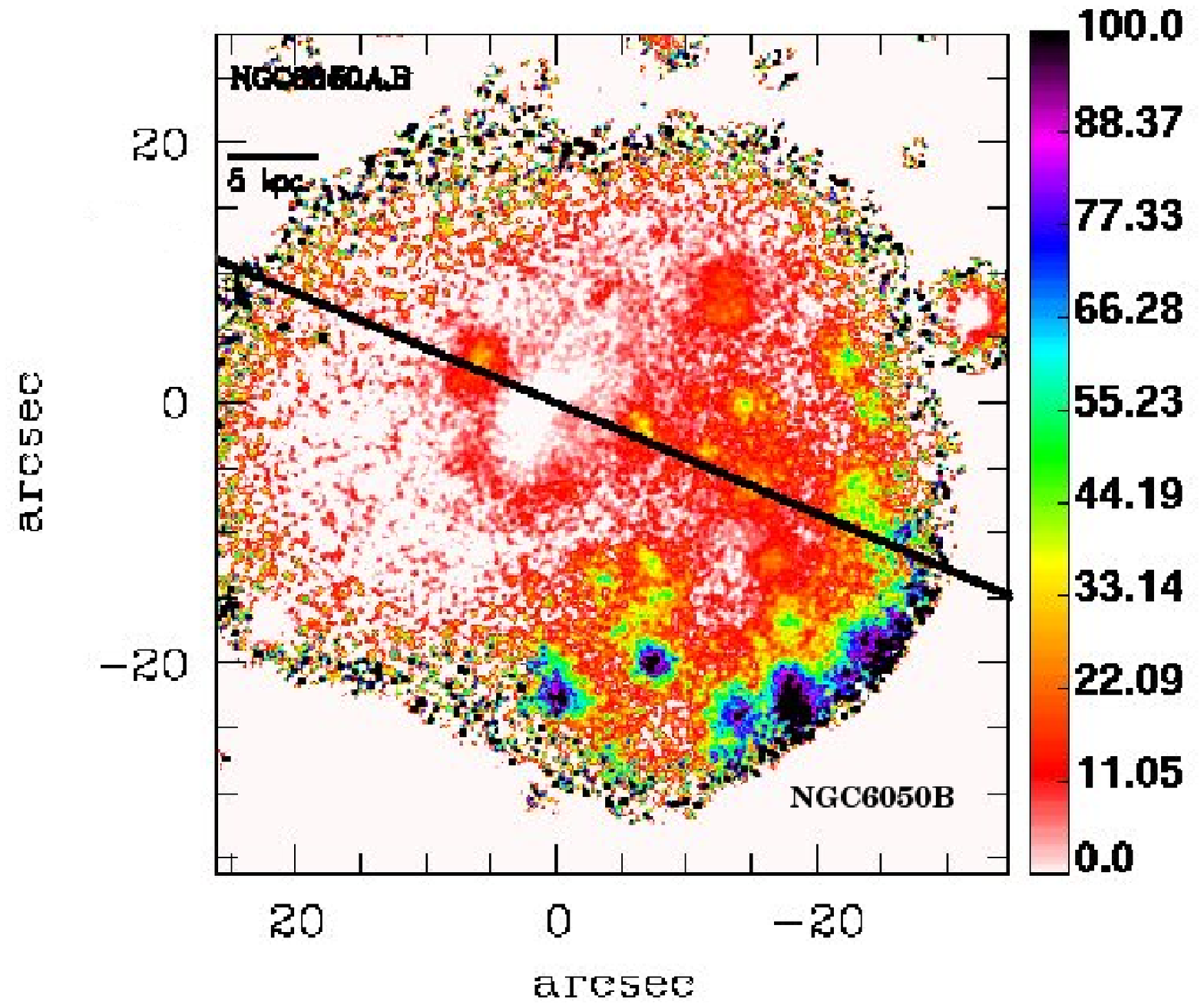}
\includegraphics[height=4.9cm,angle=0]{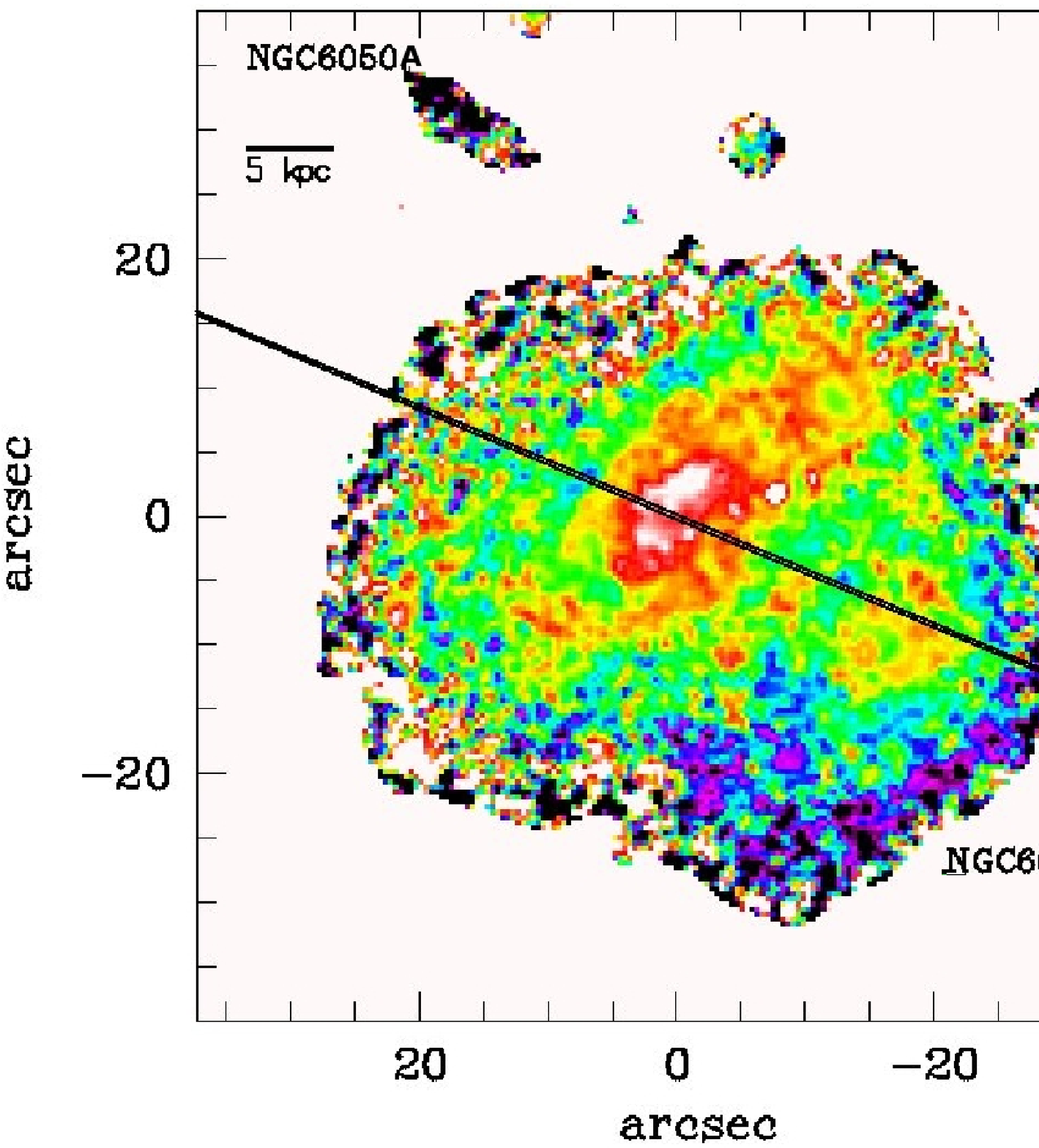}

\includegraphics[height=5cm,angle=0]{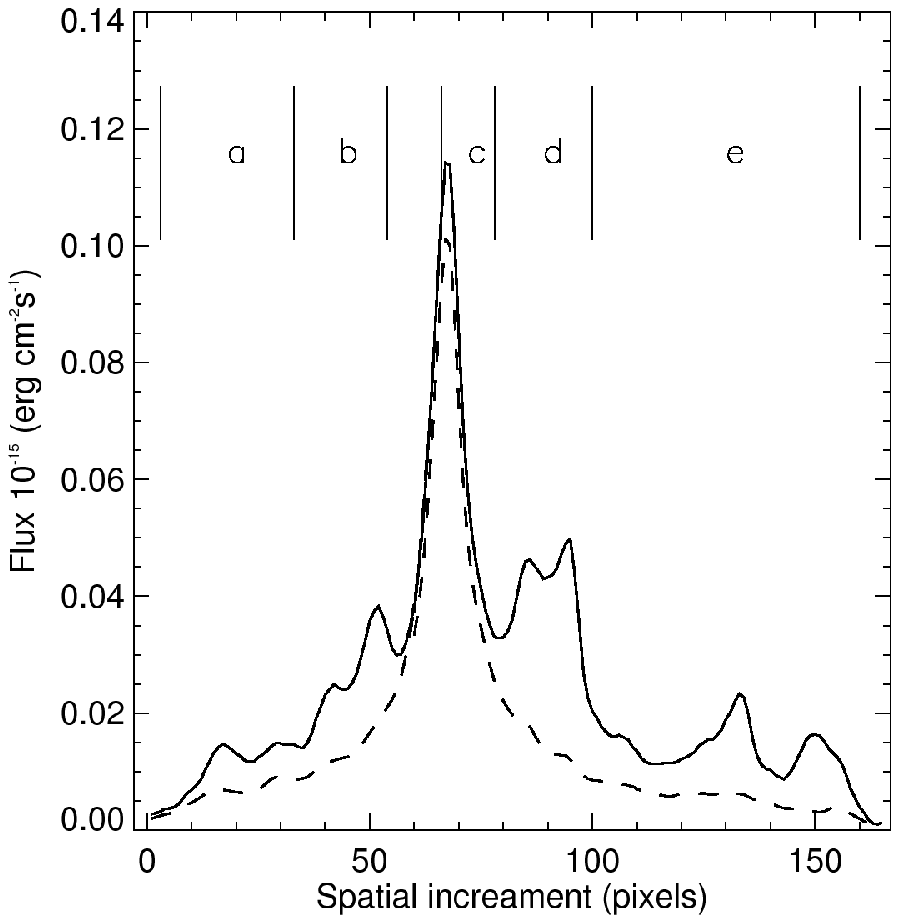}
\includegraphics[height=4.5cm,angle=0]{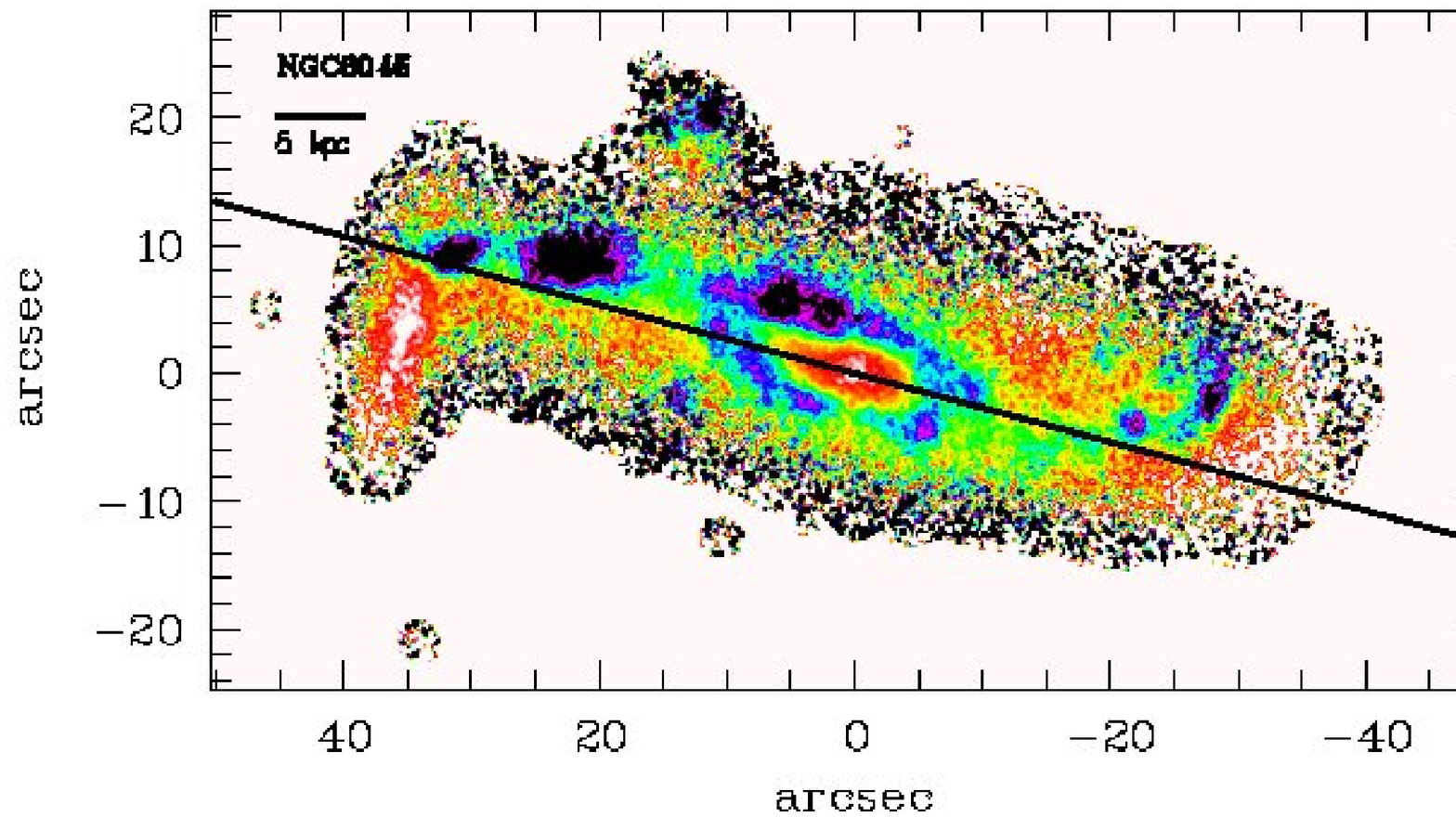}
\includegraphics[height=4.5cm,angle=0]{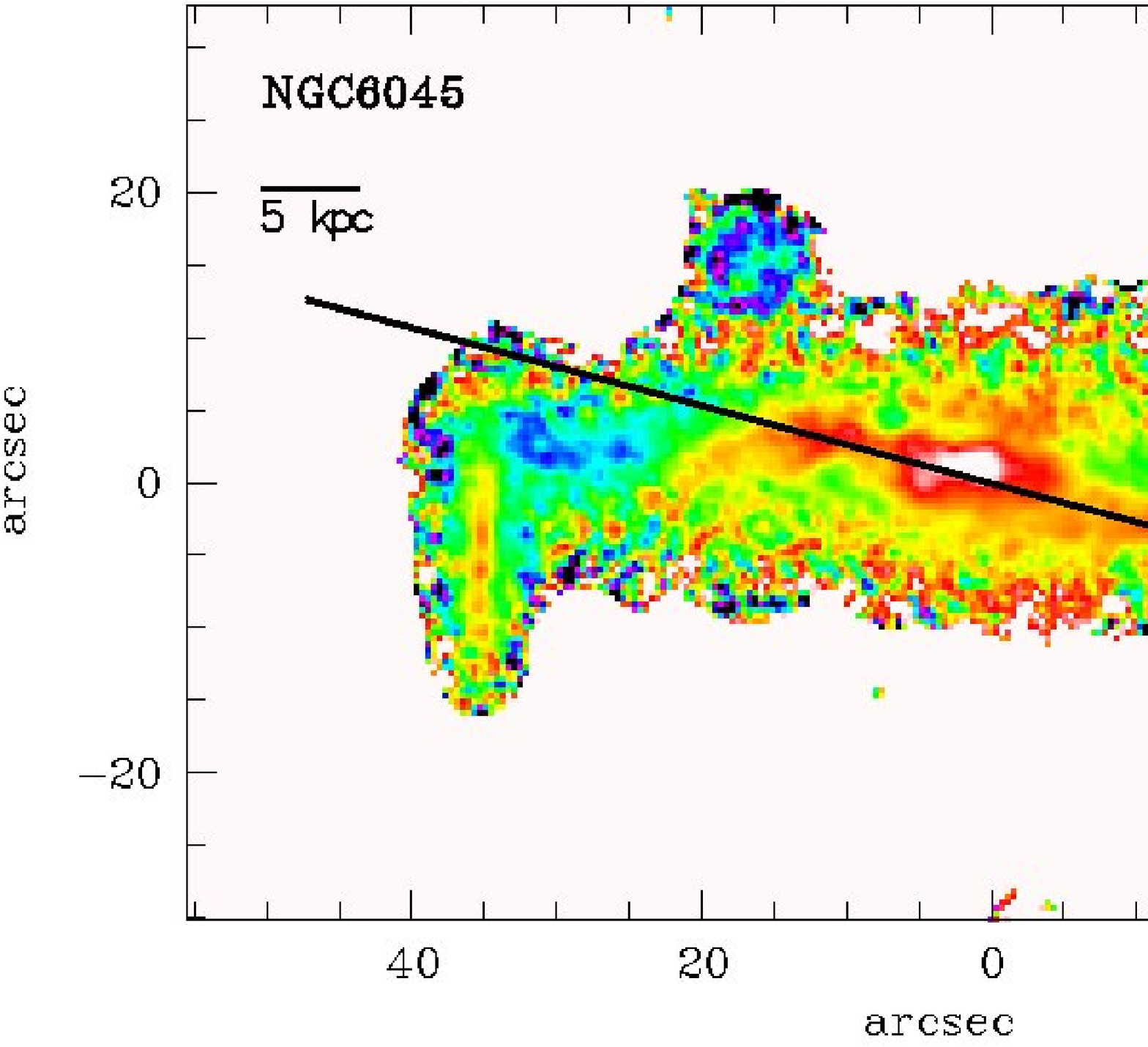}

\includegraphics[height=4.9cm,angle=0]{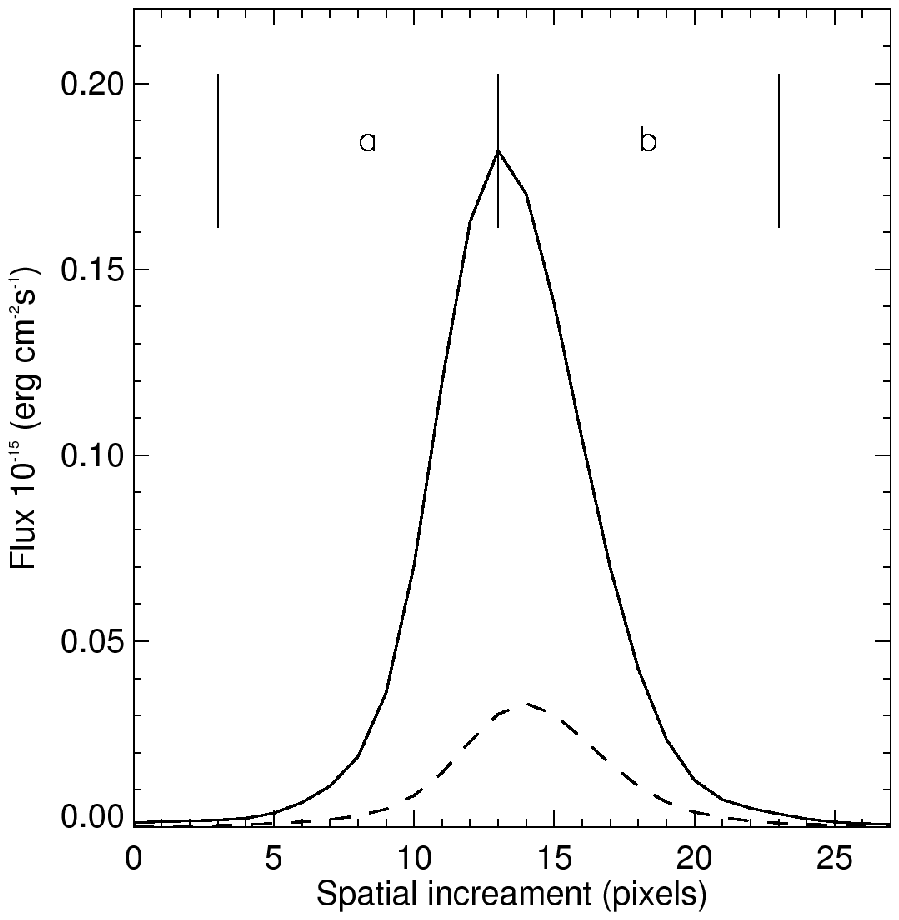}
\includegraphics[height=4.9cm,angle=0]{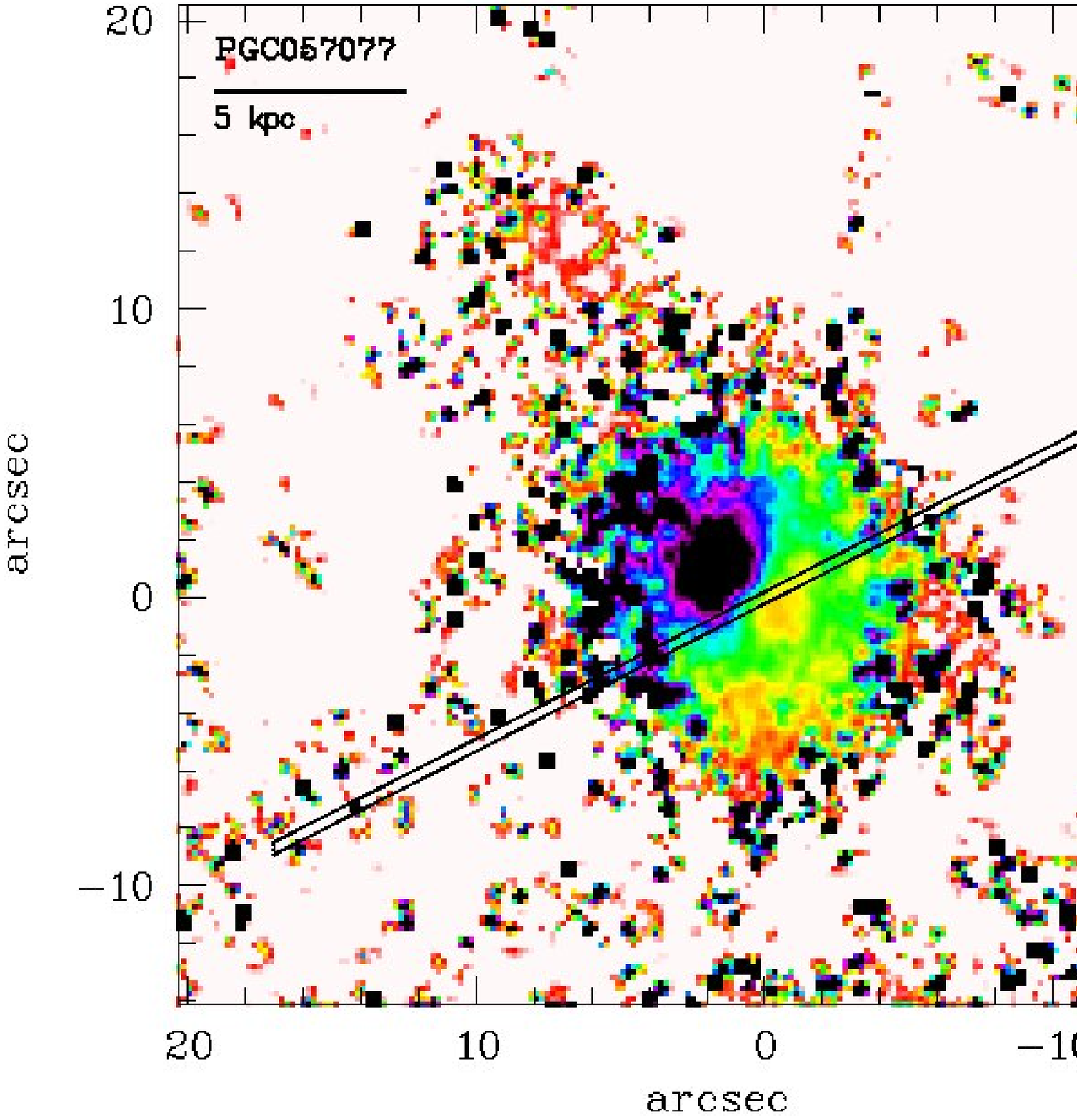}
\includegraphics[height=4.9cm,angle=0]{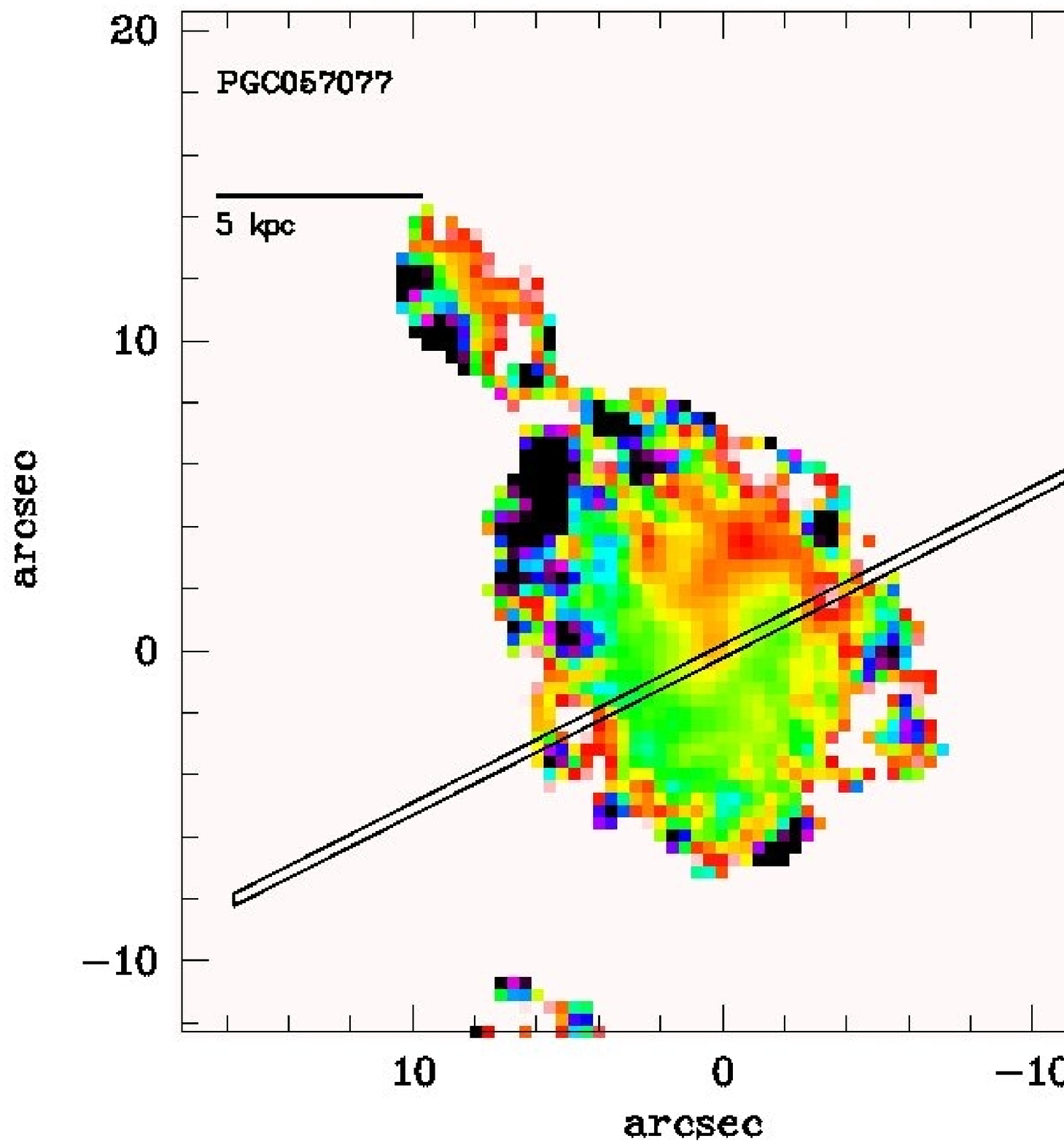}
\caption{Continued.\label{}}
\end{figure*}

\addtocounter{figure}{-1}
\begin{figure*}
\includegraphics[height=5cm,angle=0]{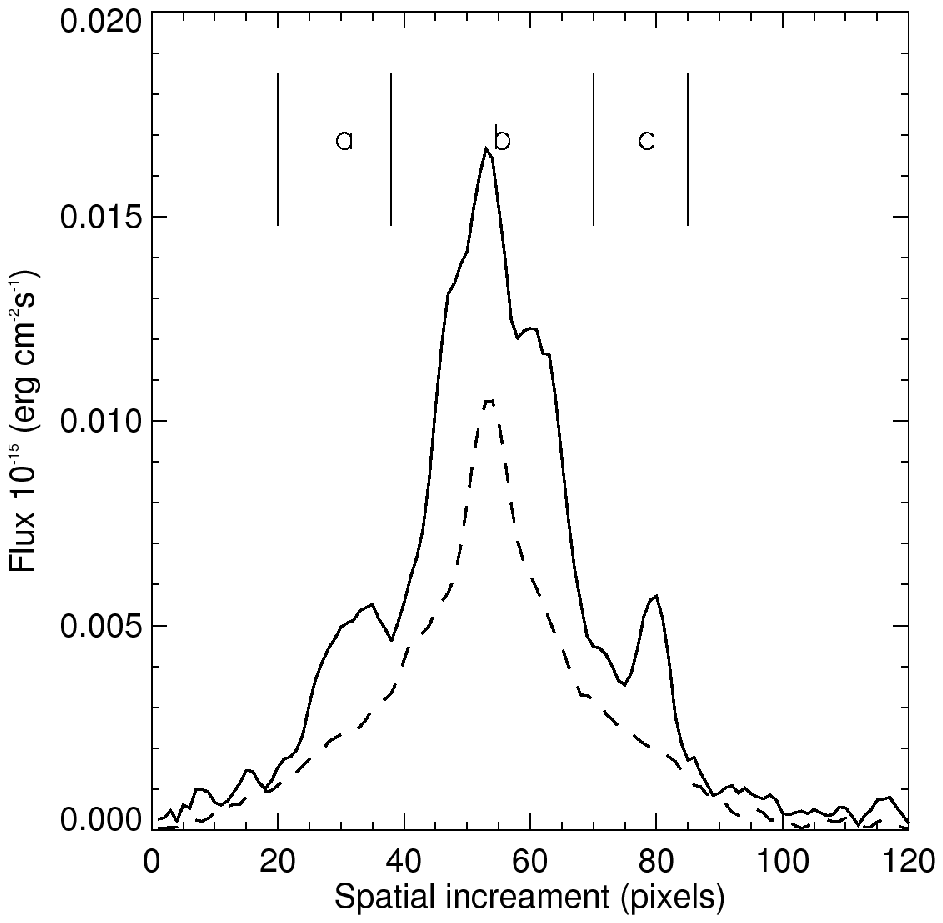}
\includegraphics[height=5cm,angle=0]{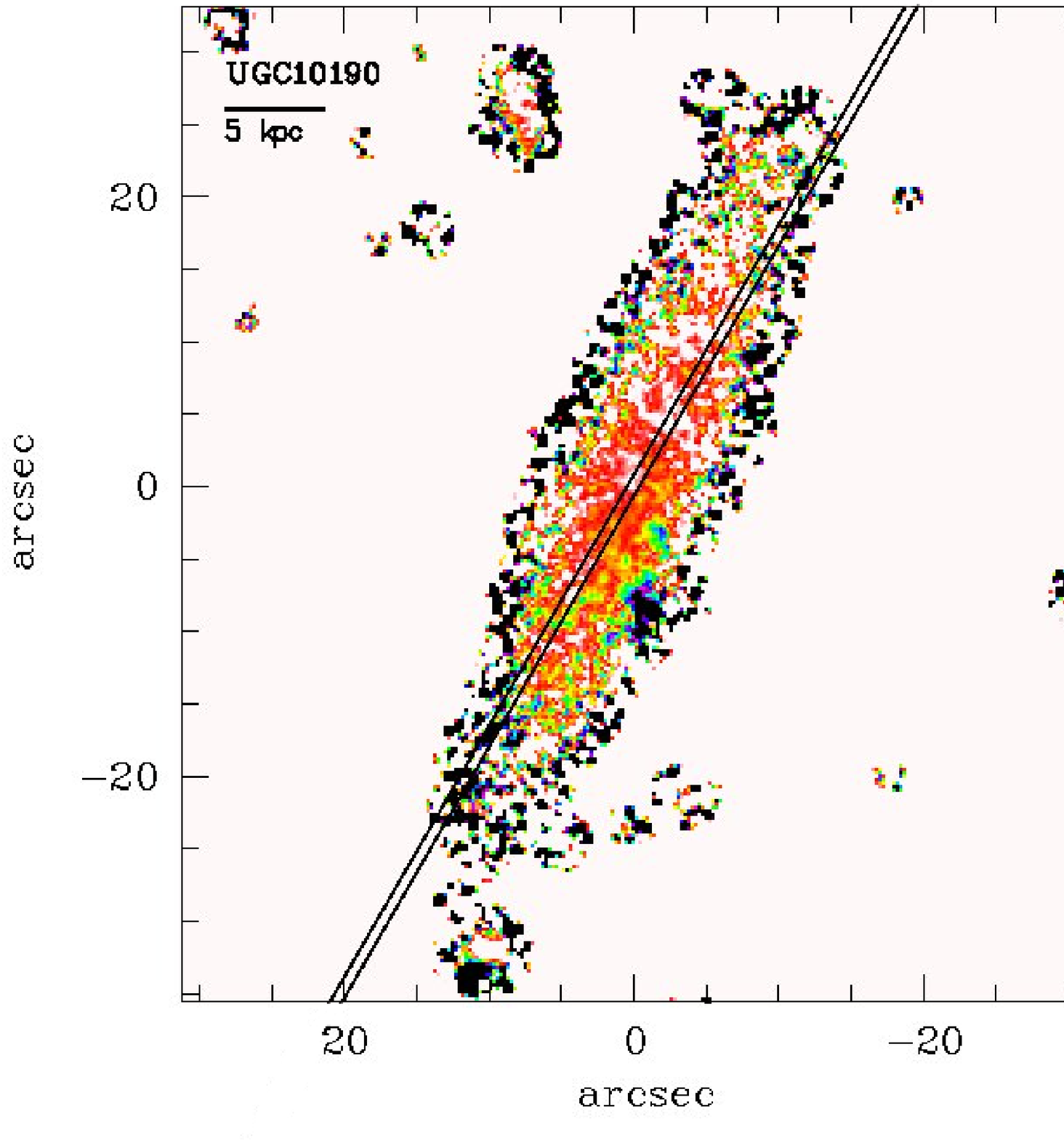}
\includegraphics[height=5cm,angle=0]{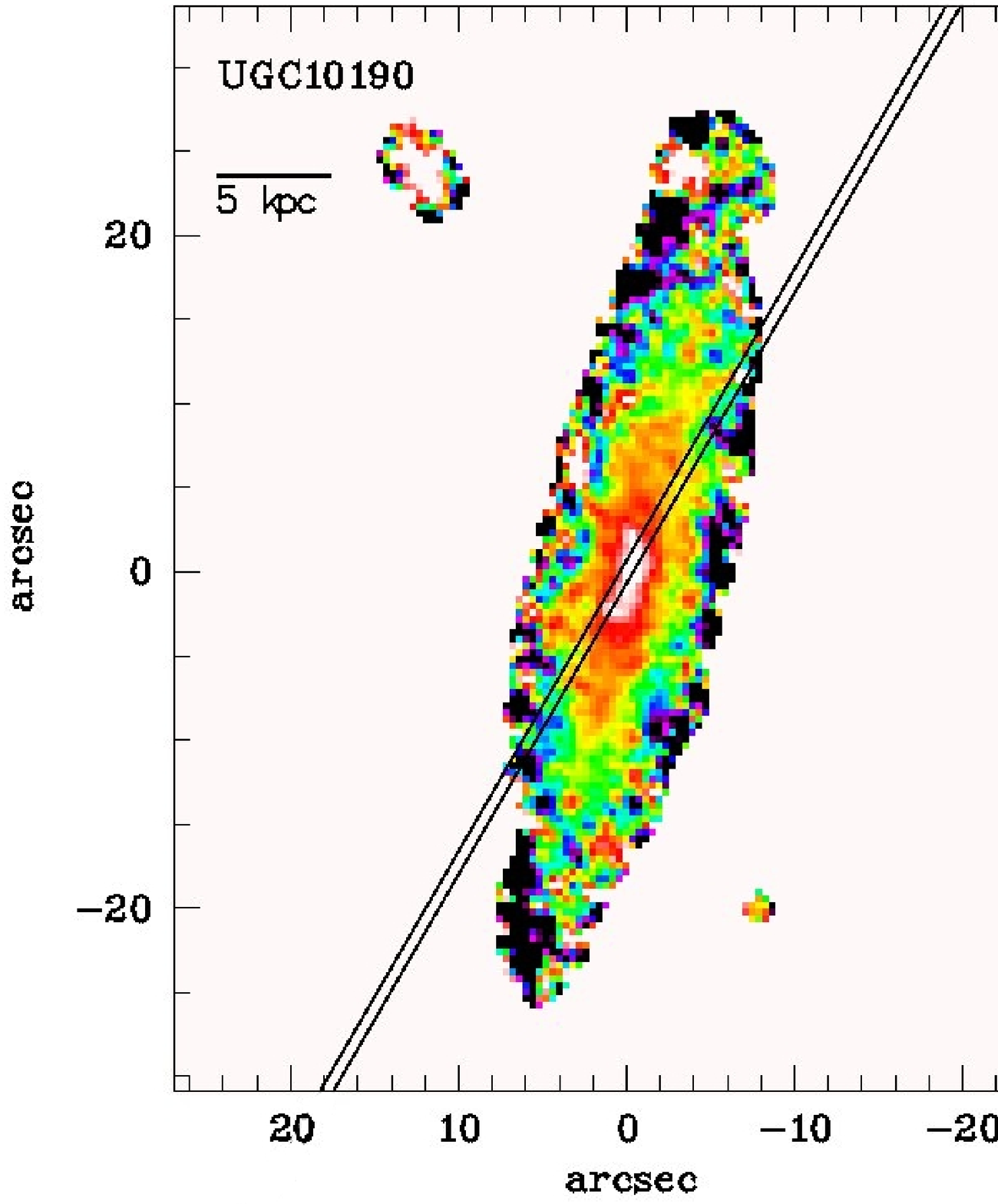}

\includegraphics[height=5cm,angle=0]{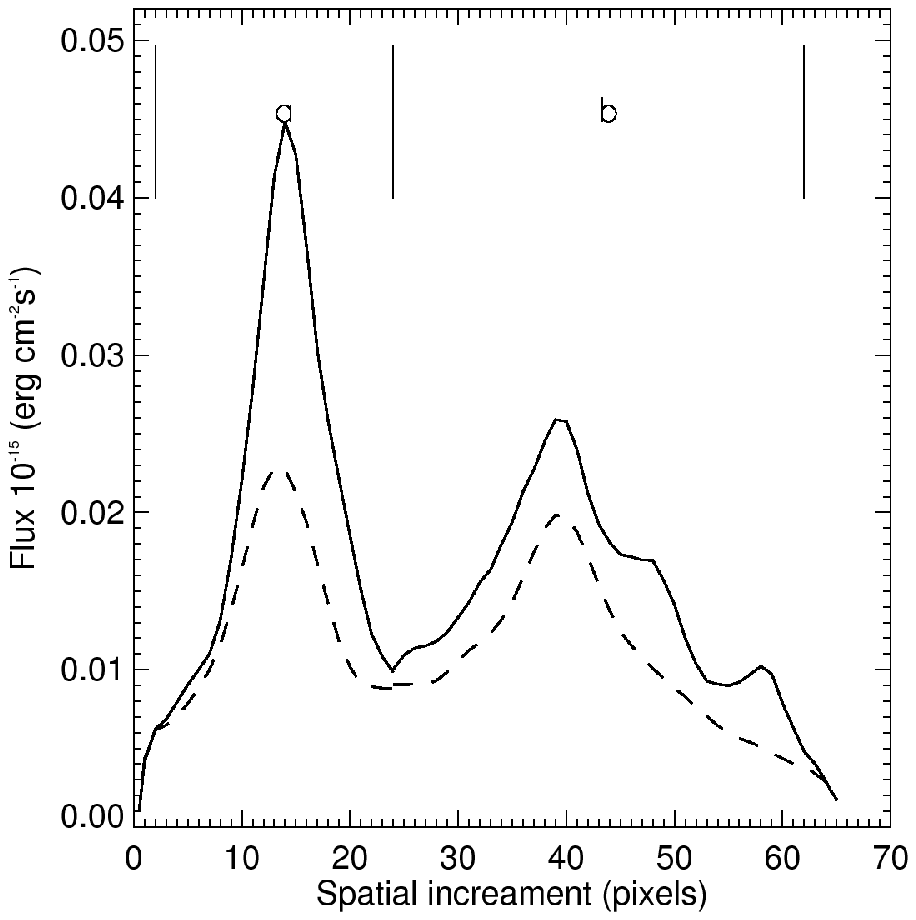}
\includegraphics[height=4.2cm,angle=0]{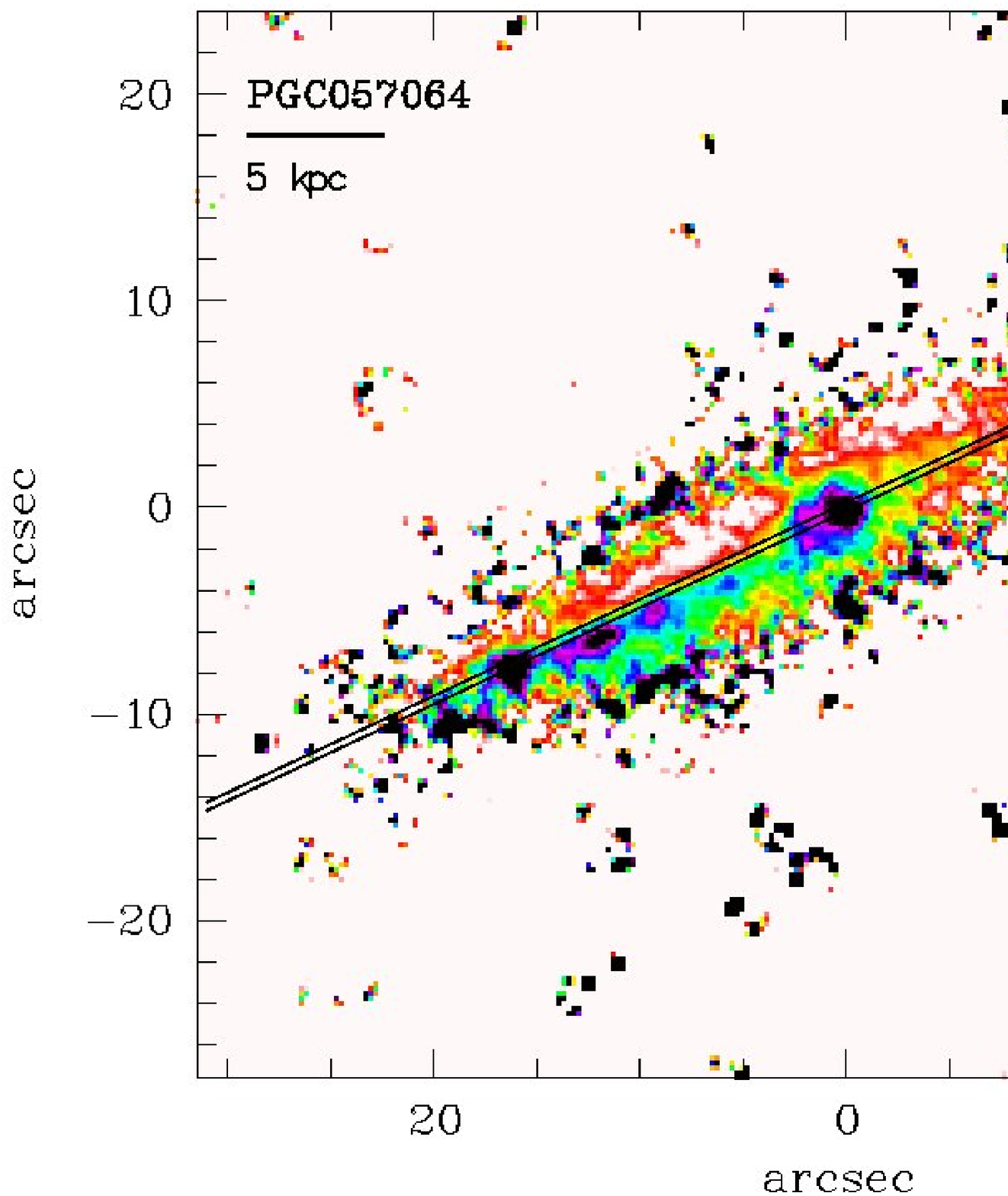}
\includegraphics[height=4.2cm,angle=0]{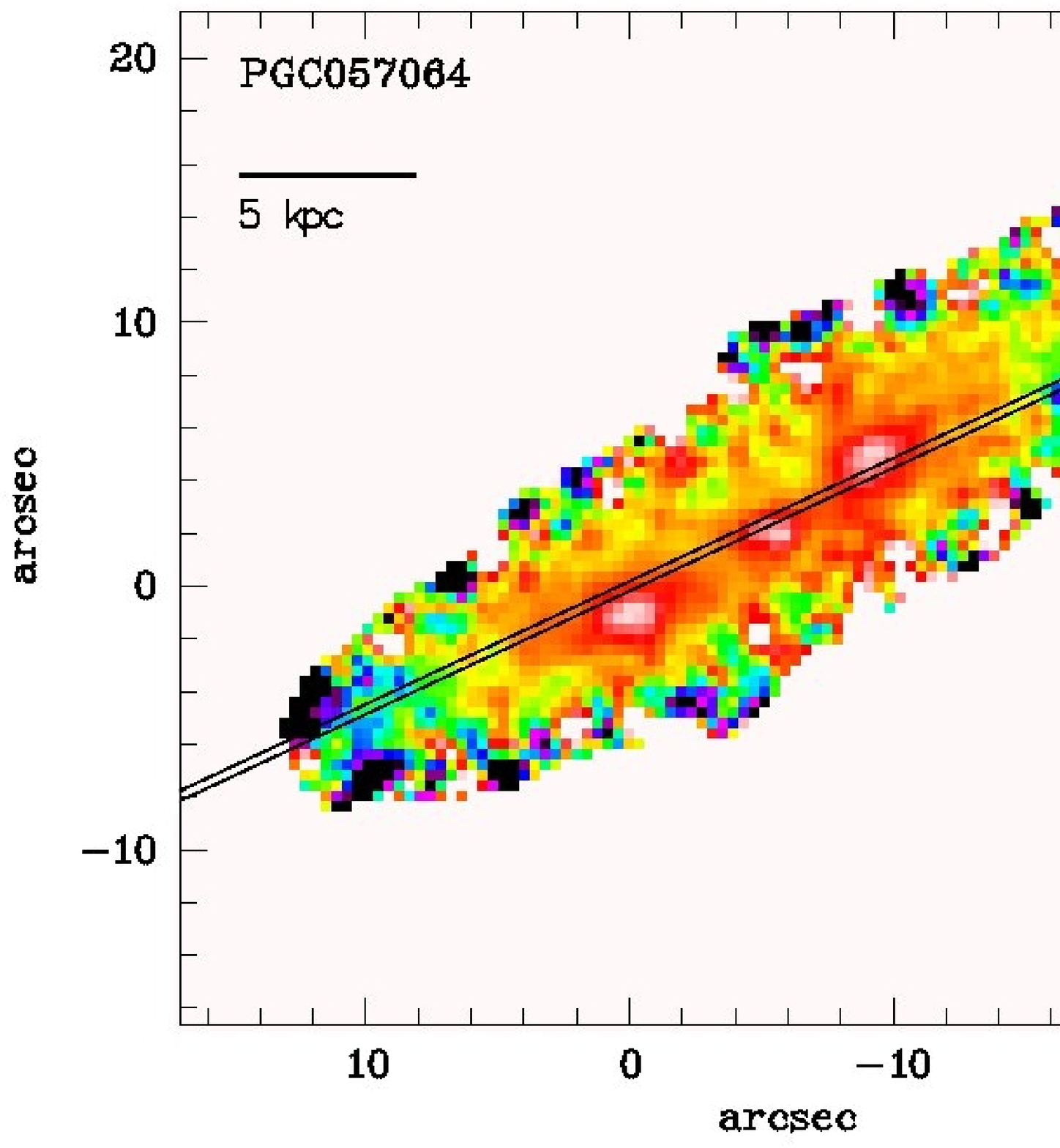}

\includegraphics[height=5cm,width=5cm,angle=0]{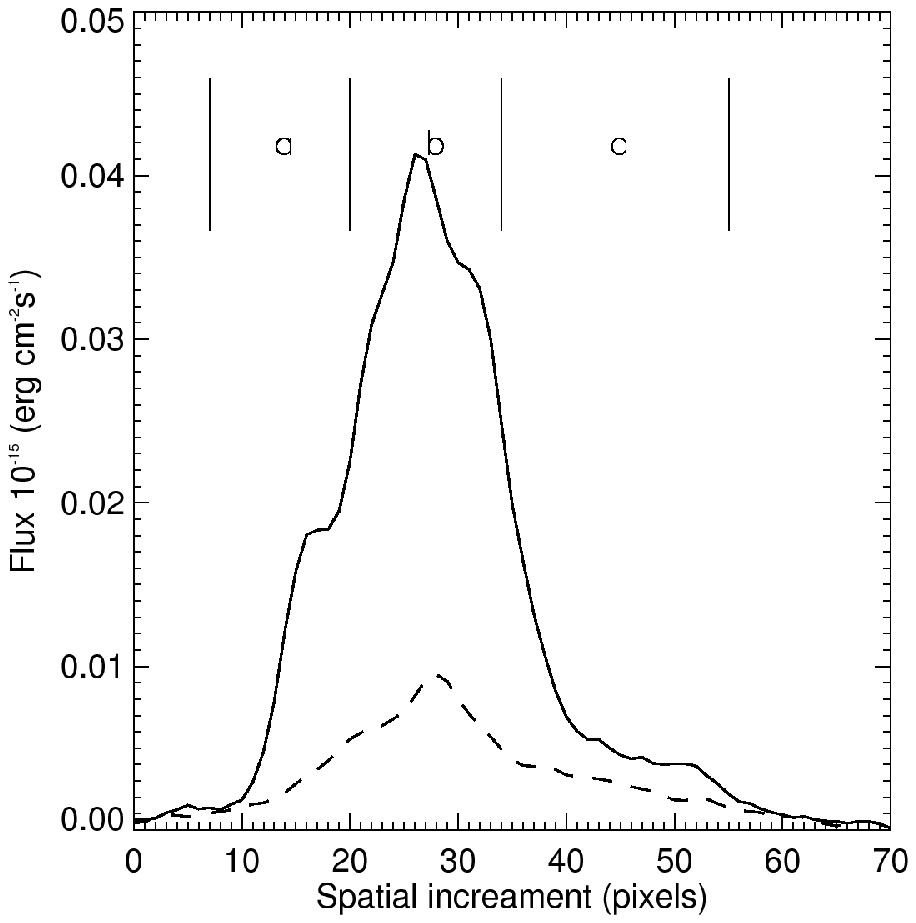}
\includegraphics[height=3.9cm,angle=0]{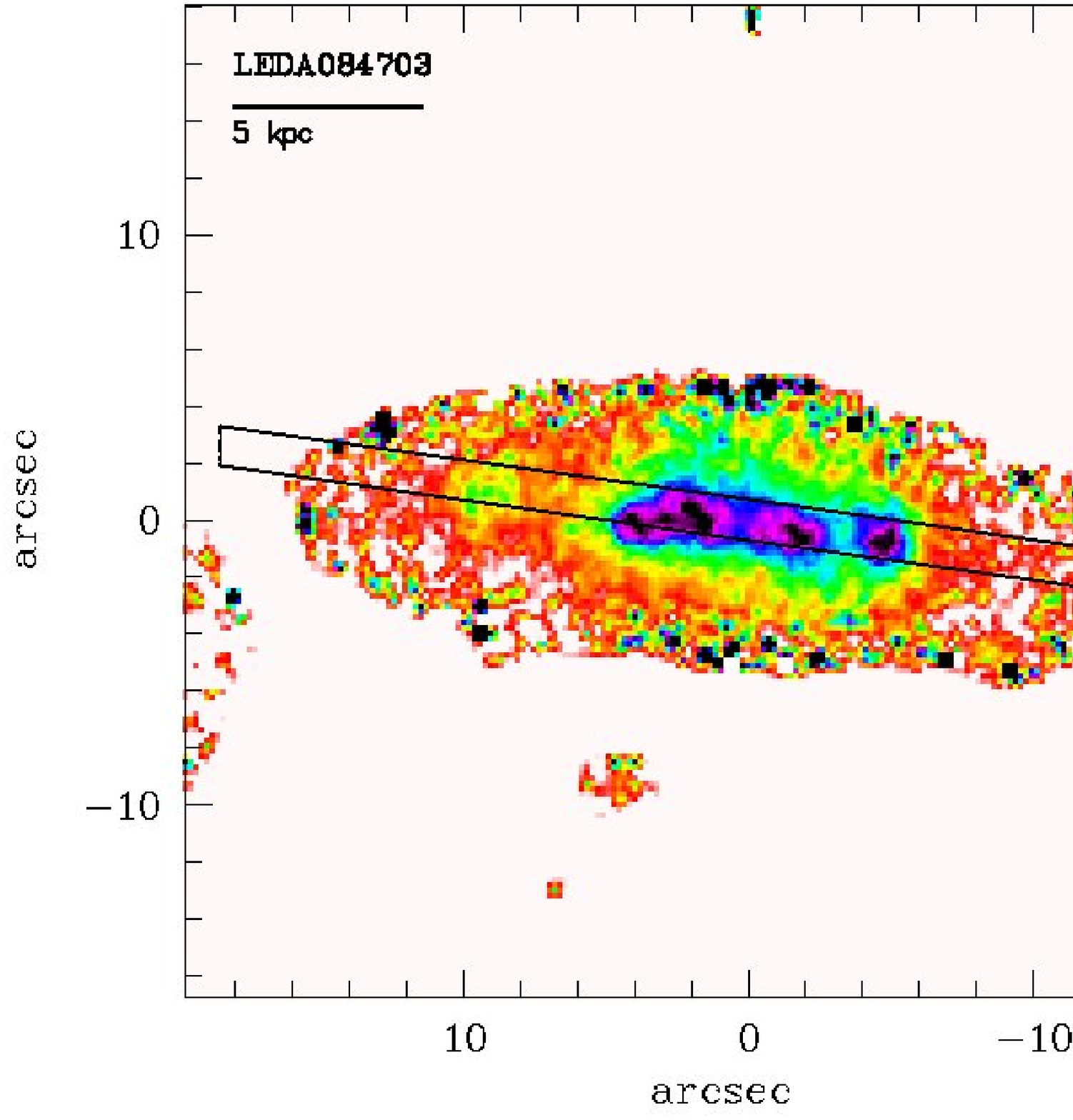}
\includegraphics[height=3.9cm,angle=0]{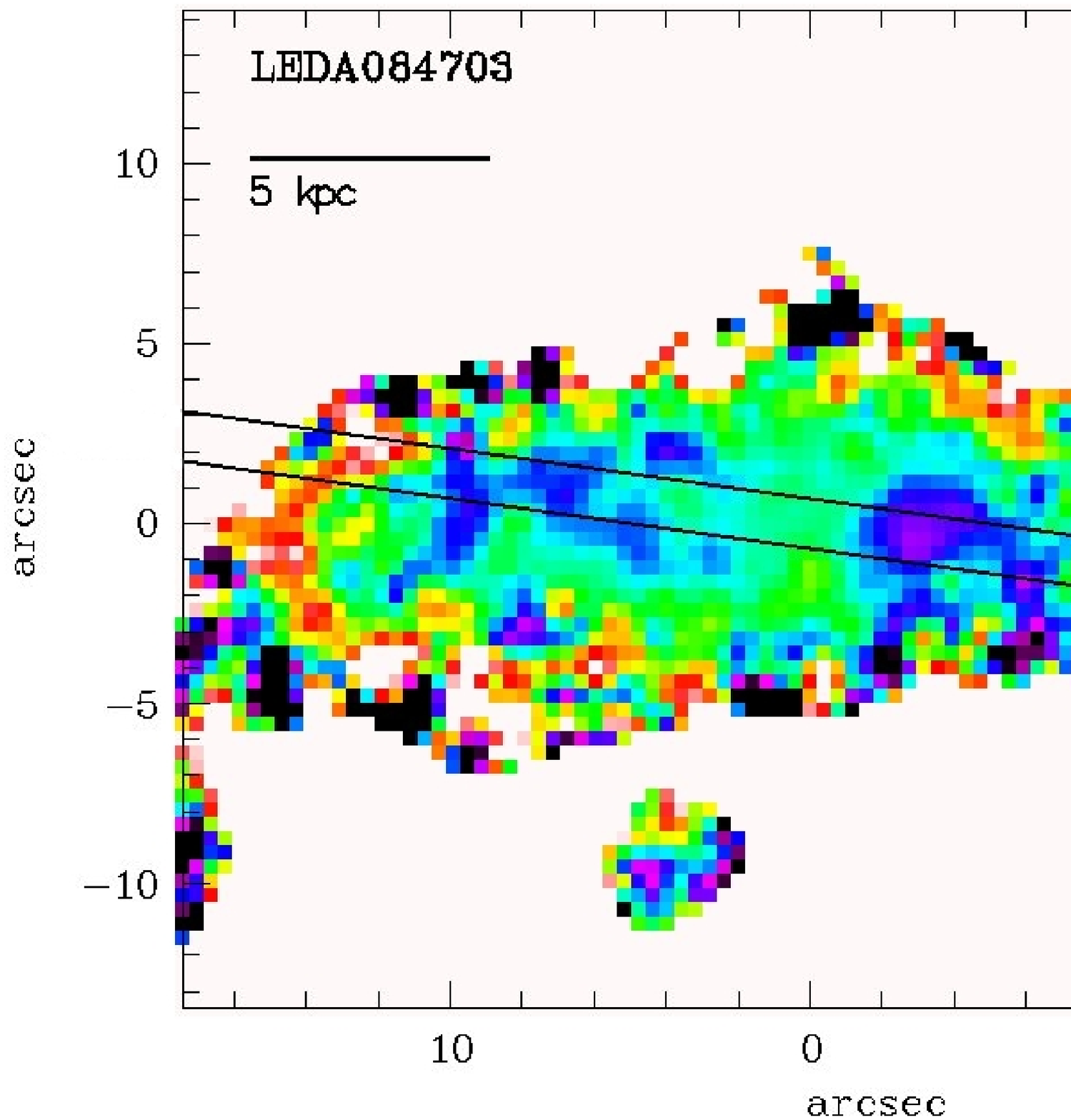}
\caption{Continued.\label{}}
\end{figure*}


\end{document}